# Interfacial piezoelectric polarization locking in printable Ti$_3$C$_2$T$_x$ MXene-fluoropolymer composites

Nick A. Shepelin,[ab] Peter C. Sherrell,[ab] Emmanuel N. Skountzos,[cd] Eirini Goudeli,[a] Jizhen Zhang,[e] Vanessa C. Lussini,[f] Beenish Imtiaz,[a] Ken Aldren S. Usman,[e] Greg W. Dicinoski,[f] Joseph G. Shapter,[g] Joselito M. Razal,[e] Amanda V. Ellis *[ab]

[a]Department of Chemical Engineering, The University of Melbourne, Parkville, Victoria 3010, Australia

[b]BioFab3D, Aikenhead Centre for Medical Discovery, St Vincent's Hospital Melbourne, Fitzroy, Victoria 3065, Australia

[c]Department of Chemical Engineering, University of Patras, Greece

[d]FORTH/ICE-HT, Patras, GR 26504, Greece

[e]Institute for Frontier Materials, Deakin University, Geelong, Victoria 3216, Australia

[f]Note Issue Department, Reserve Bank of Australia, Craigieburn, Victoria 3064, Australia

[g]Australian Institute for Bioengineering and Nanotechnology, The University of Queensland, Brisbane, Queensland 4072, Australia

Email: amanda.ellis@unimeb.edu.au

## Abstract

Piezoelectric fluoropolymers convert mechanical energy to electricity and are ideal for sustainably providing power to electronic devices. To convert mechanical energy, a net polarization must be induced in the fluoropolymer, which is currently achieved via an energy intensive electrical poling process. Eliminating this process will enable the low-energy production of efficient energy harvesters. Here, by combining molecular dynamics simulations, piezoresponse force microscopy, and electrodynamic measurements, we reveal a hitherto unseen polarization locking phenomena of poly(vinylidene fluoride–*co*–trifluoroethylene) (PVDF-TrFE) perpendicular to the basal plane of two-dimensional (2D) $Ti_3C_2T_x$ MXene nanosheets. This polarization locking, driven by strong electrostatic interactions enabled exceptional energy harvesting performance, with a measured piezoelectric charge coefficient, $d_{33}$, of -52.0 picocoulombs per newton, significantly higher than electrically poled PVDF-TrFE (approximately -38 picocoulombs per newton). This study provides a new fundamental and low energy input mechanism of poling fluoropolymers, which enables new levels of performance in electromechanical technologies.

## Introduction

For dielectric materials exhibiting piezoelectricity, inducing polarization through the alignment of the dipoles is paramount to couple mechanical and electrical energy.[1] To achieve dipole alignment, electrical poling is considered a necessary task in the post-processing of piezoelectric materials (Fig. 1a-c).[2] Electrical poling is energy intensive, with electric fields on the order of tens to hundreds of megavolts per meter commonly used (Fig. 1b,c).[3–5] In fluoropolymers such as poly(vinylidene fluoride) (PVDF), a class of semicrystalline linear-chain polymers exhibiting a dipole moment between the hydrogen and fluorine moieties perpendicular to the carbon backbone (Fig. 1a), the poling process additionally requires elevated temperature conditions.[3,4,6] The highly valorized commercial applications for piezoelectric materials, including precision motorized stages and inkjet printheads, utilize the converse piezoelectric effect,[7] converting an applied electric field to discrete mechanical outputs.[6] In contrast, emerging applications which utilize the direct piezoelectric effect[7] to convert mechanical to electrical energy, the electrical poling process is a roadblock to commercialization, requiring a higher energy input than can be harvested in the device lifespan. These emergent applications, including energy harvesting,[3,8] robotic interfaces,[9] piezocatalysis[10] and piezophotonics,[11] require revisiting of the electrical poling process and examination of the pathways for inducing polarization without high input energies.[1,12]

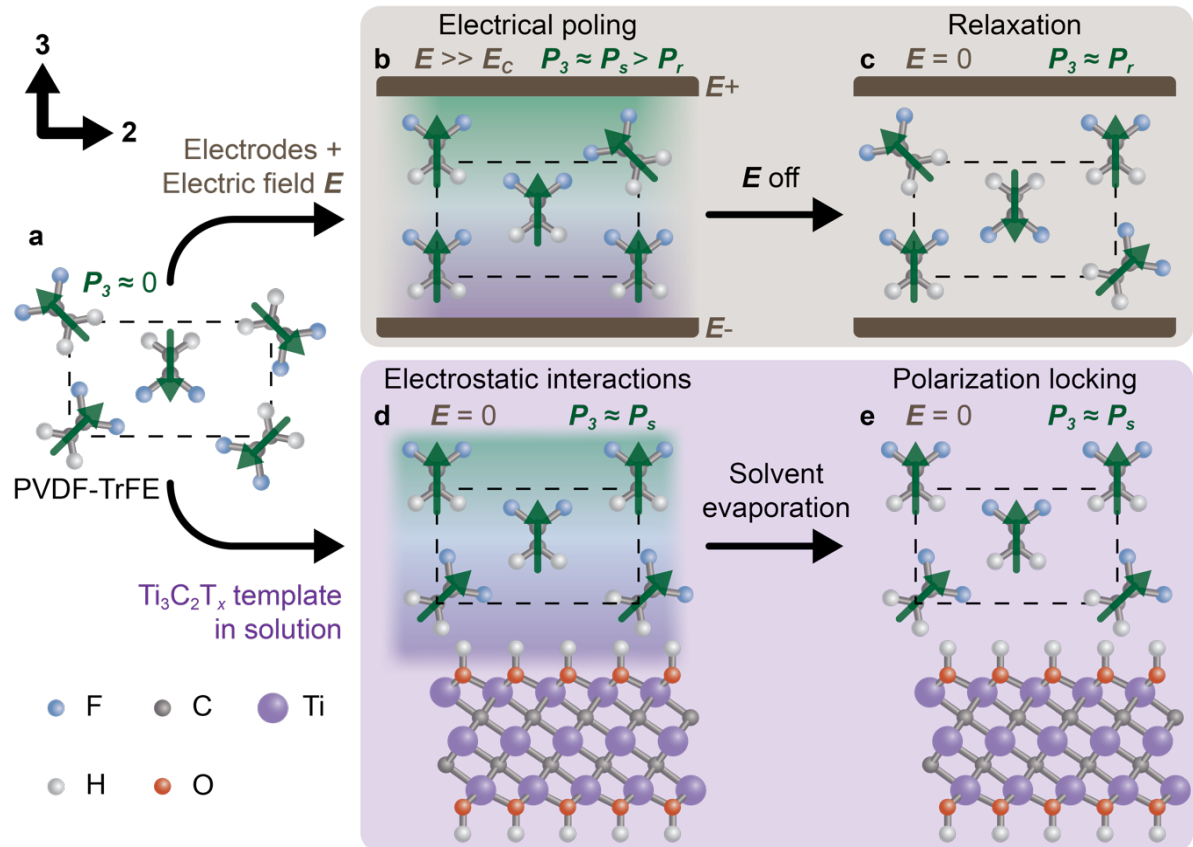

**Fig. 1: Simplified schematic outlining the nanomaterial-induced polarization locking mechanism in PVDF-TrFE as an alternative to the conventional electrical poling method. a** The β phase PVDF-TrFE chains, obtained directly following film deposition, exhibit a randomized dipolar orientation (green arrows), resulting in a negligible net polarization ($P$). **b** In the electrical poling method, electrodes are attached to the surfaces perpendicular to the desired polarization direction and an electric field ($E$) is applied, significantly higher than the coercive field ($E_C$) in order to orient the individual dipole moment vectors and maximize the $P$ to the spontaneous polarization ($P_s$). **c** Following the removal of the $E$, the PVDF-TrFE chains undergo partial relaxation from $P_s$ to the remnant polarization ($P_r$). **d** Conversely, adding the $Ti_3C_2T_x$ ($T_x \approx$ OH) nanosheets to the PVDF-TrFE in solution enables the $P$ to align, without an applied electric field, perpendicular to the basal plane of the $Ti_3C_2T_x$ nanosheets *via* electrostatic interactions at the interface. **e** Following deposition, the $Ti_3C_2T_x$ nanosheets are generally aligned parallel to the substrate, and the subsequent evaporation of the solvent locks the $P$ at $P_s$ with no relaxation.

Recent efforts have investigated alternative pathways to polarize fluoropolymers by tuning solvent composition[13] or using nanofillers.[8,14–18] The templating of polarization has been realized by nanomaterial-fluoropolymer interactions from piezoelectric $BaTiO_3$ nanoparticles,[14,15] reduced graphene oxide nanosheets,[16,17] hexagonal boron nitride nanoflakes,[18] and single-walled carbon nanotubes.[8] However, the mechanism of dipole alignment arising from templating piezoelectric polymers with nanofillers remains poorly understood. Notably, the aforementioned nanomaterials provide limited scope to probe dipole moment alignment, as they possess piezoelectric properties such as $BaTiO_3$,[14] or they alter the number of dipoles through changes in polymer conformation (e.g., by reduced graphene oxide or hexagonal boron nitride).[16,17] The single-walled carbon nanotube template has shown promise on a nanoscale,[8] although the mechanism of polarization templating was not elucidated.

To investigate the mechanism of polarization templating (Fig. 1c), a nanofiller must have out-of-plane polarizability without out-of-plane piezoelectric properties.[19] Further, it should be mechanically rigid with well-defined surface properties and functionality. To this end, the rapidly evolving class of transition metal carbides (MXenes) is an excellent candidate to probe polarization templating, with $Ti_3C_2T_x$ being the most well characterized MXene.[20–27] Importantly, it has out-of-plane polarizability[21] with symmetry perpendicular to the basal plane[23,28] and is therefore hypothesized to not possess out-of-plane piezoelectric properties.

In this work, we provide a deep mechanistic understanding of how polarization templating can be achieved in fluoropolymers from $Ti_3C_2T_x$ nanosheet templates (Fig. 1c,d). We employ molecular dynamics (MD) simulations to probe the evolution of the polarization of PVDF-TrFE in relation to the $Ti_3C_2T_x$ nanosheet, revealing that the electrostatic interactions between the Ti$_3$C$_2$T$_x$ nanosheet and the fluoropolymer are crucial to achieve effective induced local polarization locking. We then extend this induced local polarization locking to a macroscale net polarization using solvent-evaporation assisted (SEA) 3D printing to impart enhanced shear alignment.[8,29] The resultant composites show a piezoelectric charge coefficient ($d_{33}$), voltage coefficient ($g_{33}$) and figure of merit of -52.0 pC N$^{-1}$, 402 mV m N$^{-1}$ and 20.9 x 10$^{-12}$ Pa$^{-1}$ respectively, higher than electrically poled PVDF-TrFE co-polymer (approximately -38 pC N$^{-1}$, 380 mV m N$^{-1}$ and 14.4 x 10$^{-12}$ Pa$^{-1}$),[3,30] which demonstrates that our composites are fully polarized within the electroactive crystalline phase. The advancements herein enable rapid, cost-effective, energy-efficient, and scalable production of fluoropolymers for emerging applications utilizing the direct piezoelectric effect, including as a power supply for broad scale wearable electronics.

## Results

**Preparation of Ti$_3$C$_2$T$_x$/PVDF-TrFE composite SEA 3D printing inks**

Ti$_3$C$_2$T$_x$ nanosheets are a 2D material with the point group of P63/mmc, which is symmetric (and therefore non-piezoelectric) in the z-direction (i.e., no out-of-plane polarization).[23,28] The Ti$_3$C$_2$T$_x$ nanosheets were exfoliated from a Ti$_3$AlC$_2$ MAX phase (Supplementary Fig. S1a) using the minimally intensive layer delamination (MILD) exfoliation method with LiF and HCl, exhibiting an average lateral size of 310 nm (Supplementary Fig. S1b, c) and a thickness of ~1 nm (Supplementary Fig. S1d).[24,27] This soft exfoliation technique resulted in a dominant OH surface termination (T$_x$), with trace amounts of F and O functionality (Supplementary Information Fig. S1e-g).[25,26]

Composites of Ti$_3$C$_2$T$_x$ nanosheets and PVDF-TrFE were prepared as inks by a simple mixing process, whereby a small volume of the Ti$_3$C$_2$T$_x$ nanosheets in DMF were added to a PVDF-TrFE (40 wt%) solution in acetone and homogenized *via* stirring at room temperature. Concentrations between 0.00 wt% and 0.50 wt% $Ti_3C_2T_x$, relative to the mass of PVDF-TrFE, were produced as viscous inks (Supplementary Fig. S2). Interestingly, all inks showed sustained stability, with retention of their initial color and flow properties for up to 5 months post-mixing (Supplementary Fig. S3). This stability is primarily due to the retardation of either/or the surface oxidation and agglomeration of the $Ti_3C_2T_x$ nanosheets that can typically occur.[26]

**MD simulations of the $Ti_3C_2T_x$/PVDF-TrFE interface**

To understand the interface between the $Ti_3C_2T_x$ nanosheets and the PVDF-TrFE co-polymer, MD calculations were performed using the periodic lattice of $Ti_3C_2T_x$ (where $T_x$ = OH) and 70 'mer' chains of PVDF-TrFE (Fig. 2a, left). This was then compared to a periodic lattice of graphene (Fig. 2b, left) with an equivalent polymer film. These simulations revealed an extremely strong electrostatic interaction between the PVDF-TrFE chains and the $Ti_3C_2T_x$ nanosheet, requiring ~ 4.17 pN of force to desorb one PVDF-TrFE chain from the $Ti_3C_2T_x$. This strong interaction limits the motion of the polymer (Supplementary Video S1), with chains adjacent to the surface elongated and unable to move, forming a tightly packed structure (Fig. 2a) with local density of approximately 1.6 g cm$^{-3}$ (Supplementary Fig. S5). Over 4 ns, more chains preferentially fill the free space on the $Ti_3C_2T_x$ lattice, leading to the decrease in the density of the second layer. Interestingly, there is no statistical difference between the H and F positions on the PVDF-TrFE chains relative to the $Ti_3C_2T_x$ nanosheet (Supplementary Fig. S6, shown for a film of 14 chains). There is a clear difference in the proportion of *trans* (63%) and *gauche* (37%) conformations within the PVDF-TrFE chains at the interface of the $Ti_3C_2T_x$ nanosheet (Supplementary Fig. S7, Supplementary Fig. S8). This ratio of bond conformations suggests the inhibition of local electroactive phase within the fluoropolymer.

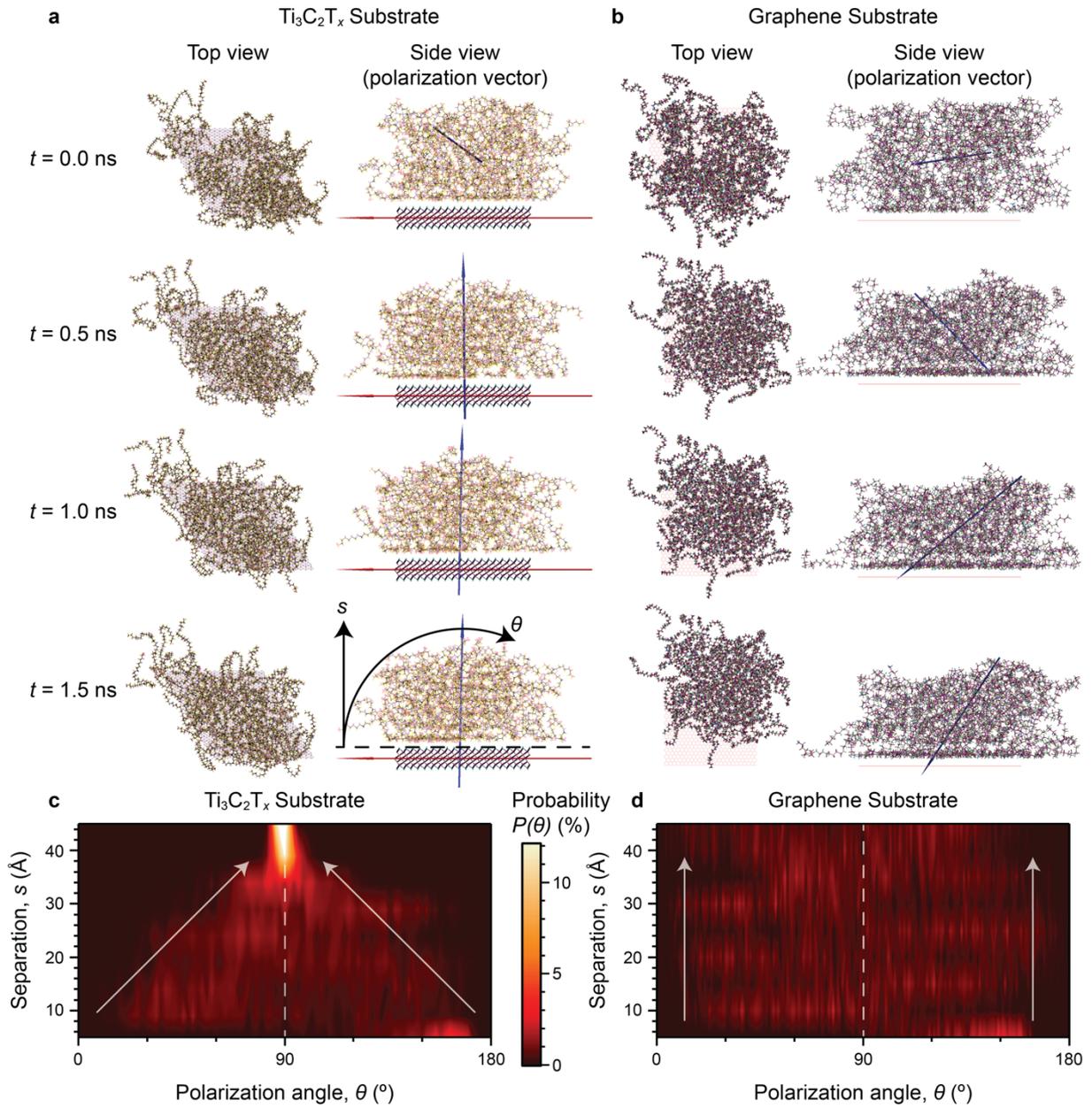

**Fig. 2: Comparative MD simulations of the PVDF-TrFE co-polymer film polarization on $Ti_3C_2T_x$ and graphene substrates. a,b** The top view (left column) and side view (right column) snapshots at $t$ = 0, 0.5, 1 and 1.5 ns, with the resultant dipole moment vectors (right column) of the PVDF-TrFE co-polymer film (blue arrow) and the **a** $Ti_3C_2T_x$ substrate (red arrow) or **b** graphene substrate. The annotated $s$ and $\theta$ in **a** represent the separation of the fluoropolymer from the substrate and the PVDF-TrFE co-polymer dipole angle relative to the substrate, respectively. **c,d** Probability distributions of the angle ($\theta$) between the dipole moment vector of the PVDF-TrFE co-polymer film layers and the **c** $Ti_3C_2T_x$ or **d** graphene substrates as a function of separation ($s$), as calculated from the equilibrated region of the obtained MD simulation.

The orientation of the net dipole moment vector of the PVDF-TrFE co-polymer film was compared to that of the $Ti_3C_2T_x$ nanosheet, quantified through the angle $\theta$ (Fig. 2a, right). The spatial and temporal evolution of $\theta$ against the separation ($s$) reveals the fluoropolymer layers adjacent to the $Ti_3C_2T_x$ nanosheet (within 5 Å from the $Ti_3C_2T_x$ nanosheet surface, directly experiencing the strong electrostatic attraction) have a preferential orientation parallel to the $Ti_3C_2T_x$ nanosheet (Fig. 2c). However, with increasing separation from the $Ti_3C_2T_x$ nanosheet, this PVDF-TrFE co-polymer net dipole moment becomes increasingly perpendicular to the basal plane of the $Ti_3C_2T_x$ nanosheet substrate. Interestingly, thicker co-polymer films (within 39 Å from the $Ti_3C_2T_x$ nanosheet surface) exhibit an extremely tight distribution of the polarization orientation (Fig. 2c), which remains perpendicular to the $Ti_3C_2T_x$ nanosheet for the entire simulation (Supplementary Fig. S9, Supplementary Video S1). These results show that on a local scale, a net polarization of the PVDF-TrFE co-polymer is formed perpendicular to the basal plane of the $Ti_3C_2T_x$ nanosheet (Supplementary Video S2). Given that the electrostatic screening length in 2D materials is between 1 nm and 10 nm,[31] and considering the strength of the electrostatic interaction observed between the PVDF-TrFE co-polymer and the $Ti_3C_2T_x$ nanosheet, it is highly probable that this polarization locking is occurring as a consequence of electrostatic forces.[19]

For comparison, the net dipole moment vector of the same PVDF-TrFE co-polymer film on a graphene sheet was investigated under identical simulation conditions (Fig. 2b, right). As an atomically thin 2D sheet, graphene does not exhibit out of plane polarizability and is thus a suitable comparative substrate system to $Ti_3C_2T_x$.[19] In contrast to the fluoropolymer film on $Ti_3C_2T_x$ nanosheet, the fluoropolymer film on graphene is able to migrate on the periodic lattice easily (Supplementary Video S1) and requires a significantly lower force (~2.78 pN) to detach a single fluoropolymer chain from the lattice, indicating a weaker interaction between the components. The polarization vector of the fluoropolymer film always exhibits random orientations with respect to the graphene surface, regardless of the layer thickness (Fig. 2b, right, Fig. 2d, Supplementary Video S1), and this phenomenon is observed for individual PVDF-TrFE co-polymer chains (Supplementary Video S2).

The MD simulations show a clear and strong binding interaction between the $Ti_3C_2T_x$ nanosheet and the PVDF-TrFE co-polymer chains, driven by electrostatic interactions. This binding results in the subsequent self-assembly of the fluoropolymer film, oriented in such a way as to tightly lock the polarization of the PVDF-TrFE perpendicular to the $Ti_3C_2T_x$ basal plane, providing guidance towards an experimental tool for self-assembly driven polarization in bulk fluoropolymer materials.

**Towards printing and net polarization**

For the translation of this fundamental understanding of the induced local net polarization, developed using MD simulations, into macroscale piezoelectric generators (PEGs), a careful selection of the $Ti_3C_2T_x$/PVDF-TrFE composite processing route is required. SEA 3D printing can enable shear alignment of both nanofillers[27,29] and polymer chains,[8,32] parallel to the direction of printing. Given the formation of induced local polarization, perpendicular to the basal plane of the $Ti_3C_2T_x$ nanosheets, this parallel alignment is hypothesized to give out-of-plane piezoelectricity without the need for electric poling.

To understand the interaction between the $Ti_3C_2T_x$ nanosheets and the PVDF-TrFE co-polymer in solution, as well as assess the suitability for relevant solution processing routes including extrusion printing, the shear strain ($\gamma_s$) response of the $Ti_3C_2T_x$ loaded PVDF-TrFE (40 wt%) inks was studied (Fig. 3a). The extended rheological characterization of the pristine PVDF-TrFE ink is presented in the Supplementary Information (Supplementary Fig. S10, Supplementary Fig. S11, Supplementary Fig. S12). Both the pristine PVDF-TrFE co-polymer and the $Ti_3C_2T_x$/PVDF-TrFE inks demonstrated exceptional flow properties for printing. This was clearly demonstrated by the storage modulus ($G'$) being greater than the loss modulus ($G''$) at low $\gamma_s$, indicating the ability of the ink to retain a physical shape. With increasing $\gamma_s$, analogous to the strain applied during extrusion printing, both inks exhibited yielding and liquid-like behavior, indicated by the cross-over of $G'$ and $G''$, and the subsequent region with $G' < G''$. Similarly, the angular frequency ($\omega$) response (Fig. 3b) confirmed the formation of a strong physical gel in both the pristine PVDF-TrFE co-polymer and the $Ti_3C_2T_x$/PVDF-TrFE inks, highly desirable for extrusion printing.[33] A weakening of the PVDF-TrFE/acetone interaction was apparent through the presence of a larger shoulder at $\omega \approx 0.15$ rad s$^{-1}$ in the $G''$ of the $Ti_3C_2T_x$/PVDF-TrFE ink. The weakening of the physical gel was attributed to a strong interaction between the PVDF-TrFE co-polymer and the $Ti_3C_2T_x$ nanosheets. However, the $Ti_3C_2T_x$/PVDF-TrFE ink, nonetheless, exhibited solid-like behavior ($G' > G''$) at $\omega$ as low as 0.10 rad s$^{-1}$. The PVDF-TrFE co-polymer ink was subjected to consecutive small amplitude oscillatory shear (SAOS, 1 Pa) and large amplitude oscillatory shear (LAOS, 5 kPa) cycling at 1 Hz to replicate the shear stress of extrusion printing, shown in Fig. 3c.[34] It was found to consistently flow under 5 kPa shear stress ($\sigma_s$), with immediate G'' recovery and >60% G' recovery after 70 s at $\sigma_s$ = 1 Pa, retaining similar characteristics with >85% G' recovery on the second $\sigma_s$ cycle. This testing method showed the ability of the PVDF-TrFE co-polymer and $Ti_3C_2T_x$/PVDF-TrFE inks to print continuous and complex structures without void formation or defects in the resultant structures.

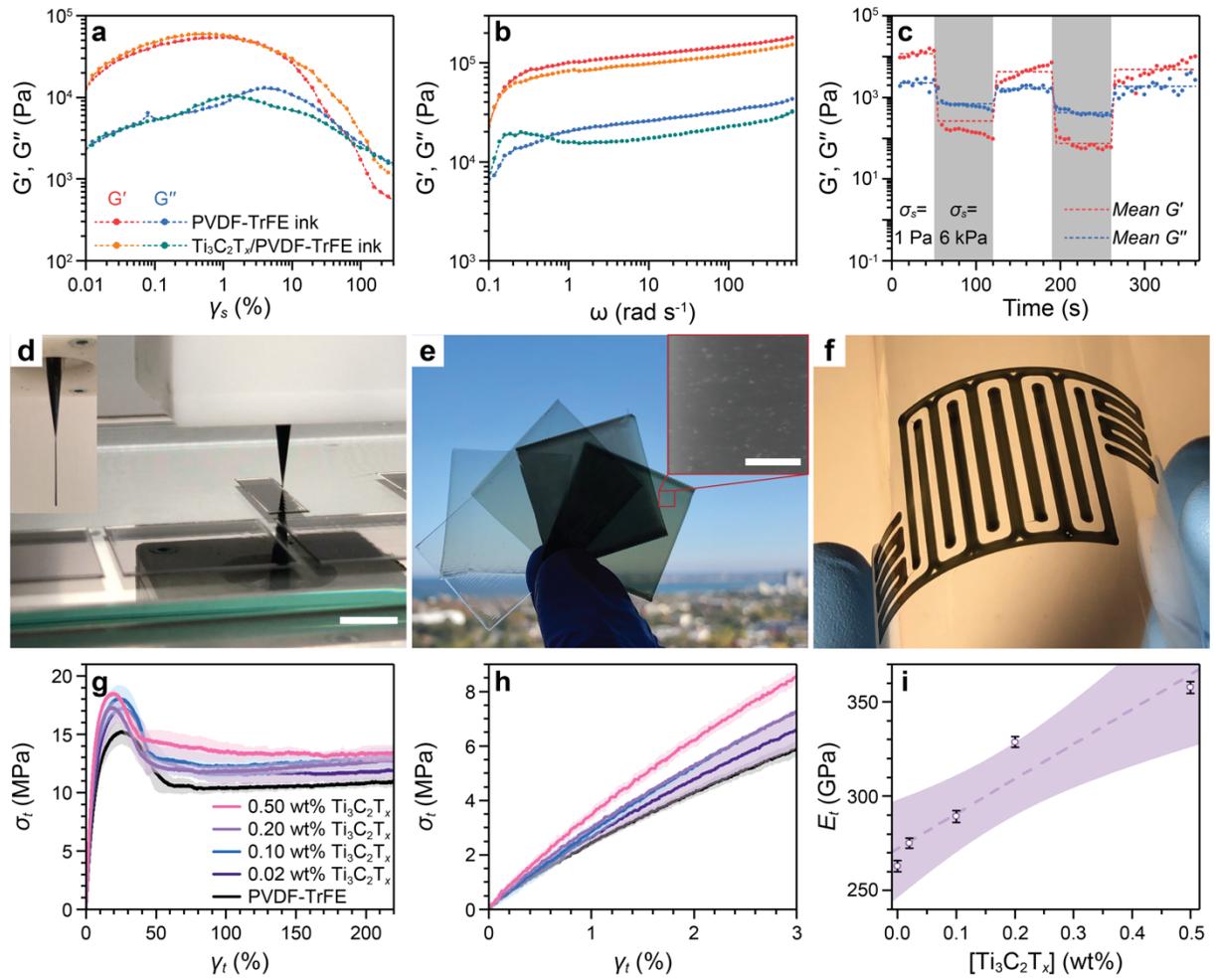

**Fig. 3: SEA extrusion printing of strengthened Ti₃C₂Tₓ/PVDF-TrFE films. a** Oscillatory $\gamma_s$ sweep ($\omega$ = 1 Hz) and **b** oscillatory $\omega$ sweep ($\gamma_s$ = 1%) of the 40 wt% PVDF-TrFE ink with and without 0.20 wt% Ti₃C₂Tₓ nanosheets. **c** The temporal evolution of the $G'$ and $G''$ during $\sigma_s$ cycling between 1 Pa (white regions, representing the stationary state) and 5 kPa (grey regions, representing printing stress). **d** Photograph showing the SEA extrusion printing process for the Ti₃C₂Tₓ/PVDF-TrFE (0.50 wt%) ink, scale bar 10 mm. Inset shows a self-supporting single filament. **e** Photograph of the free-standing Ti₃C₂Tₓ/PVDF-TrFE films at various Ti₃C₂Tₓ nanosheet loadings (from left to right: 0.00 wt%, 0.02 wt%, 0.10 wt%, 0.20 wt% 0.50 wt%). Inset shows helium ion beam microscopy (HIM) image of Ti₃C₂Tₓ/PVDF-TrFE (0.50 wt%) film surface, demonstrating the distribution of Ti₃C₂Tₓ nanosheets near the surface of the film. Scale bar represents 40 μm. **f** Photograph demonstrating an interdigitated structure of 0.50 wt% Ti₃C₂Tₓ/PVDF-TrFE composite printed onto a flexible poly(ethylene terephthalate) (PET) substrate. **g** Tensile strain ($\gamma_t$)-stress ($\sigma_t$) profiles of the printed Ti₃C₂Tₓ/PVDF-TrFE films (shaded region shows the error from replicates). **h** The elastic region of **g**. **i** The Young's modulus ($E_t$) obtained from **h**. The overlay in **g** and **h**, and the error bars in **i** represent the mean ± SD. The overlay in **i** represents a 95% CI for a linear fit.

# SEA extrusion printed and mechanically robust $Ti_3C_2T_x$/PVDF-TrFE composite films

With a clear understanding of the rheological performance of the $Ti_3C_2T_x$/PVDF-TrFE inks, as well as evidence of weakening of the PVDF-TrFE/acetone interaction due to the strong $Ti_3C_2T_x$/PVDF-TrFE interface, thin films were prepared for studies of the macroscale polarization and energy harvesting properties.

$Ti_3C_2T_x$/PVDF-TrFE inks at various $Ti_3C_2T_x$ nanosheet loadings (0.00 wt%, 0.02 wt%, 0.10 wt%, 0.20 wt% 0.50 wt%) were SEA extrusion printed as single-layer thin films (Fig. 3d) through a nozzle (internal diameter (ID) = 200 μm) onto a clean glass plate and then subsequently removed to form free-standing films (Fig. 3e). The distribution of the $Ti_3C_2T_x$ nanosheets was homogeneous in the resultant films with the basal plane appearing aligned parallel to the printing direction (Fig. 3e, inset), as expected from an extrusion process. This SEA extrusion printing process could be further extended to print complex shapes on flexible substrates, such as poly(ethylene terephthalate) (PET), as shown in Fig. 3f. The versatility of the SEA extrusion printing technique enables the deposition of multi-layer systems,[32] on conformal[35] and moving substrates,[36] allowing our PEG to be deployed in broad and highly specialized applications such as in the point-of-care printing of *in-vivo* energy harvesters.[37,38]

The mechanical properties of the SEA extrusion printed $Ti_3C_2T_x$/PVDF-TrFE (0.00 wt%, 0.02 wt%, 0.10 wt%, 0.20 wt% 0.50 wt%) composite films were studied *via* tensile extension (Fig. 3g). All films were shown to yield at $\gamma_t$ <50% and did not break at $\gamma_t$ = 220% (displacement limit of the instrument), confirming the high ductility of all tested materials. Subsequent elongation at uncontrolled force by hand showed a maximum $\gamma_t$ >1,000% prior to failure (Supplementary Fig. S13). The Young's modulus ($E_t$), obtained from the low-$\gamma_t$ region (Fig.

3h), increased linearly with an increase in $Ti_3C_2T_x$ nanosheet loadings in the PVDF-TrFE co-polymer up to 358 MPa at a $Ti_3C_2T_x$ nanosheet loading of 0.5 wt% (Fig. 3i). This linear increase showed that there is a homogenous dispersion of the $Ti_3C_2T_x$ nanosheets through the fluoropolymer at all concentrations, as aggregation would lead to poor load transfer between the components. The $E_t$ of the pristine PVDF-TrFE co-polymer film (263 MPa) was comparable to literature values for 3D printed PVDF (419 MPa), although lower than other processing routes.[39]

**PVDF-TrFE phase dependence in $Ti_3C_2T_x$/PVDF-TrFE composites**

The semi-crystalline PVDF-TrFE co-polymer exists in three favorable conformations, the symmetric and therefore non-electroactive α phase, the semi-polar and moderately electroactive γ phase, and the polar and highly electroactive β phase.[3] These phases arise from the presence and order of *trans* (T) and *gauche* (G) bond conformations, whereby the α phase consists of alternating conformations (TGTG'), the γ phase is an intermediate (TTTG) and the β phase is all-*trans* (TTTT).[4] Inherently, changes in the crystallinity and phase composition affect the maximum polarization of fluoropolymers. Understanding how these parameters change is crucial to elucidating a mechanism for enhanced piezoelectric response. Ultimately, entropy within the polymer during deposition results in negligible polarization of the polymer films, thus requiring electrical poling to align the dipole moment vectors.[8] To achieve this understanding of crystallinity and phase composition, thorough phase characterization (differential scanning calorimetry (DSC) and Raman confocal microscopy) was performed (Fig. 4). These tools enable the determination of both the total crystallinity of the fluoropolymer, and the relative proportion of β and γ phases, enabling the exhaustive understanding of the material for the assessment of energy harvesting capabilities.

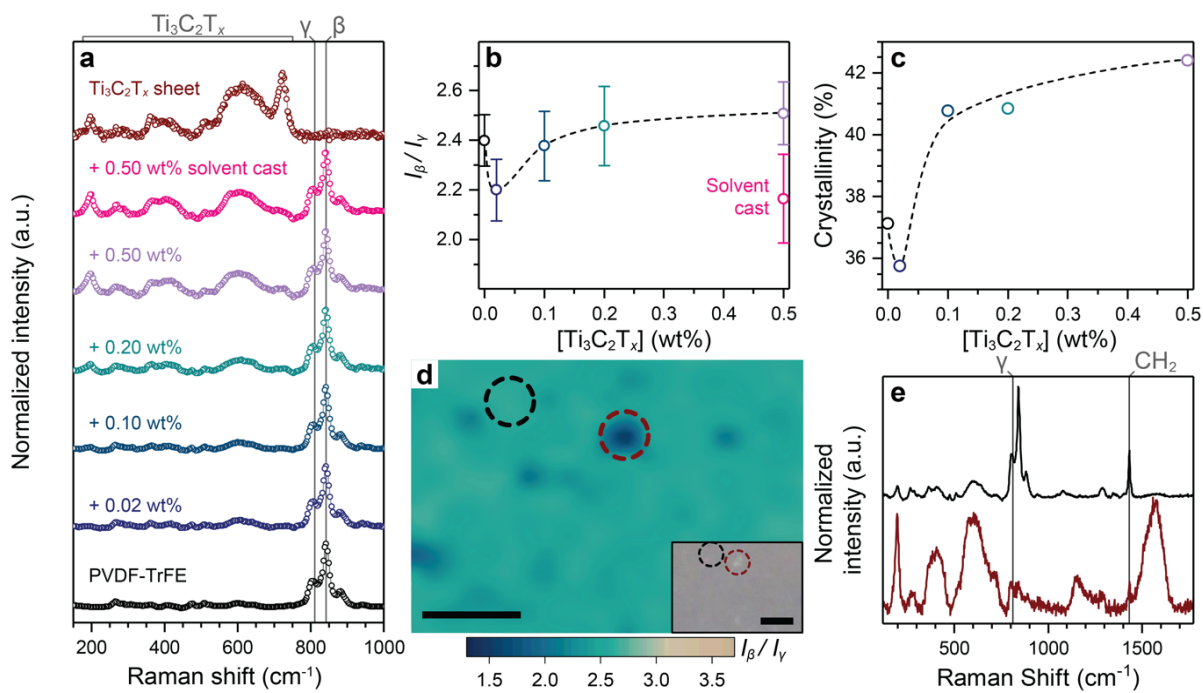

**Fig. 4: Material characteristics of Ti$_3$C$_2$T$_x$/PVDF-TrFE films. a** Raman spectra of Ti$_3$C$_2$T$_x$/PVDF-TrFE composites (at various Ti$_3$C$_2$T$_x$ loadings) and Ti$_3$C$_2$T$_x$ nanosheets. **b** The β/γ phase ratio, $I_β/I_γ$, calculated from the intensities of the β phase (840 cm$^{-1}$) and γ phase (811 cm$^{-1}$) from Raman confocal mapping (Supplementary Fig. S15). The error bars represent the mean ± SD in $I_β/I_γ$ over each Raman map. **c** The effect of Ti$_3$C$_2$T$_x$ nanosheet loading on the PVDF-TrFE co-polymer crystallinity, as measured by DSC. The dashed lines in **b** and **c** are visual guides. **d** The Raman $I_β/I_γ$ map of the Ti$_3$C$_2$T$_x$/PVDF-TrFE (0.50 wt%) film, where the dark point corresponds to a large Ti$_3$C$_2$T$_x$ nanosheet at the film surface [X,Y = 1 μm, 0 μm]. The inset shows a confocal microscope image of the area analyzed using Raman confocal mapping. Both scale bars represent 5 μm. **e** The representative Raman spectra for the area designated by the black circle [X,Y = -5 μm, 2 μm] and the red circle [X,Y = 1 μm, 0 μm] in **d**, showing the clear inhibition of β phase locally at the Ti$_3$C$_2$T$_x$ nanosheet surface.

Raman spectroscopy showed a clear increase in the Ti$_3$C$_2$T$_x$ nanosheet spectrum vibrational modes (150 cm$^{-1}$ – 740 cm$^{-1}$) with increasing concentration (Fig. 4a). An initial decrease was observed in the intensity ratio $I_β/I_γ$ (β phase at 840 cm$^{-1}$ and γ phase at 811 cm$^{-1}$) from 2.4 for the pristine PVDF-TrFE co-polymer to 2.2 for the 0.02 wt% Ti$_3$C$_2$T$_x$/PVDF-TrFE composite, before an increase to 2.5 for the 0.5 wt% Ti$_3$C$_2$T$_x$/PVDF-TrFE composite (Fig. 4b).[3]

DSC revealed that small Ti$_3$C$_2$T$_x$ nanosheet loadings (0.02 wt%) resulted in a notable decrease in the overall crystallinity of the PVDF-TrFE co-polymer matrix (Fig. 4c).

Confocal Raman mapping was also performed (Fig. 4d, e) to probe the decrease in both the $I_\beta/I_\gamma$ and the crystallinity, as shown by the average Raman spectra (Fig. 4b) and DSC (Fig. 4c), respectively. Notably, the analysis of these maps found a clear difference in the $I_\beta/I_\gamma$ between the bulk PVDF-TrFE co-polymer (Fig. 4d, black circle) and the Ti$_3$C$_2$T$_x$/PVDF-TrFE interface (Fig. 4d, red circle), decreasing from 2.65 to 1.45, respectively. The individual spectra for these regions (Fig. 4e) showed a disproportionate suppression of the β phase peak intensity at the Ti$_3$C$_2$T$_x$ nanosheet surface, while the intensity ratio between γ and the CH$_2$ stretch (1432 cm$^{-1}$) maintains the same ratio.[40] Importantly, these findings suggest a local inhibition of the electroactive β phase crystallization at the interface between the Ti$_3$C$_2$T$_x$ nanosheet and the PVDF-TrFE co-polymer (Supplementary Fig. S7), as well as the local densification of the fluoropolymer film (Supplementary Fig. S5).

Thus, the decrease in the crystallinity at 0.02 wt% Ti$_3$C$_2$T$_x$/PVDF-TrFE (Fig. 4c) was attributed to a lower fraction of the electroactive β phase in the PVDF-TrFE co-polymer forming directly on the Ti$_3$C$_2$T$_x$ nanosheet surface.

Given the understanding developed through MD simulations (Fig. 2), the fluoropolymer densification was hypothesized to occur in solution prior to printing, rather than during printing. There are two factors to consider here: (1) the effect of shear alignment on the fluoropolymer itself,[32,41] and (2) the effect of shear on orienting the Ti$_3$C$_2$T$_x$ nanosheets.[27,29] At 0.02 wt% Ti$_3$C$_2$T$_x$ nanosheets, it is proposed that the decrease in the β phase at the PVDF-TrFE/Ti$_3$C$_2$T$_x$ interface, coupled to minimal shear orientation of the Ti$_3$C$_2$T$_x$ nanosheet, has a stronger negative effect compared to shear aligning of the fluoropolymer molecules. For higher Ti$_3$C$_2$T$_x$ nanosheet loadings, extruded through the nozzle, a greater shear alignment phenomenon is observed within the material, resulting in a very slight net increase in β phase and overall crystallinity. This hypothesis is supported by the Raman spectra of the solvent cast 0.50 wt% $Ti_3C_2T_x$/PVDF-TrFE composite film (Fig. 4b, Supplementary Fig. S15f), which does not undergo shear induced crystallization and has a $I_\beta/I_\gamma$ comparable to the printed 0.02 wt% $Ti_3C_2T_x$/PVDF-TrFE composite film.

**Energy harvesting of the $Ti_3C_2T_x$/PVDF-TrFE composite films**

Prior to printing, the strong electrostatic binding interactions between the $Ti_3C_2T_x$ nanosheets and the PVDF-TrFE co-polymer chains, as described by the MD simulations, enables the $Ti_3C_2T_x$ nanosheet to remain sterically stabilized in the inks, without aggregation (Supplementary Fig. S3). This lack of $Ti_3C_2T_x$ nanosheet aggregation enables the enhancement in the $E_t$ (Fig. 3i). To understand how this induced local polarization locking translates to bulk and macroscopic energy harvesting, both piezoelectric force microscopy (PFM) and bulk electromechanical testing were performed.

PFM was carried out by applying a bias between -20 V and +20 V to a conductive platinum (Pt) cantilever in contact with the $Ti_3C_2T_x$/PVDF-TrFE films (at various $Ti_3C_2T_x$ nanosheet loadings) and the subsequent local changes in thickness (arising from the expansion or contraction of the unit cell in the polarized electroactive phases of the PVDF-TrFE co-polymer) were measured.[6] At voltages below the poling electric field (<50 MV m$^{-1}$), piezoelectric materials exhibit a strong correlation between the induced strain ($\gamma_3$) and the $d_{33}$.[3,4] In PFM, the correlation is qualitative[42] and the $d_{33} = A cos(\varphi)/Q_f V$, where $A$ is the amplitude, $\varphi$ is the phase, $Q_f$ is the Q-factor of the cantilever and $V$ is the applied bias.[8] The extended discussion surrounding the PFM is presented in the Supplementary Information.

To confirm the lack of piezoelectric contribution arising in the z-direction of the P63/mmc point group of the $Ti_3C_2T_x$, a $Ti_3C_2T_x$ nanosheet on a gold-coated silicon (Au@Si) substrate was probed using dual AC resonance tracking (DART) PFM, whereby a constant bias (1 V) was applied to the cantilever (Supplementary Fig. S20). The $Ti_3C_2T_x$ nanosheet, approximately 300 nm in the lateral dimension, was visible topographically, however no discernible changes were observed in the $A$ and $\varphi$ traces between the $Ti_3C_2T_x$ nanosheet and the underlying substrate (Fig. 5a). Therefore, the $Ti_3C_2T_x$ nanosheet exhibited no observable out-of-plane piezoelectric effect (perpendicular to the nanosheet basal plane).

To measure the $d_{33}$ of the composite $Ti_3C_2T_x$/PVDF-TrFE films, a variable bias was applied, with the spatial map of the applied bias shown in Supplementary Fig. S21. The representative PFM response for the SEA extrusion printed $Ti_3C_2T_x$/PVDF-TrFE (0.00 wt%, 0.02 wt%, 0.10 wt%, 0.20 wt% and 0.50 wt%) films is shown in Fig. 5b (extended data presented in Supplementary Fig. S22, Supplementary Fig. S23). The amplitude, corrected for the directionality of strain and the Q-factor of the cantilever ($A\cos(\varphi)/Q_f$), exhibited a maximum at -41.6 pm for the pristine PVDF-TrFE co-polymer film at -20 V (Supplementary Fig. S24), confirming a spatial alignment effect arising from shear stresses during the extrusion printing process, as previously reported.[8,32] The $A\cos(\varphi)/Q_f$ value was shown to increase sharply with an increase in $Ti_3C_2T_x$ nanosheet loading within the PVDF-TrFE co-polymer (Supplementary Fig. S24d). A substantial increase was observed in the maximum $A\cos(\varphi)/Q_f$ at -20 V for the $Ti_3C_2T_x$/PVDF-TrFE (0.50 wt%) SEA extrusion printed film (-182.3 pm), compared to the pristine PVDF-TrFE co-polymer film (-41.6 pm), indicating an intensified piezoelectric response upon the addition of the $Ti_3C_2T_x$ nanosheets.

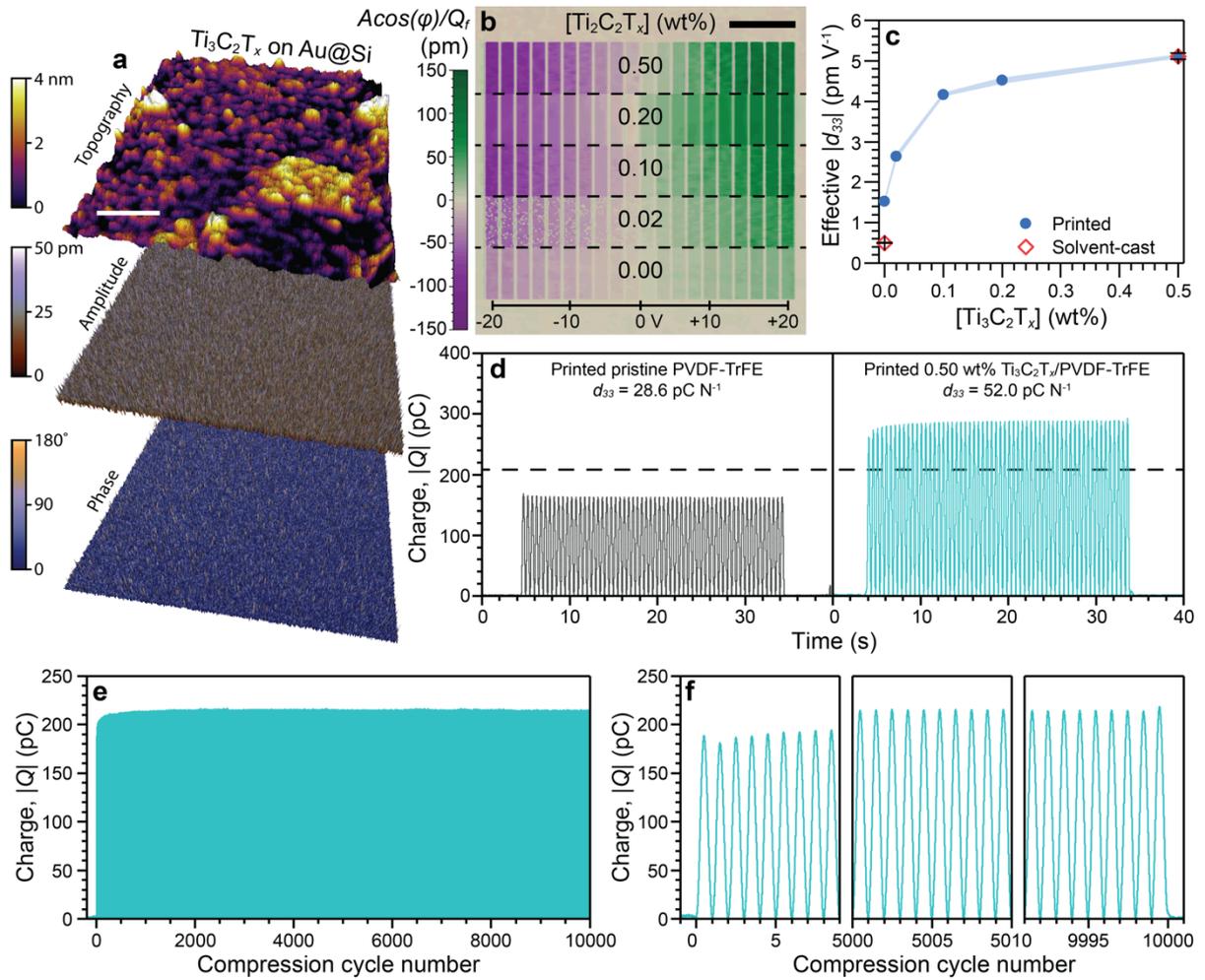

**Fig. 5: Polarization and energy harvesting of Ti$_3$C$_2$T$_x$/PVDF-TrFE composites. a** Dual AC resonance tracking (DART) PFM of a Ti$_3$C$_2$T$_x$ nanosheet on a gold-coated silicon (Au@Si) substrate, showing the topography (top), piezoelectric amplitude (middle) and piezoelectric phase trace (bottom). Scale bar represents 200 nm. **b** PFM of Ti$_3$C$_2$T$_x$/PVDF-TrFE (0.00 wt%, 0.02 wt%, 0.10 wt%, 0.20 wt% and 0.50 wt%) SEA extrusion printed films, showing the piezoresponse ($Acos(\varphi)/Q_f$) under an applied voltage between -20 V and +20 V. Scale bar represents 1 μm. **c** The effective piezoelectric charge coefficient ($d_{33}$) calculated from the PFM data, including for solvent-cast PVDF-TrFE and Ti$_3$C$_2$T$_x$/PVDF-TrFE (0.50 wt%) films as controls. The error bars represent the mean ± SE. **d-f** The macroscale energy harvesting characterization of the Ti$_3$C$_2$T$_x$/PVDF-TrFE (0.00 wt% and 0.50 wt%) SEA extrusion printed PEGs with input force ($\Delta F$) at 10 N following a sinusoidal input signal. **d** The generated surface charge as a function of time for 60 compression cycles at 2 Hz. The horizontal dashed line represents the charge generated from completely polarized ($d_{33}$ = -38 pC N$^{-1}$) PVDF-TrFE co-polymer for a PEG with similar dimensions. **e** The stability of the generated charge as a function of cycle count over 10,000 cycles for the Ti$_3$C$_2$T$_x$/PVDF-TrFE (0.50 wt%) PEG at 10 Hz and **f** the expanded initial, middle and end regions of the charge stability data in e showing 1 s of data (10 cycles) in each panel.

The measured $d_{33}$ as a function of Ti$_3$C$_2$T$_x$ nanosheet concentration in the printed films is shown in Fig. 5c. The effective $d_{33}$ for the pristine PVDF-TrFE co-polymer SEA extrusion printed film was -1.53 pm V$^{-1}$, 207% higher than that of the solvent-cast PVDF-TrFE co-polymer ($d_{33}$ = -0.50 pm V$^{-1}$). These results are consistent with reports on solvent-cast fluoropolymer films and shear stress-induced partial polarisation.[8,32,41,43,44] The $d_{33}$ of the Ti$_3$C$_2$T$_x$/PVDF-TrFE (0.50 wt%) SEA extrusion printed film increased up to a maximum at -5.11 pm V$^{-1}$, an ultimate improvement of 234% over the printed pristine PVDF-TrFE co-polymer film and 926% over the solvent-cast PVDF-TrFE co-polymer film.

Unexpectedly, considering the lower $I_\beta/I_\gamma$ ratio (Fig. 4b) and net crystallinity (Fig. 4c) of the Ti$_3$C$_2$T$_x$/PVDF-TrFE (0.02 wt%) film, an increase in the $d_{33}$ of 72% relative to the pristine PVDF-TrFE co-polymer film was observed (Fig. 5c). This would suggest that even at low Ti$_3$C$_2$T$_x$ nanosheet loadings there is a sufficient increase in induced polarization locking of the PVDF-TrFE co-polymer to offset the net decrease in the electroactive phase within the composite.

Surprisingly, the effective $d_{33}$ for the solvent-cast Ti$_3$C$_2$T$_x$/PVDF-TrFE (0.50 wt%) film matched that of the extrusion printed film with the same Ti$_3$C$_2$T$_x$ nanosheet loading (-5.11 pm V$^{-1}$), suggesting the effect of induced polarization locking from the Ti$_3$C$_2$T$_x$ nanosheet surface is favored over the shear-induced polarization with high Ti$_3$C$_2$T$_x$ nanosheet loading. It is hypothesized that this arises due to surface area minimization of the Ti$_3$C$_2$T$_x$ nanosheets during solvent casting, where the sheets settle perpendicular to gravity in order to maximize the area upon which the force is acting.[27] This result, while unintuitive, confirms that the driving force for the polarization enhancement is the fundamental interaction and local dipole locking occurring in solution between the Ti$_3$C$_2$T$_x$ nanosheet and the PVDF-TrFE co-polymer, rather than fluoropolymer chain alignment induced through SEA 3D printing.

**Macro energy harvesting using Ti$_3$C$_2$T$_x$/PVDF-TrFE composite PEGs**

Macroscale electromechanical testing confirmed the trends in the $d_{33}$ observed by PFM and quantified the energy output from the Ti$_3$C$_2$T$_x$/PVDF-TrFE PEGs. For the measurements, the PEGs were compressed following a sinusoidal force pattern as a function of time with amplitude ($\Delta F$) at 10 N (Supplementary Fig. S26a, b), pre-loaded to 5 N to minimize artefacts from contact electrification (i.e., contact-separation and lateral-sliding triboelectric modes).[44,45] The harvested energy was measured by a charge amplifier, negating any inherent effects of capacitance from the Ti$_3$C$_2$T$_x$/PVDF-TrFE composite films and the electrical cables, commonly unaccounted for during voltage measurements.

Macroscale $d_{33}$ measurements were performed on the SEA extrusion printed pristine PVDF-TrFE and Ti$_3$C$_2$T$_x$/PVDF-TrFE (0.5 wt%) PEGs (Fig. 5d). The $d_{33}$ of the pristine SEA printed pristine PVDF-TrFE PEG (-28.6 pC N$^{-1}$) was lower than literature reports for poled PVDF-TrFE (approximately -38 pC N$^{-1}$), however, this is not unexpected.[3,30] More importantly, the $d_{33}$ of the Ti$_3$C$_2$T$_x$/PVDF-TrFE (0.50 wt%) was significantly larger than any prior reports on unpoled fluoropolymers at -52.0 pC N$^{-1}$.[4] Notably, the ratio observed between the $d_{33}$ of pristine PVDF-TrFE to the Ti$_3$C$_2$T$_x$/PVDF-TrFE (0.50 wt%) *via* the direct piezoelectric effect (macroscale measurements, $D_3 = d_{33}\sigma_3$) of ∼1:2 correlated well to the trend observed *via* the converse piezoelectric effect (PFM, $\gamma_3 = d_{33}E_3$) of ∼1:3. The improvement in measured charge arises from the piezoelectric effect, connecting the induced local polarization locking described by MD to the macroscale polarization and subsequent energy harvesting.

The $g_{33}$ was determined from the measured dielectric permittivity ($\varepsilon_{33}$), which is shown in Supplementary Information Fig. S30. The relative permittivity was observed to increase slightly with the addition of the Ti$_3$C$_2$T$_x$ nanosheets, from 11.6 in the pristine PVDF-TrFE film to 14.4 in the Ti$_3$C$_2$T$_x$/PVDF-TrFE (0.50 wt%) film, in close agreement with prior reports on the dielectric properties of MXene/fluoropolymer composites.[46] The $g_{33}$ of the SEA extrusion printed PEGs exhibited similar characteristics upon the addition of Ti$_3$C$_2$T$_x$ nanosheets to the PVDF-TrFE co-polymer, improving 18% at 0.50 wt% Ti$_3$C$_2$T$_x$ (402 mV m N$^{-1}$) relative to the pristine PVDF-TrFE (341 mV m N$^{-1}$). This enhancement is important, as the $g_{33}$ possesses an inverse relationship with the $\varepsilon_{33}$ and most reports enhance only one of the two piezoelectric coefficients.[47] Thus, the observed $g_{33}$ value of the 0.50 wt% Ti$_3$C$_2$T$_x$/PVDF-TrFE PEG is higher than that of literature values for electrically poled PVDF-TrFE PEGs (380 mV m N$^{-1}$).[3] The piezoelectric figure of merit, $d_{33}g_{33}$, is consequently larger for the 0.50 wt% Ti$_3$C$_2$T$_x$/PVDF-TrFE PEG (20.9 x 10$^{-12}$ Pa$^{-1}$) relative to both the pristine PVDF-TrFE PEG prepared here (9.7 x 10$^{-12}$ Pa$^{-1}$) and literature values for poled PVDF-TrFE PEGs (14.4 x 10$^{-12}$ Pa$^{-1}$),[3] as shown in Supplementary Information Fig. S31.

To demonstrate the stability of the PEGs, a compressive cycling stability study on the Ti$_3$C$_2$T$_x$/PVDF-TrFE (0.50 wt%) PEG was performed for 10,000 cycles with a $\Delta F$ = 10 N and frequency of 10 Hz (Fig. 5e). Fig. 5f shows selected regions at the beginning, middle and end of the measurement. Aside from the initial period of cycling, which exhibited a rising generated charge from compressive stress to approximately -185 pC, the measured charge remained stable for the entire cycling period at -215 pC. The stability of the generated charge infers these completely solid-state PEGs can be used for long-term energy harvesting from human motion.

## Discussion

In summary, we have developed a mechanistic understanding of how nanofillers with no out-of-plane piezoelectricity can influence the local and macroscale polarization of fluoropolymers using 2D Ti$_3$C$_2$T$_x$ nanosheets as templates. We show conclusively that Ti$_3$C$_2$T$_x$ nanosheets have a strong electrostatic interaction with the PVDF-TrFE co-polymer, resulting in the evolution of a locked polarization in the fluoropolymer, perpendicular to the basal plane of the $Ti_3C_2T_x$ nanosheet. This effect is not observed using graphene nanosheets. The unique 2D geometry of $Ti_3C_2T_x$ nanosheets means that by either SEA extrusion printing or solvent casting we can elegantly and simply translate this induced local net polarization into macroscale polarization, demonstrating an exceptional $d_{33}$ of -52.0 pC N$^{-1}$, without the need for arduous and energy intensive electrical poling. The strong electrostatic interactions between the nanofiller and fluoropolymer resulted in a mechanically robust and flexible PEG device, capable of harvesting energy over 10,000 cycles without any degradation in performance.

Tuning surface terminations on MXenes and other 2D materials could afford enhanced electrostatic interactions leading to further improvements in piezoelectric outputs in fluoropolymers. Leveraging this new understanding of nanoscale phenomena at the interface of a fluoropolymer and a 2D sheet now opens up a plethora of research opportunities to design piezoelectric composites with broad applicability, including in wearable energy harvesting,[8] piezo-catalysis,[10] piezo-photonics,[11] and anisotropic sensors.[48] Coupled to the elegant and versatile fabrication methods, our sustainable system can enable bespoke device design in emerging fields for robotic interfaces,[9] biomedical implants,[9] and direct-on-organ printed electronics.[37]

## Methods

### Synthesis of $Ti_3C_2T_x$ dispersion

$Ti_3C_2T_x$ nanosheets were synthesized by selectively etching aluminium atoms out of a $Ti_3AlC_2$ (MAX phase) parent ternary carbide precursor.[49] Briefly, 1 g of $Ti_3AlC_2$ powder (Carbon-Ukraine Ltd., <40 μm) was slowly added into hydrochloric acid (HCl, 20 mL, 9 M) containing lithium fluoride (LiF, 1.6 g, 99.5%). The dispersion was then stirred at room temperature for 24 h to etch out the aluminium from the $Ti_3AlC_2$ MAX phase. Afterwards, the dispersion was repeatedly centrifuged at 3500 rpm (10 min each time) and washed using ultra-pure water to raise the pH of the dispersion. When the pH approached 6, the $Ti_3C_2T_x$ suspension was probe sonicated for 10 min (2 s on and 2 s off) in an ice-bath under an argon gas flow. The dispersion was then centrifuged at 1,500 rpm for 30 min to remove any multi-layer $Ti_3C_2T_x$ and unetched $Ti_3AlC_2$. The supernatant containing single-layer $Ti_3C_2T_x$ nanosheets was concentrated by further centrifuging at 8,000 rpm for 30 min and the sediment was dispersed into DMF. This process was repeated three times to prepare dispersions of $Ti_3C_2T_x$ nanosheets in DMF (at 4.4 mg mL$^{-1}$).

**Molecular dynamics simulations**

All-atom molecular dynamics simulations were used to elucidate the interactions between PVDF-TrFE co-polymer films and the $Ti_3C_2T_x$ nanosheet substrate. Each PVDF-TrFE chain contained 21 VDF and 9 TrFE monomers, corresponding to molar concentrations of 70 mol% VDF and 30 mol% TrFE, similar to our previous work.[8] The co-polymer chains were introduced at 2 nm from the substrate having an initial film density of 1.3 g cm$^{-3}$ using the materials and processes simulations (MAPS) 4.3 platform (Scienomics). The $Ti_3C_2T_x$ monolayer nanosheet-PVDF-TrFE co-polymer film interactions were simulated in the NVT (constant number of atoms, volume, and temperature) ensemble at 298.15 K in a simulation box with periodic boundary conditions. All subsequent simulations were performed on the University of Melbourne's High-Performance Computing system using the large-scale atomic/molecular massively parallel simulator (LAMMPS).[50] The equations of motion were integrated using the velocity-Verlet algorithm[51] with a time step of 1 fs using the transferable, extensible, accurate and modular forcefield (TEAM-FF) for both the $Ti_3C_2T_x$ nanosheet and the PVDF-TrFE co-polymer. Partial charges were assigned to each atom of the PVDF-TrFE using the bond increments method, while the charges for the atoms of the $Ti_3C_2T_x$ surface were adopted from first-principle calculations.[52] The interatomic potential was validated with the density ($\rho$) of bulk PVDF-TrFE melts of various molecular weights at 230 °C, converging to a plateau at approximately 1.42 g cm$^{-3}$ (Supplementary Fig. S2). The obtained density agreed well (deviation <5%) with the experimental value ($\rho_{bulk}$ = 1.49 g cm$^{-3}$) of the bulk PVDF-TrFE co-polymer, validating the TEAM-FF interatomic potential used in the present MD simulations. It should be mentioned that in our simulations the $Ti_3C_2T_x$ nanosheet substrate was kept frozen, interacting with the atoms of the PVDF-TrFE co-polymer chains only *via* van der Waals and electrostatic interactions.

To quantify the strength of the interaction between the $Ti_3C_2T_x$ nanosheet substrate and a single PVDF-TrFE chain, individual NVT MD simulations were carried out by applying a constant force normal and opposite to the $Ti_3C_2T_x$ nanosheet surface to each atom of the PVDF-TrFE chain and gradually increasing it from 0.00 pN to 6.95 pN with a step of 0.695 pN. At each one of these simulations, we monitored the position of the co-polymer chain and recorded at which applied force the chain was fully desorbed from the $Ti_3C_2T_x$ nanosheet surface.

**Extrusion printing ink preparation**

PVDF-TrFE co-polymer powder (75 mol% VDF and 25 mol% TrFE, $M_w$ at 420 kDa, density at 1.49 g cm$^{-3}$, Solvay) and acetone (AR grade, Chem-Supply Pty Ltd) were used without further purification. The stock $Ti_3C_2T_x$ nanosheet dispersion ($Ti_3C_2T_x$ nanosheets in DMF at 4.4 mg mL$^{-1}$) was diluted in acetone to provide solutions with $Ti_3C_2T_x$ nanosheet concentrations of 0.00 mg mL$^{-1}$, 0.10 mg mL$^{-1}$, 0.52 mg mL$^{-1}$, 1.05 mg mL$^{-1}$ and 2.61 mg mL$^{-1}$. PVDF-TrFE co-polymer powder was added to the $Ti_3C_2T_x$ nanosheet/acetone solutions at 40 wt% relative to acetone and stirred at 23 °C until homogeneous inks formed containing $Ti_3C_2T_x$ nanosheet concentrations relative to PVDF-TrFE at 0.00 wt%, 0.02 wt%, 0.10 wt%, 0.20 wt% and 0.50 wt%, respectively. The PVDF-TrFE co-polymer concentration optimization in acetone is presented in the Supplementary Information.

**Extrusion printing and PEG device fabrication**

The $Ti_3C_2T_x$/PVDF-TrFE composite inks (0.00 wt%, 0.02 wt%, 0.10 wt%, 0.20 wt% and 0.50 wt%) were transferred into 3D printing dispensing barrels (30 mL, Optimum, Nordson EFD), then sealed and stored at -5 °C prior to printing. A 3D printer (Bioplotter 3D, Envisiontec) was preset to print a single-layer square film (3 cm x 3 cm) onto a clean glass plate substrate using a raster pattern with a line spacing of 350 μm. During printing, the cartridge containing the ink was kept at 5 °C and the glass plate substrate at 23 °C, the xy speed was 30 mm s$^{-1}$ and extrusion pressure was 1.7 bar through a tapered nozzle with an internal diameter of 200 μm (SmoothFlow, Nordson EFD). After printing the films were placed in a vacuum oven set at 23 °C for 20 min to ensure the complete extraction of the solvent.

To fabricate the PEGs, the SEA extrusion printed films were removed from the glass plate, and coated with a seeding layer of chromium (Cr) and an electrode layer of gold (Au) (total thickness of 60 nm) *via* sputter deposition (Nanochrome I, Intlvac Thin Film Corporation) through a shadow mask on both sides of the film with a total overlapping electrode area of 2.4 cm$^2$. Wires (FLEXI-E 0.15, Stäubli Electrical Connectors AG) were soldered to copper (Cu) foil with conductive adhesive (1181, 3M) and adhered to each side of the printed film without overlap, ensuring strong contact between the Cu tape and Cr/Au coating. The printed films were finally encapsulated in insulating Kapton polyimide tape on both surfaces to complete the PEGs. The extended fabrication details are presented in the Supplementary Information.

## Materials characterization

### Rheology

The rheological characterization of the pristine PVDF-TrFE co-polymer and the $Ti_3C_2T_x$/PVDF-TrFE (0.20 wt%) inks in acetone was undertaken using a strain-controlled rheometer (MCR 702, Anton Paar) with cone-plate geometry at 5 °C. The gap was set at 102 μm, with a cone diameter of 25 mm and an angle of 2°. Frequency and shear strain measurements were undertaken in oscillatory mode with an upwards logarithmic ramp of the frequency under fixed shear strain at 1% and shear strain under fixed frequency at 1 Hz, respectively. The printing simulation measurement was performed in oscillatory mode with the frequency fixed at 1 Hz and controlled shear stress. Initially, 1 Pa shear stress was applied to the inks for 50 s to demonstrate the properties of the inks at rest. Subsequently, 6 kPa shear stress was applied to the sample for 70 s, followed by 1 Pa for 70 s to demonstrate recovery of the ink, repeated for an additional high shear stress cycle.

### Helium Ion Microscopy

The surface of the SEA extrusion printed $Ti_3C_2T_x$/PVDF-TrFE (0.50 wt%) film was imaged using helium ion microscopy (HIM) (Orion NanoFab, Zeiss) in order to determine the orientation and distribution of the $Ti_3C_2T_x$ nanosheets in the PVDF-TrFE co-polymer. The micrographs were obtained with a 100 μm field of view using a dwell time of 0.5 μs and accelerating voltage of 30 kV.

### Tensile Testing

Tensile tests were performed on a mechanical tester (Electroforce 5500, Bose). The $Ti_3C_2T_x$/PVDF-TrFE SEA extrusion printed films (0.00 wt%, 0.02 wt%, 0.10 wt%, 0.20 wt% and 0.50 wt%) were cut into strips (27 mm x 5 mm) and mounted in the grips, with an exposed length at 5 mm. The samples, with average thickness at 45 μm, were extended at 0.01 mm s$^{-1}$ up to a maximum of 11.4 mm.

**Crystallinity**

The crystallinity of the Ti$_3$C$_2$T$_x$/PVDF-TrFE SEA extrusion printed films (0.00 wt%, 0.02 wt%, 0.10 wt%, 0.20 wt% and 0.50 wt%) was obtained using differential scanning calorimetry (DSC). The samples were placed in a ceramic crucible at 25 °C and heated at 10 °C min$^{-1}$ to 200 °C under a nitrogen gas flow (20 mL min$^{-1}$). The crystallinity calculation methods are presented in the Supplementary Information.

**Vibrational Spectroscopy**

The distribution of the Ti$_3$C$_2$T$_x$ nanosheets in the PVDF-TrFE co-polymer, along with the β:γ intensity ratio was confirmed using Raman confocal microscopy (inVia, Renishaw), equipped with a 532 nm laser, an 1,800 line mm$^{-1}$ grating and a 50x objective. Typical maps were obtained for a surface area of 20 μm length and width, at a pixel density of 1 px μm$^{-1}$. Each pixel of the map consisted of a spectrum, centered at 1,300 cm$^{-1}$ and obtained with an exposure time of 0.15 s accumulated over 1,000 scans.

**Piezoresponse Force Microscopy**

Nanoscale polarization of the Ti$_3$C$_2$T$_x$/PVDF-TrFE SEA extrusion printed films (0.00 wt%, 0.02 wt%, 0.10 wt%, 0.20 wt% and 0.50 wt%) was measured using an Asylum Research Cypher ES atomic force microscope. The data was obtained at 23 °C in air using conductive platinum cantilevers (12PT400B, Rocky Mountain Nanotechnology) with spring constant at 0.3 N m$^{-1}$ and tip radius below 20 nm. The scans were undertaken on an area of 5 μm x 5 μm at 256 pixels per line, corresponding to approximately 20 nm per pixel. The piezoresponse force microscopy (PFM) was carried out in contact lithography, whereby a potential was applied by the tip to the sample following a pre-defined pattern, observing the amplitude and phase changes relative to the unbiased areas. The applied potential ranged between -20 V and +20 V in increments of 2 V on each scan line at a scan rate of 0.2 Hz, with each voltage step applied to a total area of 0.195 μm in the x direction and 3.906 μm in the y direction. The resultant amplitude and phase values were then correlated to the corresponding voltage for each pixel of the scan. Extended details of the PFM technique are presented in the Supplementary Information.

**Macroscale Electromechanical Quantification**

Macroscale polarization of the $Ti_3C_2T_x$/PVDF-TrFE PEGs (0.00 wt% and 0.50 wt%) was measured *via* the input of cyclic compressive force and measurement of the resulting surface charges. The force was applied by a mechanical tester (Electroforce 5500, Bose) to the active area of the PEG using a sinusoidal waveform with frequency at 2 Hz, the minimum force set at 5 N and maximum force set at 15 N, corresponding to $\Delta F$ at 10 N and stress at 25 kPa. The surface charge was measured with a charge amplifier (Nexus 2692, Brüel & Kjær) and recorded through a data acquisition system (9223, National Instruments). The cycling stability testing was undertaken at 10 Hz. The extended data and discussion are presented in the Supplementary Information.

# Acknowledgements

This research was supported by the Australian Government through the Australian Research Council's Linkage Projects funding scheme (LP160100071), Future Fellowships funding scheme (FT130100380) and Industry Transformation Research Hub funding scheme (IH140100018). This work was performed in part at the Materials Characterization and Fabrication Platform (MCFP) at the University of Melbourne and the Melbourne Centre for Nanofabrication (MCN) in the Victorian Node of the Australian National Fabrication Facility (ANFF). The authors wish to thank Dr James Bullock and Dr Brett Johnson for performing the dielectric characterization, and Robert Delaney for helpful discussions on energy harvesting methods.

## Author contributions

Conceptualization, N.A.S., P.C.S., J.M.R., and A.V.E.; Methodology, N.A.S., E.N.S., E.G., and J.Z.; Formal Analysis, N.A.S., P.C.S., E.N.S., and E.G.; Investigation, N.A.S., P.C.S., E.N.S., E.G., J.Z., V.C.L., B.I., and K.A.S.U.; Data Curation, N.A.S.; Writing - Original Draft, N.A.S., P.C.S., and A.V.E; Writing - Review and Editing, all authors; Visualization, N.A.S., P.C.S., E.N.S., P.C.S., J.Z., and B.I.; Supervision, A.V.E., J.M.R., G.W.D., and J.G.S.; Project Administration, A.V.E., G.W.D., J.G.S., and J.M.R.; Funding acquisition, A.V.E., G.W.D., J.G.S., and J.M.R.

## Competing interests

There are no competing interests to declare.

## Data availability

The data that support the findings of this study are available from the corresponding author upon reasonable request.

**Supplementary Information**

# Interfacial piezoelectric polarization locking in printable Ti$_3$C$_2$T$_x$ MXene-fluoropolymer composites

Nick A. Shepelin,[a,b] Peter C. Sherrell,[a,b] Emmanuel N. Skountzos,[c,d] Eirini Goudeli,[a] Jizhen Zhang,[e] Vanessa C. Lussini,[f] Beenish Imtiaz,[a] Ken Aldren S. Usman,[e] Greg W. Dicinoski,[f] Joseph G. Shapter,[g] Joselito M. Razal,[e] Amanda V. Ellis *[a,b]

[a]Department of Chemical Engineering, The University of Melbourne, Parkville, Victoria 3010, Australia

[b]BioFab3D, Aikenhead Centre for Medical Discovery, St Vincent's Hospital Melbourne, Fitzroy, Victoria 3065, Australia

[c]Department of Chemical Engineering, University of Patras, Greece

[d]FORTH/ICE-HT, Patras, GR 26504, Greece

[e]Institute for Frontier Materials, Deakin University, Geelong, Victoria 3216, Australia

[f]Note Issue Department, Reserve Bank of Australia, Craigieburn, Victoria 3064, Australia

[g]Australian Institute for Bioengineering and Nanotechnology, The University of Queensland, Brisbane, Queensland 4072, Australia

Email: amanda.ellis@unimeb.edu.au

# Contents

# Ti$_3$C$_2$T$_x$ MXene nanosheets and properties

The X-ray powder diffraction (XRD) patterns of the Ti$_3$AlC$_2$ MAX phase and the Ti$_3$C$_2$T$_x$ MXene nanosheets were obtained using a powder diffractometer (X'Pert Powder, PANalytical) equipped with a Cu Kα radiation (40 kV, 30 mA) with an X-ray wavelength (λ) of 1.54 Å at a 2θ scan step of 0.013°. Transmission electron microscopy (TEM) (JEM-2100, JEOL, Ltd.) characterization was employed to study the exfoliated Ti$_3$C$_2$T$_x$ nanosheets. Topographical atomic force microscopy (AFM) images were obtained using Bruker's proprietary ScanAsyst scan mode (MultiMode 8-HR, Bruker) to measure the Ti$_3$C$_2$T$_x$ nanosheet thickness. AFM samples were prepared by drop casting the Ti$_3$C$_2$T$_x$ nanosheet in DMF solutions onto clean silicon wafers. Dynamic light scattering (DLS) was performed using a Zetasizer (Nano ZS, Malvern Instruments) to measure the size distribution of the Ti$_3$C$_2$T$_x$ nanosheets. The X-ray photoelectron spectroscopy (XPS) data on the Ti$_3$C$_2$T$_x$ nanosheets were acquired using an AXIS Nova (Kratos Analytical Ltd.) equipped with a monochromated Al Kα source (hν = 1486.6 eV) operating at 150 W at a step of 0.1 eV.

The Ti$_3$C$_2$T$_x$ MXene nanosheets were synthesized from the Ti$_3$AlC$_2$ parent ternary carbide precursor (MAX phase) by selective etching of the aluminium layer (A-group element) using a mixture of lithium fluoride (LiF) and hydrochloric acid (HCl) at room temperature for 24 h.[1–3] The subsequent intercalation of water (H$_2$O) molecules and Li$^+$ ions within the negatively charged surface resulted in a volume increase during washing with ultra-pure water, indicating the self-delamination of multi-layered Ti$_3$C$_2$T$_x$ to few/single layers.[4] The delamination and the removal of the aluminium was confirmed by the downshifting of the (002) peak and a disappearance of the aluminium peak at 2θ of 39° in the XRD spectra (Fig. S1a).[5] The TEM image (Fig. S1b) and DLS data (Fig. S1c) showed that the Ti$_3$C$_2$T$_x$ nanosheets exhibited an average lateral size of approximately 310 nm. The AFM image showed that the Ti$_3$C$_2$T$_x$ nanosheets exhibited a clean surface at the edge (Fig. S1d). The thickness profile of the $Ti_3C_2T_x$ nanosheets (Fig. S1d, inset) showed an average height of 1.6 nm, corresponding to single-layer $Ti_3C_2T_x$ nanosheets.[1]

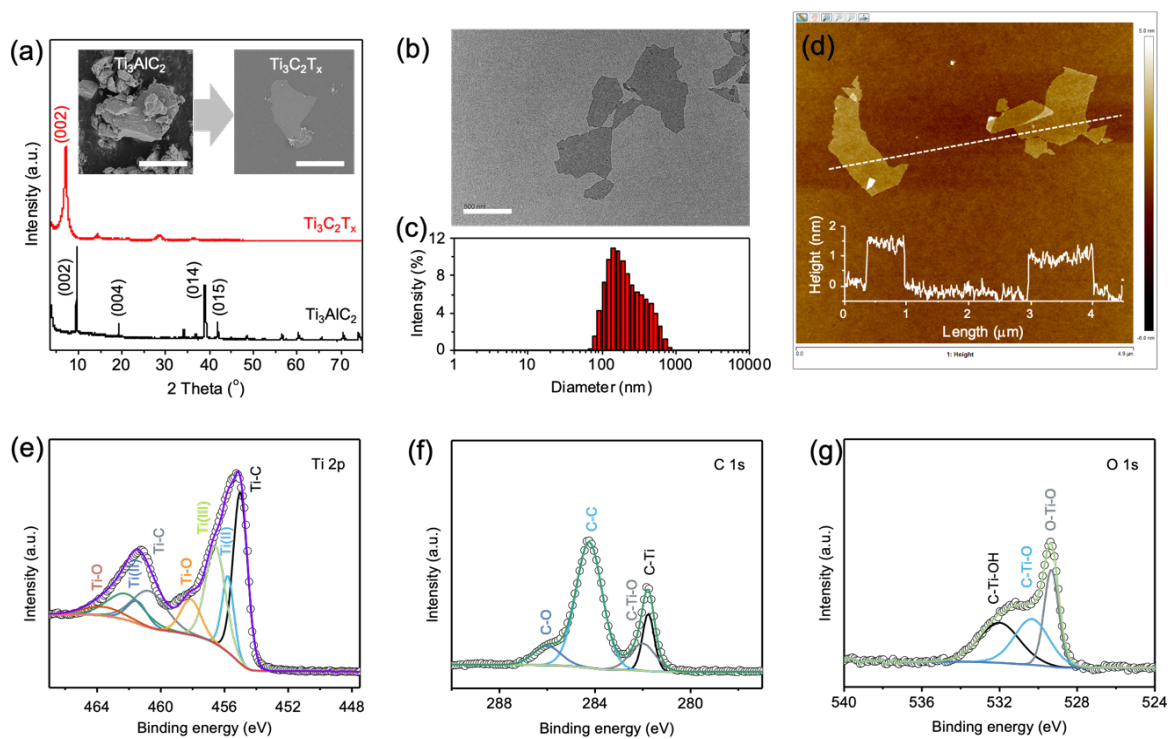

**Fig. S1:** $Ti_3C_2T_x$ MXene nanosheet characterization. **a** XRD pattern and SEM images (insets) of $Ti_3AlC_2$ MAX phase and $Ti_3C_2T_x$ nanosheets after etching. Scale bar for SEM of $Ti_3AlC_2$ represents 5 μm and that of $Ti_3C_2T_x$ nanosheets represents 1 μm. **b** TEM image of single-layer 2D $Ti_3C_2T_x$ nanosheets. Scale bar represents 500 nm. **c** The DLS result showing the hydrodynamic size of the $Ti_3C_2T_x$ nanosheets in water. **d** AFM image of the $Ti_3C_2T_x$ nanosheets deposited on a silicon wafer. Inset, represents the thickness profile along the line indicated. Deconvolution of high-resolution XPS spectra for the **e** Ti 2p, **f** C 1s and **g** O 1s orbitals, showing the surface termination and state of the $Ti_3C_2T_x$ nanosheets.

XPS of the $Ti_3C_2T_x$ nanosheets (Fig. S1e-g) revealed surface termination dominated by Ti-O and O-Ti-O bonding. The Ti 2p region (Fig. S1e) showed doublets corresponding to Ti-C, Ti (III), Ti (II) and Ti-O bonding. The dominant Ti-C peak arises from the bridging C atoms between Ti atoms, whereas the Ti-O peaks corresponds to surface termination functional groups of hydroxides (Ti-OH) or epoxide (Ti-O-Ti) structures.[6] The C 1s spectral region (Fig. S1f), shows four singlet peaks, C-Ti, C-Ti-O, C-C, and C-O. The C-Ti peak corresponds to internal bridging C atoms. The C-Ti-O peak, occurring at slightly higher binding energies, arises from the long-range influence of oxygen-based surface termination on the electronic state of the internal C atoms. The C-C signal, while anomalous given the crystal structure of $Ti_3C_2T_x$, is always observed in literature[7] and is understood to arise from residual hydrocarbons[8] in the XPS chamber. The C-O peak occurs as Ti is an extremely mobile metal, known to leave vacancies and thus slightly altered stoichiometry.[9] These vacancies result in C-O bonding in the top or bottom Ti metal layer. The O 1s region (Fig. S1g) confirms the predominant binding of O moieties to Ti atoms in the form of O-Ti-O, which can correspond to either hydroxide (favorable) or epoxide (unfavorable) surface terminations. Higher binding energy peaks for C-Ti-O and C-Ti-OH suggest hydroxy termination is dominant of the surface of the flakes.

**Preparation of PVDF-TrFE and $Ti_3C_2T_x$/PVDF-TrFE inks**

Recently, we described the dissolution and recycling of SEA extrusion printed PVDF-TrFE co-polymer films using acetone as the only solvent.[10] Here, *N,N*-dimethylformamide (DMF) was completely eliminated as a solvent for extrusion printing entirely and replaced by acetone. Acetone has inherent advantages over DMF and other solvents commonly used to dissolve fluoropolymers, with faster evaporation rates that enable rapid crystallization and drying of SEA extrusion printed polymer films.[11] In particular, DMF exhibits a high boiling point (>150 C at 101.3 kPa),[12] low vapor pressure (<0.5 kPa at 21 C)[13] and high toxicity.[12] Comparatively, acetone exhibits a low boiling point (56 C at 101.3 kPa),[14] high vapor pressure (26 kPa at 21 C)[13] and reduced toxicity, reported as one of the least toxic industrial solvents,[14] and is thus better suited for SEA extrusion printing.

Initially, pristine PVDF-TrFE inks were prepared in acetone, which were used to optimize the SEA extrusion printing parameters. These inks were prepared by a simple mixing method, whereby PVDF-TrFE powder was slowly added to acetone under mechanical stirring. The pristine PVDF-TrFE co-polymer inks were prepared at PVDF-TrFE co-polymer concentrations of 35 wt%, 40 wt% and 45 wt%, based on the concentrations of inks prepared in the previously reported DMF:acetone solvent mixture.[15] The prepared inks were viscous (Fig. S2), moving slower when tilted to a 45° angle as the concentration increased. The rheological optimization of the inks for printing is shown further in this document.

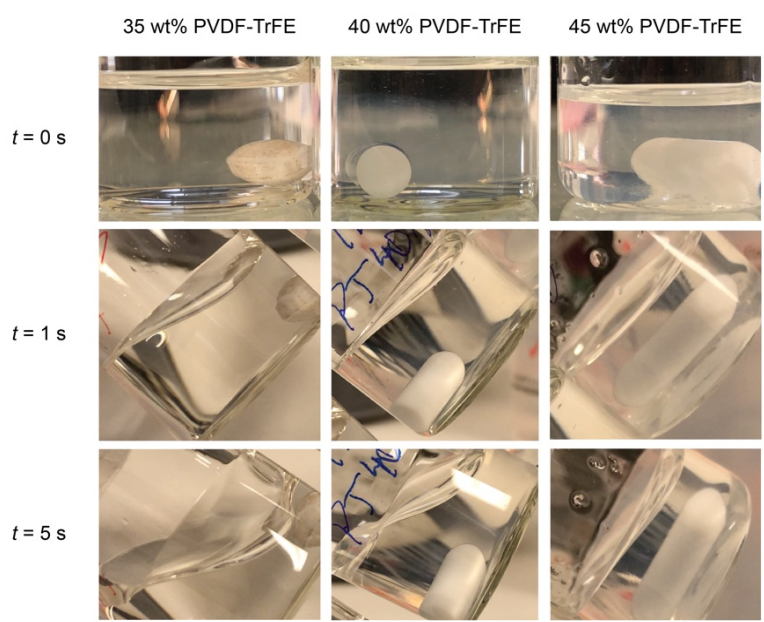

**Fig. S2:** Photographs of the 35 wt% (left column), 40 wt% (middle column) and 45 wt% (right column) PVDF-TrFE/acetone inks, demonstrating the viscosity of the inks before rotating the vial to 45° (top row, $t = 0$ s), 1 s after rotating the vial (middle row) and 5 s after rotating the vial (bottom row).

Similarly, $Ti_3C_2T_x$/PVDF-TrFE inks were prepared at $Ti_3C_2T_x$ concentrations at 0.02 wt%, 0.10 wt%, 0.20 wt% and 0.50 wt%. Here, a small aliquot of the $Ti_3C_2T_x$ stock dispersion in DMF (4.4 mg mL$^{-1}$) was added to acetone to form dispersions at 0.00 mg mL$^{-1}$, 0.10 mg mL$^{-1}$, 0.52 mg mL$^{-1}$, 1.05 mg mL$^{-1}$ and 2.61 mg mL$^{-1}$ in acetone. Subsequently, the PVDF-TrFE powder was added slowly to the $Ti_3C_2T_x$ dispersions in acetone at 23 °C while stirring, at 40 wt% relative to the mass of the dispersion, to form the $Ti_3C_2T_x$/PVDF-TrFE inks. The inks were stirred until homogeneous, then sealed with parafilm and stored at -5 °C to minimize solvent evaporation.

Throughout the experimental procedure, the stability of the $Ti_3C_2T_x$ nanosheet dispersion was monitored in the $Ti_3C_2T_x$/PVDF-TrFE ink, for up to five months (Fig. S3, middle). The $Ti_3C_2T_x$/PVDF-TrFE ink was compared to a single-walled carbon nanotube (SWCNT)/PVDF-TrFE ink, which we have recently reported (Fig. S3, right).[10] Notably, after five months of storage, all three inks exhibited similar flow properties to recently prepared inks. The SWCNTs were found to aggregate in the SWCNT/PVDF-TrFE ink, causing occasional blocking of the nozzle during printing. Conversely, the $Ti_3C_2T_x$ nanosheets showed minimal aggregation in the $Ti_3C_2T_x$/PVDF-TrFE ink due to exceptional electrostatic interactions between the $Ti_3C_2T_x$ nanosheets and the PVDF-TrFE co-polymer (Fig. S5, Supplementary Video S1). The $Ti_3C_2T_x$/PVDF-TrFE ink could be printed following long-term storage with no required changes in the extrusion printing parameters.

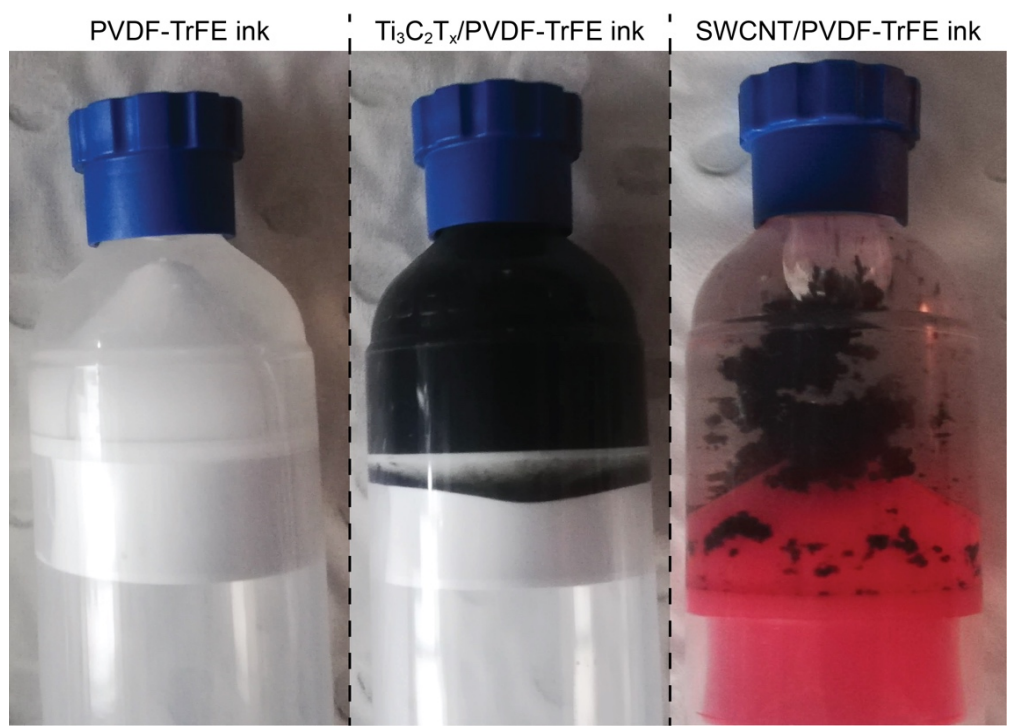

**Fig. S3:** Comparison of the stability of the PVDF-TrFE ink (left), Ti$_3$C$_2$T$_x$/PVDF-TrFE ink (middle) and SWCNT/PVDF-TrFE ink (right), five months after preparation of the ink.

# Molecular dynamics (MD) modelling of the interface between the Ti$_3$C$_2$T$_x$ nanosheets and the PVDF-TrFE co-polymer

The density of the PVDF-TrFE co-polymer melt was investigated as a function of the monomer units (alternatively the molecular weight) to validate the interatomic potential used for the simulations (Fig. S4). The density was found to increase with increasing number of monomer units, reaching an asymptotic plateau corresponding to 1.42 g cm$^{-3}$. The value obtained using MD simulations was in excellent agreement with the value of 1.49 g cm$^{-3}$ provided by the manufacturer of the PVDF-TrFE co-polymer (Solvay), validating the interatomic potential used in the MD simulations.

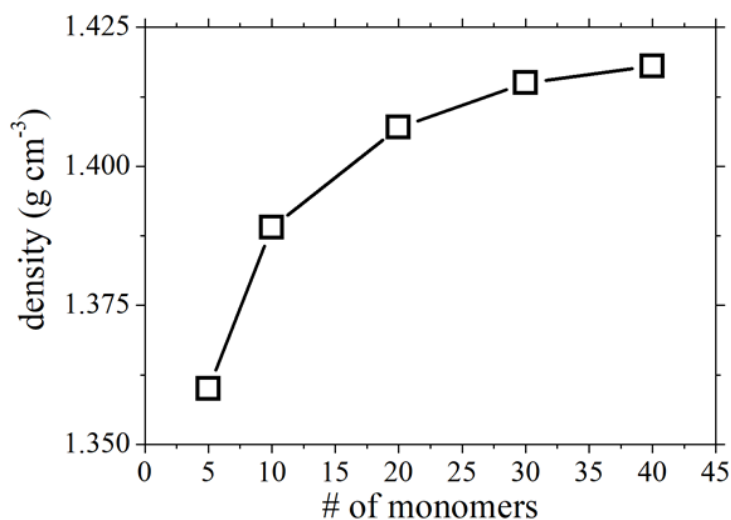

**Fig. S4:** Evolution of the density of the PVDF-TrFE co-polymer melt as a function of the number of monomers.

The distribution of the local density of the PVDF-TrFE co-polymer film (70 chains) was investigated as a function of distance from the substrate, for a graphene substrate (Fig. S5, black line) and a Ti$_3$C$_2$T$_x$ nanosheet substrate (Fig. S5, red line). The local density distribution was calculated from the mass within each separation distance interval, normalized to the volume within said interval and averaged over the duration of the simulation (1.8 ns timespan).

The shaded regions correspond to the minimum and maximum values of the local density at each separation distance interval relative to the graphene or $Ti_3C_2T_x$ nanosheet substrate. The layer adjacent to the substrate was found to adsorb to both the graphene and the $Ti_3C_2T_x$ nanosheet substrates, exhibiting a local density of approximately 2.3 g cm$^{-3}$ and 1.6 g cm$^{-3}$, respectively. The larger local density of the PVDF-TrFE co-polymer film adjacent to the graphene substrate indicates that the PVDF-TrFE co-polymer chains are more packed than those adjacent to the $Ti_3C_2T_x$ nanosheet substrate, as the latter possesses an increased surface roughness due to the OH termination, thus inducing steric effects in the PVDF-TrFE co-polymer chains. The local density of the layers further away from the substrate is practically identical in both the $Ti_3C_2T_x$/PVDF-TrFE and the graphene/PVDF-TrFE systems. Notably, the PVDF-TrFE co-polymer film adsorbs closer to the $Ti_3C_2T_x$ nanosheet relative to graphene, as the first local density peak appears at a lower separation. This decreased separation indicates a stronger attractive interaction between the $Ti_3C_2T_x$ nanosheet and the PVDF-TrFE co-polymer in comparison to the graphene substrate.

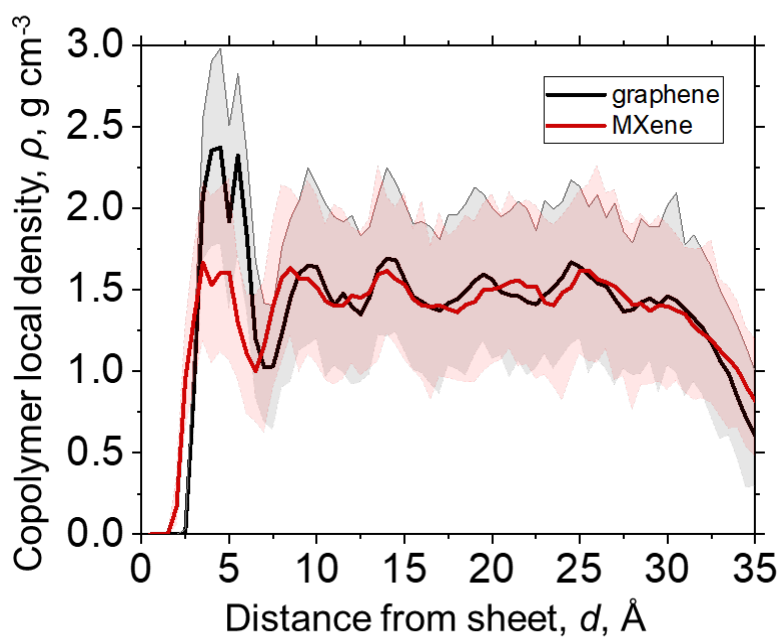

**Fig. S5:** Distribution of the local density of the PVDF-TrFE co-polymer film (70 chains) as a function of the distance from the $Ti_3C_2T_x$ nanosheet (red line) and the graphene sheet (black line) substrates.

This enhanced adhesion phenomenon was investigated by adhesion strength studies. Here, a PVDF-TrFE co-polymer chain was placed in close proximity to either the $Ti_3C_2T_x$ nanosheet or graphene sheet substrate. A force was applied to the PVDF-TrFE co-polymer perpendicular to the basal plane of the substrate to measure the desorption force. The force was increased from 0.00 pN to 6.95 pN with a step size of 0.695 pN, monitoring the position of the PVDF-TrFE co-polymer chain for the desorption from the substrate. It was found that the PVDF-TrFE co-polymer chain desorbed form the graphene substrate at approximately 2.78 pN, whereas the required desorption force increased on the $Ti_3C_2T_x$ nanosheet to approximately 4.17 pN, indicating a greater adhesion strength at the interface between the $Ti_3C_2T_x$ nanosheet and the PVDF-TrFE co-polymer.

The distribution of the H and F atoms in the PVDF-TrFE co-polymer film was further investigated as a function of the distance from the $Ti_3C_2T_x$ nanosheet substrate to investigate whether preferential orientation of these dipolar atoms in the PVDF-TrFE co-polymer were giving rise to the polarization locking mechanism (Fig. S6). The datapoints represent average values for 14 PVDF-TrFE co-polymer chains and the shaded areas represent the minimum and maximum number of H and F atoms over the entire simulation. The H and F probability distributions were observed to be approximately equal at all distances from the $Ti_3C_2T_x$ nanosheet substrate, indicating the PVDF-TrFE did not preferentially orient on the substrate. Small deviations at a low distance (up to 2 Å) were observed, whereby the H atoms were found closer to the $Ti_3C_2T_x$ nanosheet substrate relative to the F atoms. This was attributed to the shortest non-covalent hydrogen bonds between the H atoms of the PVDF-TrFE co-polymer and the hydroxyl terminations ($T_x$) of the $Ti_3C_2T_x$ nanosheet substrate.

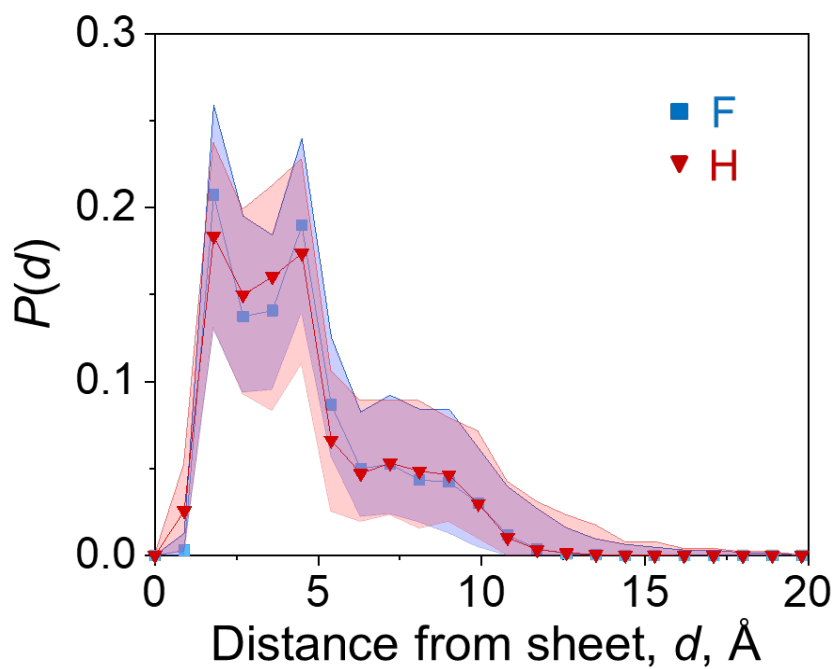

**Fig. S6:** Probability distribution ($P(d)$) of the number of F (squares) and H atoms (triangles) within the PVDF-TrFE co-polymer film (14 chains) as a function of the distance from the $Ti_3C_2T_x$ nanosheet substrate ($d$). The shaded areas represent the minimum and maximum number of F and H atoms over the entire simulation.

To understand the phase distributions of the PVDF-TrFE co-polymer film adjacent to the $Ti_3C_2T_x$ nanosheet substrate, the probability distributions of the dihedral angles were monitored for a 14-chain PVDF-TrFE co-polymer film (Fig. S7). The PVDF-TrFE co-polymer consists of three commonly found phases, namely the α phase (non-polar), γ phase (semi-polar) and the β phase (highly polar).[16] These phases correspond to spatial conformation of the bonds, either *trans* (T) or *gauche* (G). The α phase is thermodynamically favored in fluoropolymers, due to its *trans-gauche* (TGTG'TGTG') conformation, which consists of 50% *trans* bonds and 50% *gauche* bonds.[17] Conversely, the β phase is an all-*trans* (TTTTTTTT) conformation, which spatially separates the H moieties on one C atom from the F atoms on the adjacent C atom, giving rise to a strong H-F dipole moment.[18] The γ phase is a stable intermediate state between the α phase and the β phase, as evidenced by the 75% *trans* and 25% *gauche* fraction (TTTGTTTG'), giving rise to dipole moments which result in a lower maximum polarization relative to the β phase.[17] Hence, the distribution of the dihedral angles and subsequently the phase fractions can provide insight on the changes in local electroactivity of the PVDF-TrFE co-polymer film adjacent to the $Ti_3C_2T_x$ nanosheet substrate.

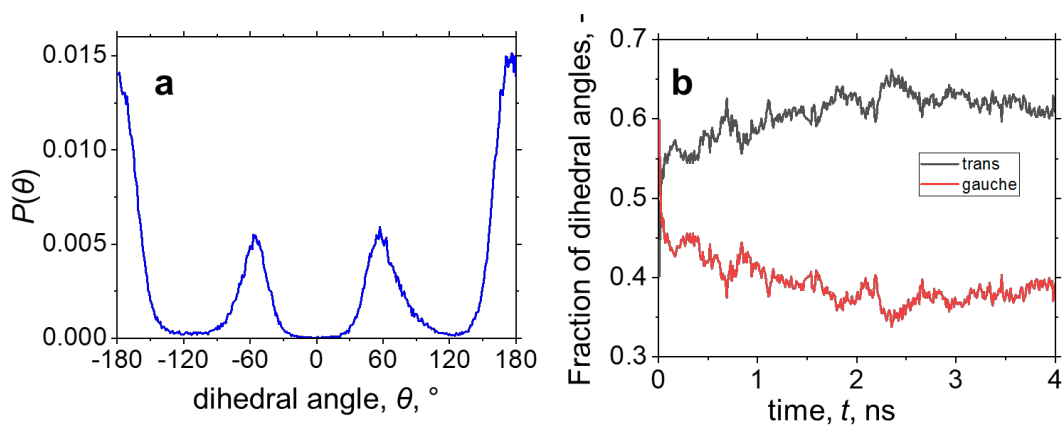

**Fig. S7:** Bond angle distributions of the PVDF-TrFE co-polymer on the $Ti_3C_2T_x$ nanosheet substrate. **a** Probability distribution of the dihedral angles averaged oven 4 ns for a film of 14 PVDF-TrFE chains interacting with an immobile $Ti_3C_2T_x$ nanosheet. **b** Time evolution of the fraction of the *trans* (black line) and *gauche* angles (red line).

The distribution of the dihedral angles (Fig. S7a) was taken as the average of all angles in a 14-chain PVDF-TrFE co-polymer film interacting with an immobile $Ti_3C_2T_x$ nanosheet substrate over the span of the simulation (4 ns). The PVDF-TrFE co-polymer chains exhibited configurations at four main dihedral angles, ±180° (*trans*) and ±60° (*gauche*).[18] As a function of simulation time, the fraction of *trans* conformation (Fig. S7b, black line) was found to increase and attain a final value of approximately 63%, whereas approximately 37% of the bonds were observed in the *gauche* conformation (Fig. S7, red line). These values correspond to either a majority of α phase (74%) with low prevalence of the β phase (26%), or a near-even distribution of the α phase (48%) and gamma phase (52%), or a combination of the two. Importantly, while the PVDF-TrFE generally crystallizes into the β phase due to the third F atom in the TrFE monomer, these values suggest a large presence of the α phase.[10] Indeed, at the local level, the experimental data observed the presence of the α and γ phases adjacent to the $Ti_3C_2T_x$ nanosheet (Fig. 4d, e); however, the FTIR (Fig. S16) and XRD (Fig. S17, Fig. S18) data presented further in this document suggest the β phase as the primary conformation in the bulk of the $Ti_3C_2T_x$/PVDF-TrFE composites.

Similarly, the temporal evolution of the dihedral angles was repeated for a 70-chain PVDF-TrFE co-polymer film on the $Ti_3C_2T_x$ nanosheet or graphene substrate (Fig. S8). Similar to the 14-chain PVDF-TrFE co-polymer films (Fig. S7b), the larger films on a $Ti_3C_2T_x$ nanosheet substrate exhibited a larger *trans* fraction (approximately 65%) relative to the *gauche* fraction (approximately 35%). Interestingly, when simulated adjacent to a graphene substrate, the same 70-chain PVDF-TrFE copolymer film exhibited a lower fraction of *trans* bonds (approximately 57%) and subsequently a higher fraction of *gauche* bonds (approximately 43%).

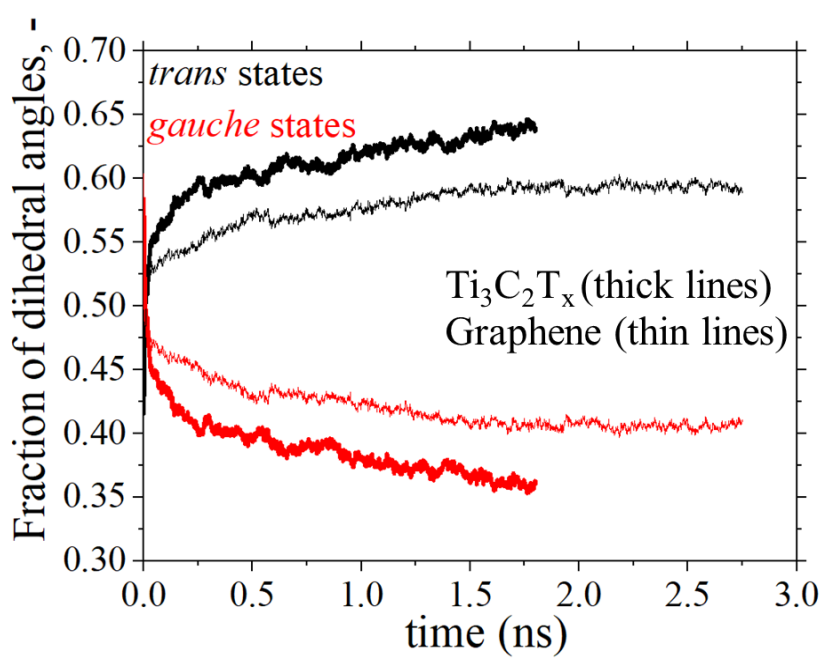

**Fig. S8:** The temporal evolution of the fraction of *trans* bond angles (black lines) and *gauche* bond angles (red lines) in the 70-chain PVDF-TrFE co-polymer film on a $Ti_3C_2T_x$ nanosheet substrate (thick lines) and graphene substrate (thin lines).

The evolution of the polarization angle ($\theta$) in the PVDF-TrFE co-polymer film relative to the basal plane of the $Ti_3C_2T_x$ nanosheet substrate was investigated as a function of the simulation time and the distance from the substrate (Fig. S9). The $\theta$ was obtained as a function of time for PVDF-TrFE co-polymer chains within 24 Å (Fig. S9, green line), 29 Å (Fig. S9, red line), 34 Å (Fig. S9, blue line) and 39 Å (Fig. S9, grey line) from the $Ti_3C_2T_x$ nanosheet substrate. The PVDF-TrFE co-polymer chains closer to the substrate (<24 Å) exhibited a broad range of $\theta$ values, found to sporadically change orientation throughout the simulation. Conversely, as the separation from the $Ti_3C_2T_x$ increased to 39 Å, the $\theta$ was found to orient perpendicular to the basal plane of the $Ti_3C_2T_x$. Additionally, the orientation of the polarization vector at 39 Å was not found to significantly deviate from the perpendicular orientation throughout the equilibrated region ($t > 0.5$ ns) of the simulation, indicating that the polarization locking spontaneously occurs near the interface of the two materials directly upon contact (in solution) and maintains the perpendicular orientation in the solid state (after deposition).

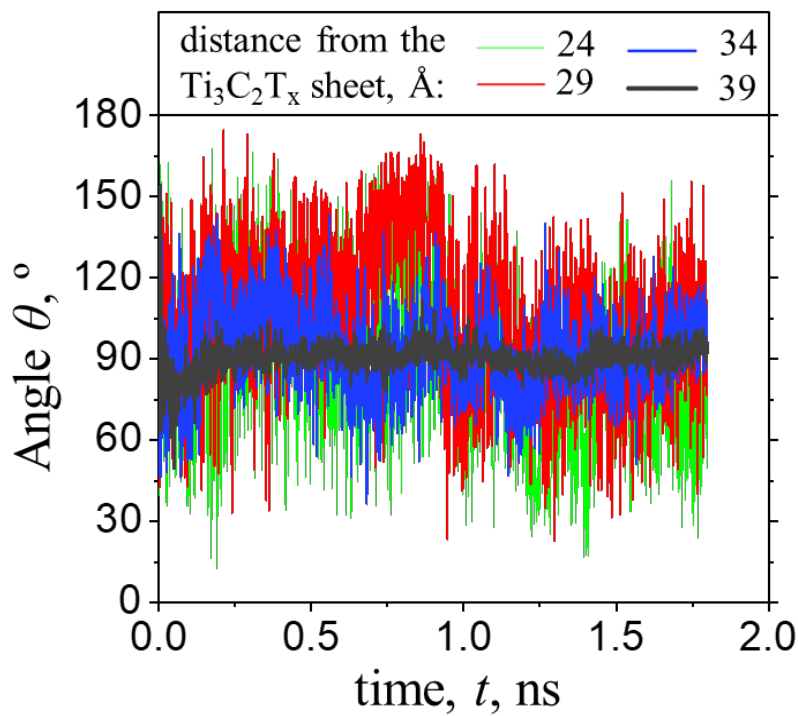

**Fig. S9:** Time evolution of the angle ($\theta$) between the total dipole vector (polarization) in the PVDF-TrFE co-polymer film and the $Ti_3C_2T_x$ nanosheet basal plane, for PVDF-TrFE co-polymer chains that are within 24 Å (green line), 29 Å (blue line), 34 Å (blue line) and 39 Å (grey line) from the $Ti_3C_2T_x$ nanosheet substrate.

The polarization vectors of five individual PVDF-TrFE co-polymer chains are shown exemplarily for the $Ti_3C_2T_x$ and graphene substrates in Supplementary Video S2, indicating a correlation between their structure and their corresponding polarization vector. Here, the rest of the PVDF-TrFE co-polymer chains (65 chains out of the 70mer) are omitted for clarity. The chains on the $Ti_3C_2T_x$ substrate are clearly more elongated than those on the graphene, attaining individual chain polarization vectors perpendicular to the basal plane $Ti_3C_2T_x$, corresponding well to the total polarization vector of the 70-chain film. Contrary to the PVDF-TrFE co-polymer chains on the $Ti_3C_2T_x$ nanosheet substrate, those on top of the graphene substrate exhibit coiled morphology with the individual polarization vectors attaining randomized orientations.

# Rheological printing optimization of pristine PVDF-TrFE in acetone

To optimize the ink system (PVDF-TrFE in acetone) for SEA 3D printing, the rheological properties of the inks were first studied for PDVF-TrFE (35 wt%, 40 wt% and 45 wt%) loadings. PVDF-TrFE powder (75 mol% VDF, 25 mol% TrFE, $M_w$ = 420 kDa) was slowly added into acetone and stirred at 23 °C until the powder completely dissolved, forming viscous inks (Fig. S2). The rheology of these inks was assessed using an MCR 702 rheometer (Anton Paar GmbH) in a cone-plate geometry, with a cone diameter of 25 mm, a cone angle of 2° and a gap at 102 μm (CP25-2, Anton Paar GmbH). The temperature in all measurements was held at 5°C.

**Steady state rheology**

Initial steady-state logarithmic shear rate ramps were used to compare the viscosity ($\eta$) of the PVDF-TrFE ink in acetone to that of the commonly reported solvent mixture of DMF and acetone (40:60 vol%), with polymer concentration at 35 wt% (Fig. S10a).[10,15] Both inks showed non-Newtonian (shear thinning) behavior, which is required for extrusion printing.[19] At a shear rate of 0.01 s$^{-1}$, which corresponded to the resting state (prior to and post printing), the $\eta$ of the ink in the DMF:acetone solvent system was measured at 670 Pa s, drastically lower than that of the ink in acetone as the solvent, measured at 430,000 Pa s. The extreme increase in the viscosity at low shear represents a three order of magnitude increase in shape retention capability of the ink directly upon printing, further aided by the faster evaporation rate of acetone relative to DMF. Interestingly, the viscosity of the ink with acetone as the only solvent exhibited a lower $\eta$ (7 Pa s) at high shear rate (1,000 s$^{-1}$, corresponding to conditions during printing) relative to the ink with DMF:acetone as the solvent system (12 Pa s). This signifies a lower pressure is required to extrude the same volume of ink, following the Hagen-Poiseuille equation.[20] This initial testing confirmed the significant improvement in the rheological properties of the PVDF-TrFE/acetone ink relative to the PVDF-TrFE/(DMF:acetone) ink and further suggested the formation of a gel, consistent with prior reports of acetone as a swelling agent for fluoropolymers.[21]

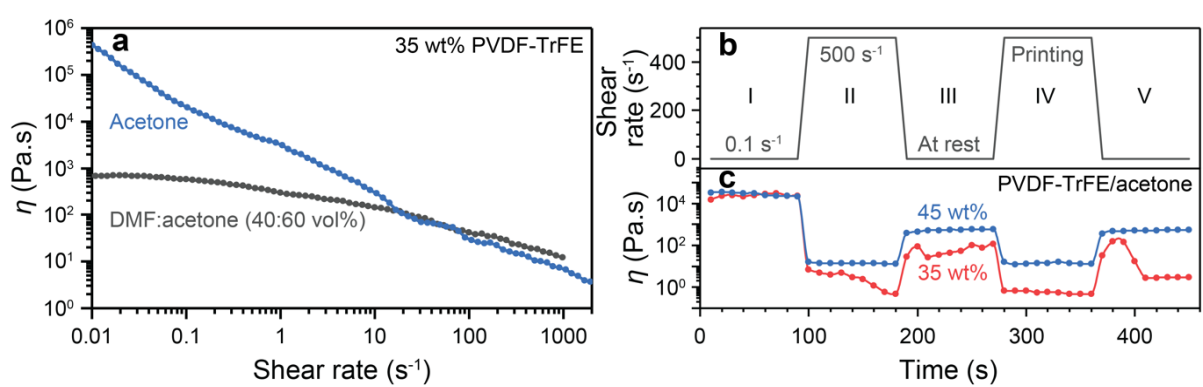

**Fig. S10:** Steady-state rheology. **a** The shear rate sweeps for inks of PVDF-TrFE/(DMF:acetone, 40:60 vol%) and PVDF-TrFE/acetone, containing PVDF-TrFE (35 wt%). **b** The applied shear rate profile as a function of time and **c** the resultant viscosity for PVDF-TrFE/acetone inks with PVDF-TrFE concentration at 35 wt% and 45 wt%.

Additional steady state testing was undertaken on the PVDF-TrFE/acetone ink at two concentrations, 35 wt% and 45 wt%, to simulate the flow and recovery parameters found during extrusion printing (Fig. S10b, c). Here, the shear rate was kept constant for 90 s, cycling between 1 $s^{-1}$ and 500 $s^{-1}$ (Fig. S10b), and measuring the $\eta$ (Fig. S10c). Both inks exhibited similar $\eta$ values in region I (between 21,000 Pa s and 23,000 Pa s at 90 s), decreasing under increased shear in region II to 16 Pa s in the 45 wt% ink and 7 Pa s in the 35 wt% ink. Notably, the $\eta$ of the 35 wt% ink exhibited further decreases at the constant shear rate to 0.5 Pa s, which is likely to arise from elongation and disentanglement of the polymer chains. Region III showed partial recovery in both inks to 588 Pa s (2.7%) and 113 Pa s (0.5%) in the 45 wt% and 35 wt% inks, respectively. This suggests the higher concentration of PVDF-TrFE assists in stabilizing the entanglement in the polymer chains; however, the application of shear nonetheless reduces the entanglement between the polymer chains, correlating to a pseudo-1D material.[22] In region IV, the 35 wt% ink was found to drop in $\eta$ to the lower value of that in region II, suggesting that the disentanglement is irreversible, whereas the $\eta$ of the 45 wt% ink was found to be consistent throughout the region, with the same values as region II. Surprisingly, upon decrease in shear rate in region V, the $\eta$ of the 35 wt% ink decreased to similar values as the high shear rate region II, unable to reliably recover to the values of region III, confirming the disentanglement effects and therefore proving unsuitable for a printing system where the printed ink must retain its shape.

**Oscillatory rheology**

Oscillatory rheology was employed at 5 °C to further probe the hypothesis of gel formation and determine the flow parameters in the PVDF-TrFE/acetone inks with PVDF-TrFE concentration at 35 wt%, 40 wt% and 45 wt% (Fig. S11).[19] Oscillatory frequency ($\omega$) sweeps were performed (Fig. S11a-c), which can give insight into the time-dependent flow properties of the inks.[23] The tests were performed with fixed shear strain ($\gamma_s$) at 1%. All measured samples exhibited a similar trend in the storage ($G'$) and loss ($G''$) moduli as a function of the $\omega$, confirming the increased $\eta$ (Fig. S10a) relative to PVDF-TrFE/(DMF:acetone) inks was due to enhanced swelling of the fluoropolymer, which has lower dependence on the fluoropolymer concentration.[24] Furthermore, minimal deviation in the slope of $G'$ and $G''$ over the entire measured $\omega$ range strongly suggested the formation of a strongly bound gel, which was solid-like ($G' > G''$) for all measured frequencies.[23]

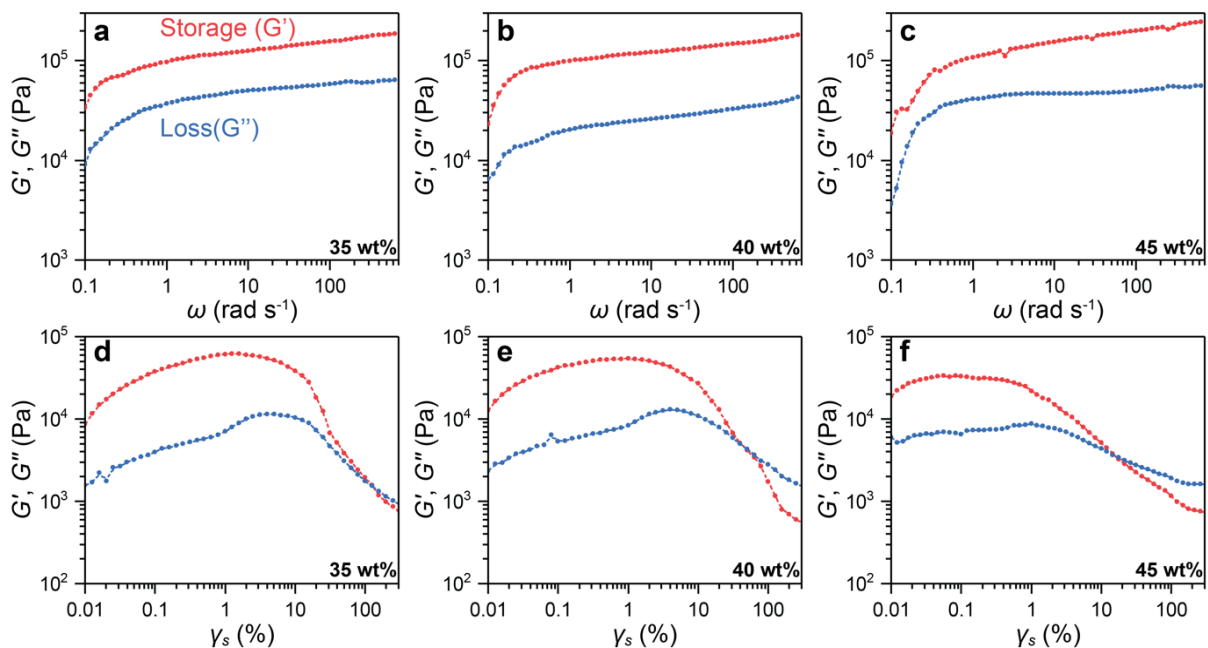

**Fig. S11:** Oscillatory rheology measurements on PVDF-TrFE/acetone inks. **a-c** The frequency sweeps for **a** 35 wt%, **b** 40 wt% and **c** 45 wt% PVDF-TrFE/acetone inks, obtained at fixed strain ($\gamma_s$ = 1%). **d-f** The shear strain sweeps for **d** 35 wt%, **e** 40 wt% and **f** 45 wt% PVDF-TrFE/acetone inks, obtained at fixed frequency ($\omega$ = 1 Hz).

Similarly, the oscillatory $\gamma_s$ sweeps (Fig. S11d-f) at low $\omega$ (1 Hz) exhibited similar characteristics between all three PVDF-TrFE concentrations in acetone. All three tested samples showed solid-like behavior ($G' > G''$) at low $\gamma_s$, followed by a liquid-like region ($G' < G''$) at high $\gamma_s$ (>100%).[19] Interestingly, the flow point ($\gamma_s$ at cross-over of $G'$ and $G''$) was found to decrease with increasing concentration, which, while counterintuitive, suggests the strong binding between the PVDF-TrFE and acetone.[25] As the concentration increases, the number of polymer-solvent contact points decreases (increasing polymer-polymer binding points), therefore the gel becomes weakened and is able to flow with a lower $\gamma_s$. In translating this theory to extrusion printing, all of the three tested inks were suitable for printing; however, a lower flow point would decrease the required pressure input to extrude the sample, meaning the inks with higher PVDF-TrFE concentration are preferred for the printing.[20]

Finally, oscillatory shear stress ($\sigma_s$) cycling was undertaken to probe the recovery parameters of $G'$ and $G''$ within the inks and determine the optimal PVDF-TrFE concentration (35 wt%, 40 wt% or 45 wt%) in acetone for extrusion printing (Fig. S12).[19] Here, the $\sigma_s$ was cycled at constant $\omega$ (1 Hz) between 1 Pa and 5 kPa, representing the induced $\sigma_s$ at rest and during printing, respectively, held constant for at least 70 s (Fig. S12a). At PVDF-TrFE (35 wt%), the ink was unable to maintain $\sigma_s$ at 6 kPa, whereas the inks containing 40 wt% and 45 wt% PVDF-TrFE exhibited consistent response to the input $\sigma_s$. The value for the *tan(δ)*, or the ratio of $G''$ and $G'$ was <1 (marked by grey horizontal line) for all samples at 1 Pa $\sigma_s$ and increased to > 1 upon the application of 5 kPa $\sigma_s$ for 70 s (Fig. S12b). For the inks containing 40 wt% and 45 wt% PVDF-TrFE, the *tan(δ)* remained constant throughout the 5 kPa $\sigma_s$ region and completely recovered for all concentrations after 70 s at 1 Pa $\sigma_s$. During the second cycle, the 35 wt% ink was found to flow with the lowest resistance, represented by a *tan(δ)* value of 20,000 (instrument limit), whereas the 40 wt% and 45 wt% inks retained similar values to the first cycle. Fig. S12c shows the complex viscosity ($\eta^*$) of the PVDF-TrFE inks. As expected, the starting $\eta^*$ was found to increase with increasing PVDF-TrFE concentration. During the first high $\sigma_s$ cycle, the $\eta^*$ was found to decrease significantly for the 35 wt% ink as a function of time and unable to recover to initial values in the subsequent low stress cycle. Conversely, the 40 wt% and 45 wt% PVDF-TrFE inks exhibited full recovery after two high stress cycles.

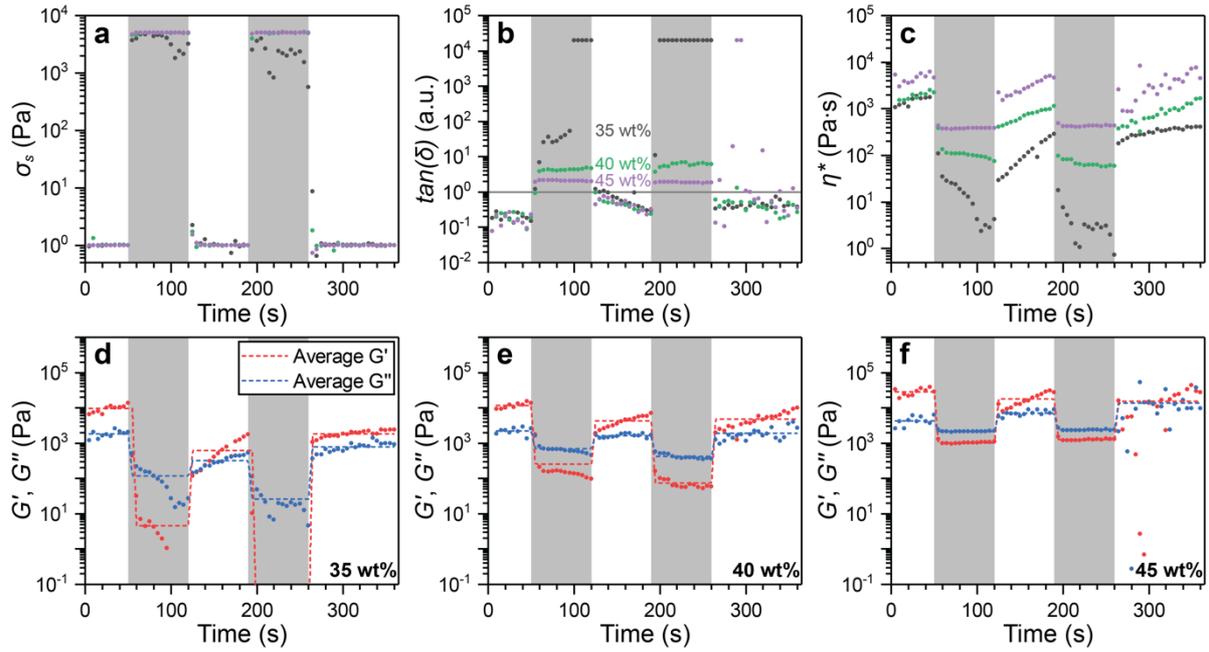

**Fig. S12:** Time-dependent oscillatory shear stress cycling for the PVDF-TrFE/acetone (35 wt%, 40 wt% and 45 wt%) inks. **a** The applied oscillatory shear stress ($\sigma_s$) with the high stress regions shaded grey. **b** The $tan(\delta)$ response to the shear stress with the liquid-solid transition point ($tan(\delta) = 1$) represented by a solid grey line. **c** The complex viscosity ($\eta^*$) response to the shear stress. **d-f** The storage ($G'$) and loss ($G''$) moduli response to the shear stress for **d** 35 wt% PVDF-TrFE/acetone ink, **e** 40 wt% PVDF-TrFE/acetone ink and **f** 45 wt% PVDF-TrFE ink, performed with $\omega$ at 1 Hz.

The temporal evolution of the $G'$ and $G''$ during the $\sigma_s$ cycling is shown in Fig. S12d-f for the three concentrations of PVDF-TrFE in acetone. For the 35 wt% ink (Fig. S12d), the $G'$ was observed to decrease rapidly as a function of time at $\sigma_s = 5$ kPa, with an average decrease over the timespan of greater than 1,000-fold. In the subsequent low-$\sigma_s$ period, the slope of $G'$ was higher than that of $G''$; however, the $G'$ was unable to recover to the initial value of 9,600 Pa, reaching a maximum of 1,736 Pa. Throughout the second $\sigma_s = 5$ kPa cycle, the $G'$ for the 35 wt% ink decreased significantly to below 1 mPa and subsequently exhibited a significantly lower slope during recovery. Conversely, the 40 wt% (Fig. S12e) and 45 wt% (Fig. S12f) PVDF-TrFE inks were stable under high-$\sigma_s$ for at least one cycle and showed considerably higher $G'$ recovery relative to the PVDF-TrFE (35 wt%) ink, from initial values of 11,800 Pa and 28,300 Pa, to final maxima of 7,000 Pa and 27,500 Pa, respectively. While the PVDF-TrFE (40 wt%) ink exhibited similar characteristics throughout the second $\sigma_s = 5$ kPa cycle (Fig. S12e), the PVDF-TrFE (45 wt%) ink was unable to consistently recover to initial values (Fig. S12f). Therefore, PVDF-TrFE (40 wt%) ink was selected for further experiments involving the incorporation of $Ti_3C_2T_x$ nanosheets.

# Properties of the SEA extrusion printed Ti$_3$C$_2$T$_x$/PVDF-TrFE films

## Mechanical properties

The tensile mechanical properties of the SEA extrusion printed Ti$_3$C$_2$T$_x$/PVDF-TrFE films were measured by a dynamic mechanical tester (ElectroForce 5500, TA Instruments). Samples, with a length of 27 mm and a width of 5 mm (Fig. S13a), were secured in grips by friction adhesive, with the distance between grips set at 5 mm (Fig. S13b). The width ($w$) and thickness ($t$) of each sample is given in Table S1. The films were extended parallel to the printing axis at a rate of 0.01 mm s$^{-1}$. Notably, the instrument displacement limit was approximately 11 mm, significantly below the breaking strain of the sample.

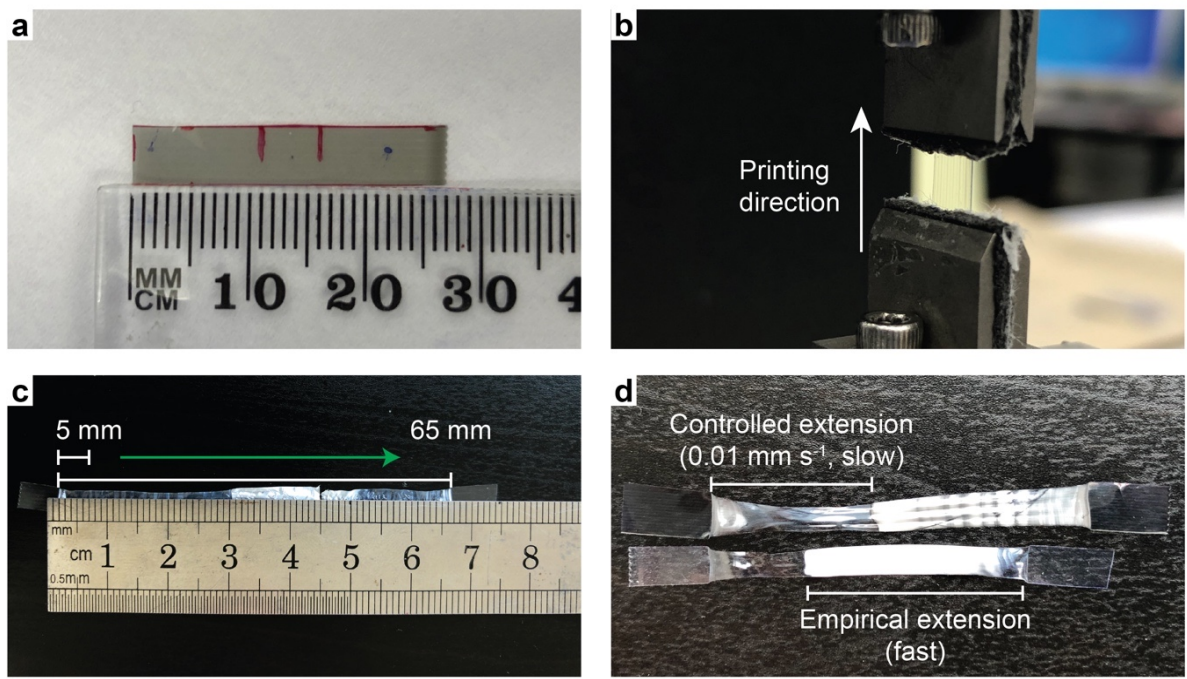

**Fig. S13:** The experimental layout of the mechanical testing. **a** Photograph of the geometry of the tested samples, with markings representing the position of the grips. **b** Photograph showing the sample inserted into the grips prior to extension. **c** Photograph showing the extent of extension required to break the SEA extrusion printed pristine PVDF-TrFE co-polymer film. **d** Photograph showing the dependence of transparency on the extension rate, shown for the SEA extrusion printed Ti$_3$C$_2$T$_x$/PVDF-TrFE (0.10 wt%) film (top) and SEA extrusion printed pristine PVDF-TrFE co-polymer film (bottom).

**Table S1:** Experimentally determined width (*w*) and thickness (*t*) for each sample measured during the tensile mechanical testing. The S values designate the sample number.

|  | Width (*w*, mm) | | | Thickness (*t*, μm) | | |
|---|---|---|---|---|---|---|
|  | S1 | S2 | S3 | S1 | S2 | S3 |
| $Ti_3C_2T_x$/PVDF-TrFE (0.00 wt%) | 4.3 | 5.0 | 4.9 | 53 | 53 | 59 |
| $Ti_3C_2T_x$/PVDF-TrFE (0.02 wt%) | 5.2 | 5.0 | 5.2 | 54 | 49 | 51 |
| $Ti_3C_2T_x$/PVDF-TrFE (0.10 wt%) | 5.1 | 5.1 | 5.2 | 45 | 46 | 47 |
| $Ti_3C_2T_x$/PVDF-TrFE (0.20 wt%) | 4.8 | 5.1 | 5.0 | 36 | 37 | 37 |
| $Ti_3C_2T_x$/PVDF-TrFE (0.50 wt%) | 5.0 | 5.0 | 5.0 | 35 | 35 | 35 |

The tensile strain ($\gamma_t$) was calculated from the data obtained during tests following Equation S1, where *L* is the displacement and $L_0$ is the distance between grips at the beginning of the test (5 mm):

$$\gamma_t = L/L_0 \quad (S1)$$

The tensile stress ($\sigma_t$) was calculated from the data obtained during tests, using the cross-sectional area of the sample ($A_{cs}$), following Equation S2:

$$\sigma_t = \frac{F}{A_{cs}} = \frac{F}{t \times w} \quad (S2)$$

Here, *F* is the measured force, *t* is the thickness of the sample, and *w* is the width of the sample (cut to approximately 5 mm).

Due to the low displacement limit of the instrument, the strain at break was approximated via empirical measurements (Fig. S13c), namely extending by hand. While these tests were merely representative, the samples were found to stretch to at least 65 mm prior to breaking, corresponding to 1,300% of the $L_0$ (5 mm). Additionally, the final transparency in the extended regions was observed to be higher when the extension rate was slower (Fig. S13d).

**Optical properties**

The optical properties of the extrusion printed $Ti_3C_2T_x$/PVDF-TrFE films were characterized by ultraviolet, visible and near-infrared (UV-vis-NIR) spectrophotometry (Lambda 950, Perkin Elmer). Spectra were obtained in the visible wavelength range (between 380 nm and 780 nm) with a step size of 5 nm. The incident light ($T_1$) was captured with no sample in place. The samples were then placed into a custom-made holder and secured to the entry port of a 150 mm diameter integrating sphere to capture all non-absorbed light ($T_2$). The total transmittance ($T_t$) of the sample was then taken as the fraction between $T_2$ and $T_1$ (Fig. S14a). The scattered light intensity in the instrument ($T_3$) was measured without the sample, using a light trap directly in the path of the beam (with diameter corresponding to 2° of the integrating sphere). Finally, the sample scatter ($T_4$) was measured with the light trap in place and the sample covering the entry port of the integrating sphere. The diffuse transmittance ($T_d$) was calculated following Equation S3 (Fig. S14b) and the haze was calculated in accordance with the ASTM D1003 standard,[26] using Equation S4 (Fig. S14c).

$$T_d = (T_4 - T_3 T_t)/T_1 \tag{S3}$$

$$Haze\ (\%) = T_d/T_t \times 100 \tag{S4}$$

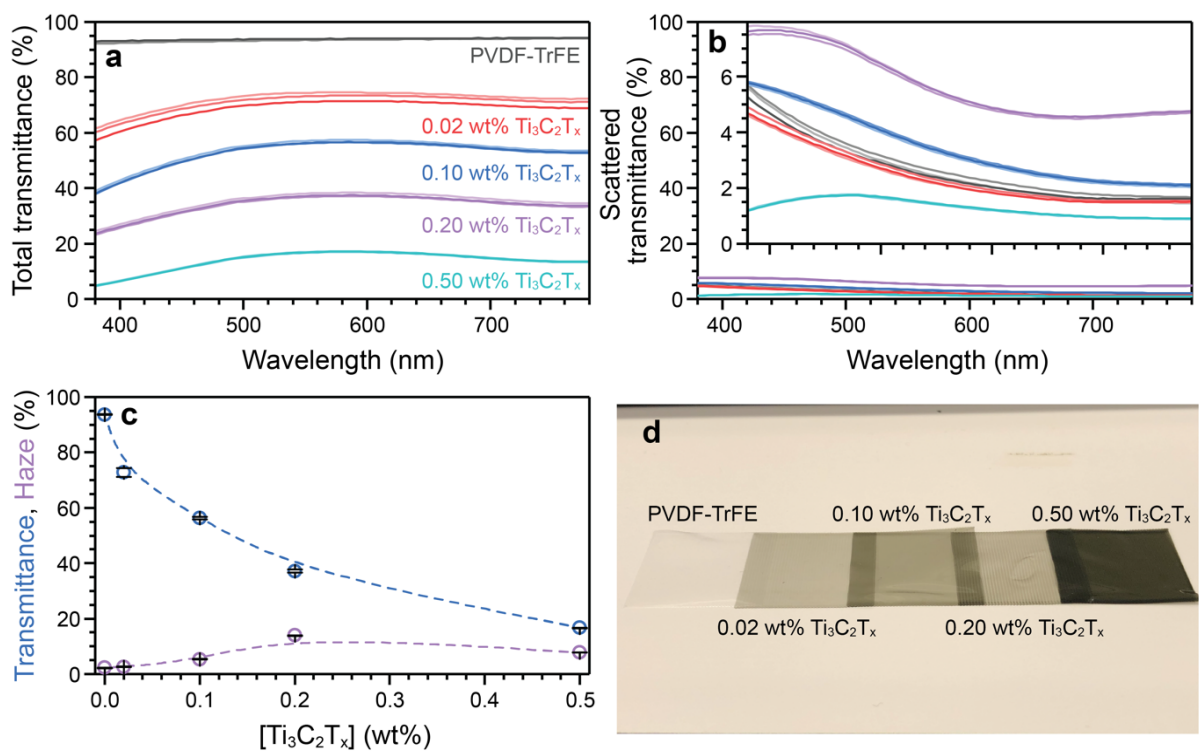

**Fig. S14:** Optical properties of SEA extrusion printed $Ti_3C_2T_x$/PVDF-TrFE (0.00 wt%, 0.02 wt%, 0.10 wt%, 0.20 wt% and 0.50 wt%) films. **a** The total light transmittance ($T_t$). **b** The scattered light transmittance ($T_d$). **c** The transmittance and haze as a function of $Ti_3C_2T_x$ nanosheet concentration. **d** Side by side photograph showing the color and transmittance of the films.

As expected, the transmittance was found to decrease as the $Ti_3C_2T_x$ nanosheet concentration increased, from 94% for pristine PVDF-TrFE co-polymer to 20% for the $Ti_3C_2T_x$/PVDF-TrFE (0.50 wt%) film (Fig. S14a,c,d). Surprisingly, the addition of $Ti_3C_2T_x$ nanosheets did not significantly increase the scattered light intensity and haze (Fig. S14b,c, Fig. 3e), with the maximum obtained haze found in the $Ti_3C_2T_x$/PVDF-TrFE (0.20 wt%) film to be 13.9%, compared to the pristine PVDF-TrFE co-polymer film (2.2%). In fact, the scattered light transmittance (Fig. S14b) was found to be lower at all wavelengths in the $Ti_3C_2T_x$/PVDF-TrFE (0.02 wt% and 0.50 wt%) films, compared to the pristine PVDF-TrFE film, suggesting the increase in haze with the incorporation of $Ti_3C_2T_x$ nanosheets was largely governed by the decrease in the transmittance, as opposed to the scattering from aggregated $Ti_3C_2T_x$ nanosheets.

**Raman analysis**

The evolution of the Raman spectra when the $Ti_3C_2T_x$ nanosheets are added to the PVDF-TrFE co-polymer reveals a clear suppression of out-of-plane vibrational modes occurring in $Ti_3C_2T_x$/PVDF-TrFE films. These modes occurring at 700 – 720 cm$^{-1}$ and 200 cm$^{-1}$ correspond to the out-of-plane $A_{1g}$ vibrational modes for oxygen functional groups bound to the $Ti_3C_2T_x$ lattice, whereas the peaks between 250 cm$^{-1}$ and 700 cm$^{-1}$ all correspond to in-plane $E_g$ vibrational modes.[27] Notably, the $A_{1g}$ modes at 700 – 720 cm$^{-1}$ disappear almost completely, even in the $Ti_3C_2T_x$/PVDF-TrFE (0.50 wt%) films, with no difference between solvent-cast and extrusion printed films (Fig. 4a). In contrast, the intensity of the higher energy $A_{1g}$ mode at 200 cm$^{-1}$ appears unchanged or even have an increased intensity relative to the main $E_{2g}$ modes (Fig. 4a). While this contrast in intensity change appears anomalous, it supports the data for well exfoliated flakes in literature.[27] Here, it should be noted that the Raman spectrum of the $Ti_3C_2T_x$ nanosheets was attained by drop-casting $Ti_3C_2T_x$ nanosheets in DMF on a silicon wafer, likely resulting in restacking and stronger $A_{1g}$ modes. The absence, or weak intensity, of these $A_{1g}$ modes in the $Ti_3C_2T_x$/PVDF-TrFE films therefore implies two key points, (1) that the PVDF-TrFE co-polymer is an excellent stabilizing agent for the $Ti_3C_2T_x$ nanosheets as there is no evidence of restacking; and (2) there is a strong binding between the PVDF-TrFE co-polymer and $Ti_3C_2T_x$ nanosheets (and subsequent polymer densification) such that the $A_{1g}$ modes are even further weakened and shifted.[28] These results confirm the strong electrostatic binding as predicted by MD simulations (Fig. 2a).

Raman mapping of the surface of the $Ti_3C_2T_x$/PVDF-TrFE films showed a significantly variable response in the $I_\beta/I_\gamma$ ratio (Fig. S15). This variation was most noticeable for the SEA extrusion printed $Ti_3C_2T_x$/PVDF-TrFE (0.02 wt%) film and decreased with an increased $Ti_3C_2T_x$ nanosheet loading up to 0.50 wt%, where the sample presented a homogenous ratio.

This improvement in sample homogeneity at higher $Ti_3C_2T_x$ nanosheet loadings is hypothesized to be due to the discrepancy in the state of the PVDF-TrFE co-polymer when it is bound to the $Ti_3C_2T_x$ nanosheet basal plane. At higher $Ti_3C_2T_x$ nanosheet loadings, we propose a high proportion of the PVDF-TrFE co-polymer is within the electrostatic sphere of influence (between 1 nm and 10 nm) of the $Ti_3C_2T_x$, thus presenting a homogenous $I_\beta/I_\gamma$.[29] The data from these maps was averaged and used for describing the average sample spectra and $I_\beta/I_\gamma$ (Fig. 4a,b). The solvent cast $Ti_3C_2T_x$/PVDF-TrFE film shows a higher variation in $I_\beta/I_\gamma$ (Fig. S14f) which is attributed to the lack of homogenization of 2D materials by the extrusion printing process.

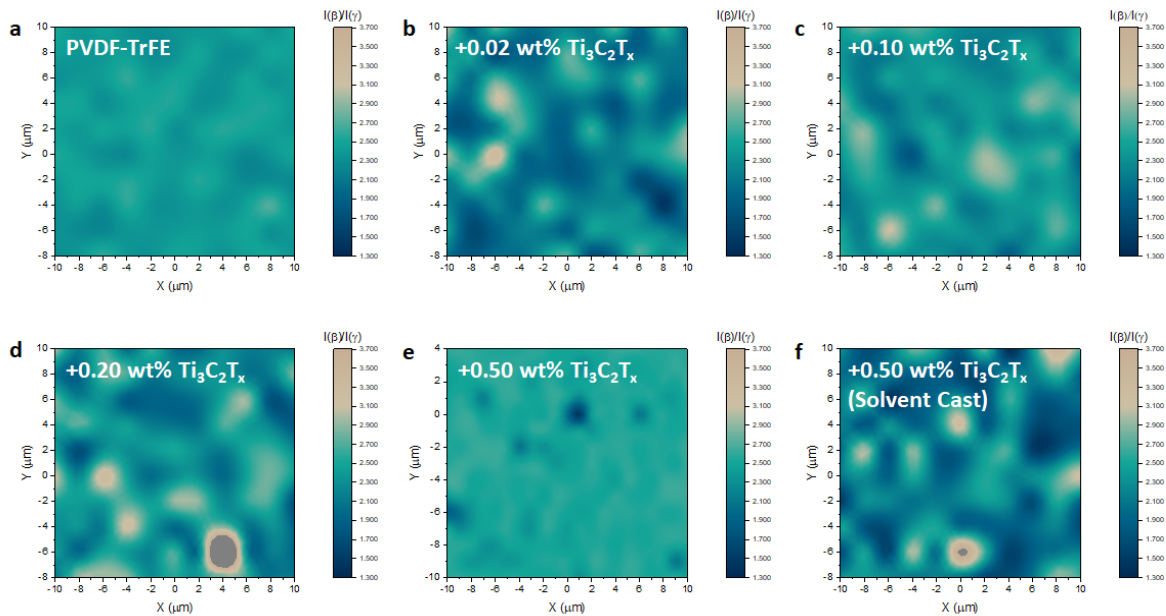

**Fig. S15:** The confocal Raman microscopy maps of the surface of the films, showing the intensity ratio ($I_\beta/I_\gamma$) between the β phase peak (842 cm$^{-1}$) and γ phase peak (811 cm$^{-1}$). **a** The SEA extrusion printed pristine PVDF-TrFE co-polymer film. **b** The SEA extrusion printed $Ti_3C_2T_x$/PVDF-TrFE (0.02 wt%) film. **c** The SEA extrusion printed $Ti_3C_2T_x$/PVDF-TrFE (0.10 wt%) film. **d** The SEA extrusion printed $Ti_3C_2T_x$/PVDF-TrFE (0.20 wt%) film. **e** The SEA extrusion printed $Ti_3C_2T_x$/PVDF-TrFE (0.50 wt%) film. **f** The solvent cast $Ti_3C_2T_x$/PVDF-TrFE (0.50 wt%) film.

**Attenuated total reflection Fourier transform infrared (ATR-FTIR) spectroscopy**

ATR-FTIR spectroscopy was performed on the samples using an ALPHA II spectrometer (Bruker). Absorbance spectra were collected by taking an average of 128 individual scans at a resolution of 1 cm$^{-1}$, between 600 cm$^{-1}$ and 4000 cm$^{-1}$.

ATR-FTIR spectroscopy was used to estimate the fraction of phases in the SEA extrusion printed Ti$_3$C$_2$T$_x$/PVDF-TrFE films, for Ti$_3$C$_2$T$_x$ nanosheet concentrations at 0.00 wt%, 0.02 wt%, 0.10 wt%, 0.20 wt% and 0.50 wt% (Fig. S16a). The peak commonly attributed to the α phase (766 cm$^{-1}$) was not distinctly visible in all the measured spectra, suggesting the low fraction of the α phase in the bulk of the samples.[17] Notably, as was determined by Raman microscopy (Fig. 4d,e), the α phase was present in close proximity to the Ti$_3$C$_2$T$_x$ nanosheet surface; however, the ATR-FTIR spectra suggested a low fraction of the α phase in the bulk of the Ti$_3$C$_2$T$_x$/PVDF-TrFE films. The peak at 840 cm$^{-1}$, indicative of the total electroactive phase (consisting of β and γ phases, denoted as β+γ), was present in all measured samples.[16] The separate peaks γ phase (1235 cm$^{-1}$) and β phase (1290 cm$^{-1}$) were both observed, confirming the presence of both electroactive phases (β+γ); however, the separate peaks could not be deconvoluted as the γ phase peak was present as a shoulder.

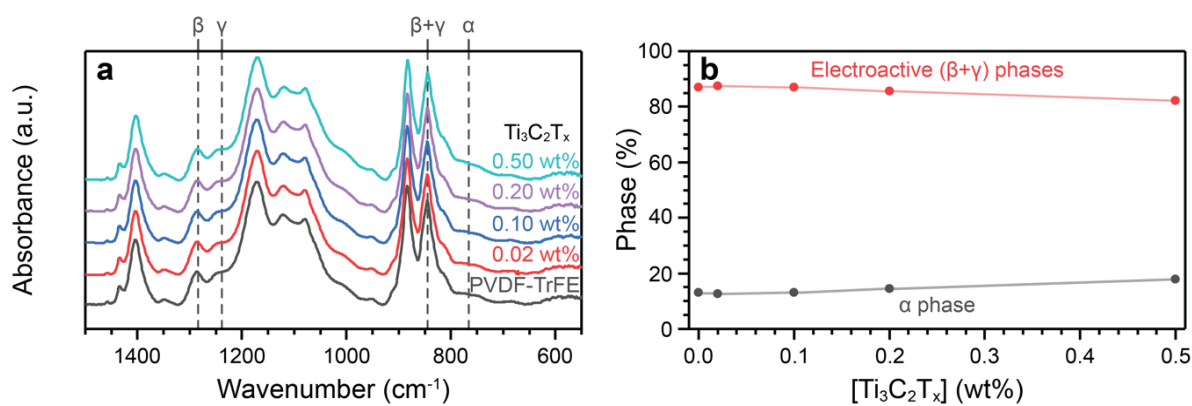

**Fig. S16:** ATR-FTIR spectroscopy characterization of SEA extrusion printed $Ti_3C_2T_x$/PVDF-TrFE (0.00 wt%, 0.02 wt%, 0.10 wt%, 0.20 wt% and 0.50 wt%) films. **a** The FTIR spectra, offset for clarity. **b** The relative fractions of the electroactive (β+ γ) phases and the α phase as a function of $Ti_3C_2T_x$ concentration.

The total electroactive phase fraction ($F_{ea}$) was calculated using ATR-FTIR data by Equation S5:

$$F_{ea}(\%) = \frac{I_{ea}}{\left(K_{ea}/K_{\alpha}\right)I_{\alpha} + I_{ea}} \tag{S5}$$

Here, $I_{ea}$ is the intensity of β+γ peak, $I_\alpha$ is the intensity of α peak, $K_\alpha$ and $K_{ea}$ are the absorption coefficients for the peaks at 766 cm$^{-1}$ and 840 cm$^{-1}$, with values of 6.1 x 10$^4$ cm$^2$ mol$^{-1}$ and 7.7 x 10$^4$ cm$^2$ mol$^{-1}$, respectively.[16] The $F_{ea}$ of the pristine PVDF-TrFE co-polymer film was 87.0% (Fig. S16b), significantly higher relative to the pristine PVDF-TrFE films SEA extrusion printed from a solvent mixture of DMF and acetone (40:60 vol%).[10] The highest $F_{ea}$ value was observed at 87.5% for the $Ti_3C_2T_x$/PVDF-TrFE (0.02 wt%) film, although this value exhibited little deviation from that of the pristine PVDF-TrFE film. Notably, at $Ti_3C_2T_x$/PVDF-TrFE (0.50 wt%), the $F_{ea}$ was found to decrease to 82.0%, consistent with the local α phase formation in the PVDF-TrFE on the surface of the $Ti_3C_2T_x$ nanosheets (Fig. 4e).

**X-ray powder diffractometry (XRD)**

XRD spectra were obtained for the SEA extrusion printed $Ti_3C_2T_x$/PVDF-TrFE films (0.00 wt%, 0.02 wt%, 0.10 wt%, 0.20 wt%, 0.50 wt%) using Bragg–Brentano geometry (D8 Advance, Bruker) using Cu-Kα radiation (λ = 1.54060 Å). The Bragg angle, 2θ, was varied between 5° and 70° with a step size of 0.02° and 1 s per step, with the sample rotated at 15 rpm. The films were placed on silicon low background holders for the measurements.

The phase distribution in the SEA extrusion printed films was analyzed with XRD (Fig. S17). Two main peaks were visible in the spectra. The broad peak at 18.1° was attributed to the (110/200) paraelectric γ phase reflection. The sharp peak at 20.2° corresponds to the (110/200) ferroelectric β phase reflection in PVDF-TrFE.[30] Due to the low concentration of $Ti_3C_2T_x$ nanosheets, no peaks were observed for the additive (Fig. S1a) in the spectra. Additional peaks were found at 35.0° and 40.8°, attributed to the (001) ferroelectric phase reflection and (111/201), (400/220) ferroelectric phase reflections, respectively, confirming the primary presence of the β phase.[31,32]

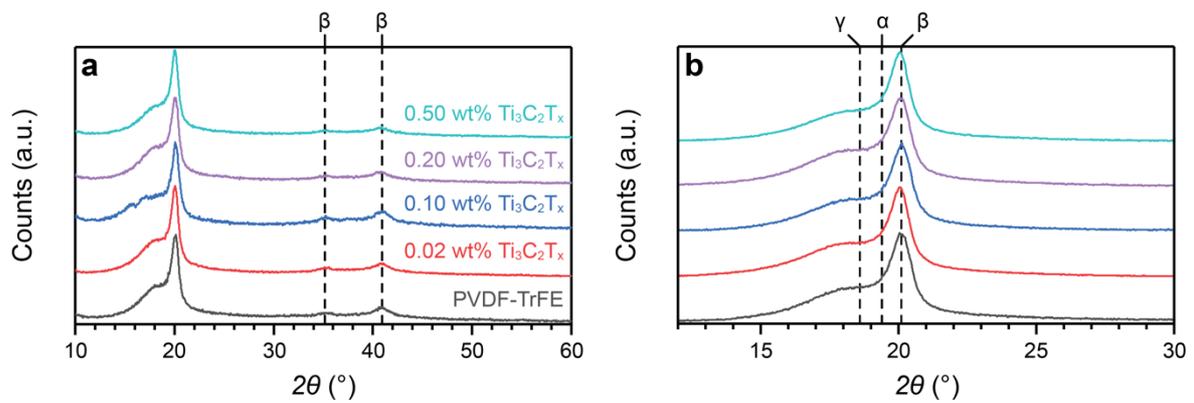

**Fig. S17:** XRD spectra of the SEA extrusion printed $Ti_3C_2T_x$/PVDF-TrFE films. **a** Survey scan. **b** Phase fingerprint region. Spectra offset for clarity.

The primary fingerprint region (Fig. S17b) region was further deconvoluted to investigate the distribution of phase fractions (Fig. S18). The region required four peaks to ensure the correct fit. The strongest peak was assigned to the β phase (blue), the broad second peak was assigned to the γ phase (purple), the third peak was attributed to the α phase (yellow) and the final peak corresponded to the amorphous regions of the polymer (gray).

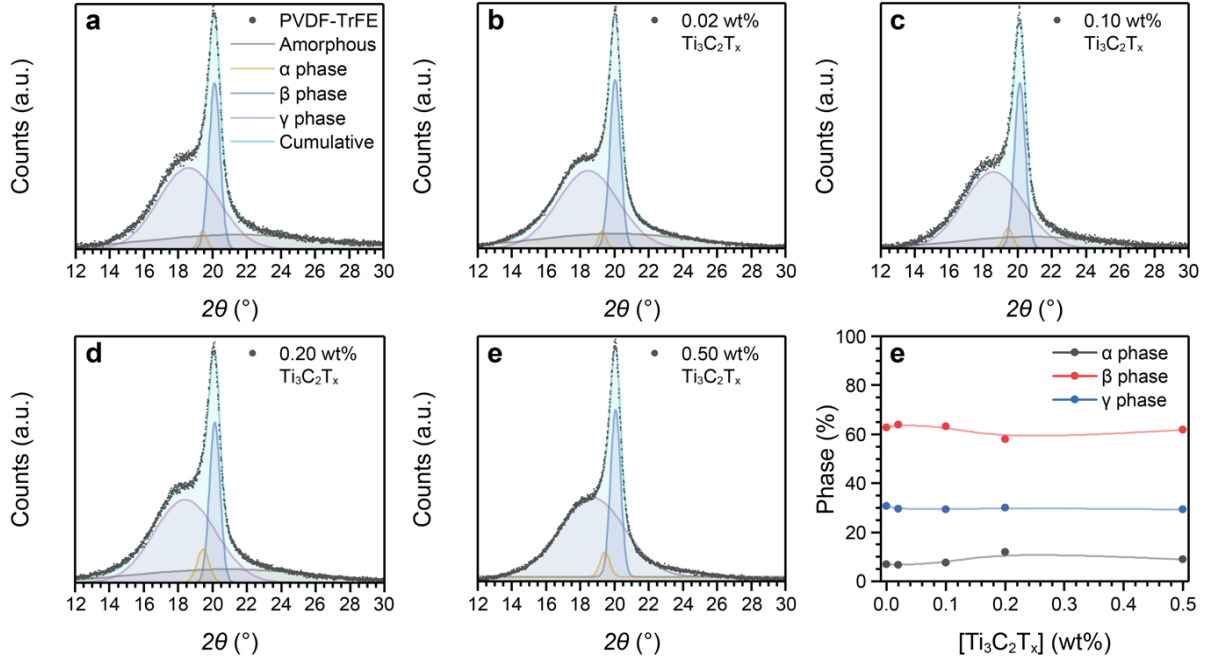

**Fig. S18:** XRD deconvolution for the phases in SEA extrusion printed films. **a** Pristine PVDF-TrFE copolymer. **b** Ti$_3$C$_2$T$_x$/PVDF-TrFE (0.02 wt%). **c** Ti$_3$C$_2$T$_x$/PVDF-TrFE (0.10 wt%). **d** Ti$_3$C$_2$T$_x$/PVDF-TrFE (0.20 wt%). **e** Ti$_3$C$_2$T$_x$/PVDF-TrFE (0.50 wt%). **f** Phase fractions determined from XRD as a function of Ti$_3$C$_2$T$_x$ nanosheet concentration.

The deconvoluted XRD spectra (Fig. S18a-e) were used to calculate the phase fractions (Fig. S18f) within the PVDF-TrFE co-polymer in the SEA extrusion printed Ti$_3$C$_2$T$_x$/PVDF-TrFE (0.00 wt%, 0.02 wt%, 0.10 wt%, 0.20 wt%, 0.50 wt%) films from the intensities for the respective peaks following Equations S6a-c:

$$F_\alpha = \frac{I_\alpha}{I_\alpha + I_\beta + I_\gamma} \tag{S6a}$$

$$F_\beta = \frac{I_\beta}{I_\alpha + I_\beta + I_\gamma} \tag{S6b}$$

$$F_\gamma = \frac{I_\gamma}{I_\alpha + I_\beta + I_\gamma} \tag{S6c}$$

Here, $I_\alpha$, $I_\beta$ and $I_\gamma$ correspond to the intensities for the peaks found at 19.4°, 20.2° and 18.1°, respectively. The phase distributions were found to correlate well with the data obtained from Raman spectroscopy (Fig. 4b) and FTIR spectroscopy (Fig. S16b). In particular, the $F_{ea}$ calculated from the FTIR spectra for the Ti$_3$C$_2$T$_x$/PVDF-TrFE films at Ti$_3$C$_2$T$_x$ nanosheet concentrations below 0.50 wt% (87%) matched closely with the sum of $F_\beta$ and $F_\gamma$ calculated from the deconvoluted XRD (between 85% and 90%). Furthermore, the Raman spectroscopy has suggested the primary phases in the bulk are the β and γ phases (Fig. 4a), with the β phase as the primary component, which is in close agreement with the XRD data.

**Differential scanning calorimetry (DSC)**

The crystallinity of the PVDF-TrFE co-polymer in the SEA extrusion printed $Ti_3C_2T_x$/PVDF-TrFE (0.00 wt%, 0.02 wt%, 0.10 wt%, 0.20 wt%, 0.50 wt%) films was measured by DSC (STA 5000, Perkin Elmer). The samples were heated from 25 °C in a ceramic sample pan (N5200040, Perkin Elmer) to 200 °C at a rate of 10 °C min$^{-1}$ under a nitrogen flow at 20 mL min$^{-1}$ (99.999% purity, BOC).

The DSC thermograms (Fig. S19) showed two endothermic peaks for all analyzed samples, centered at approximately 105 °C and 142 °C.[10] The peak at 105 °C corresponded to the ferroelectric to paraelectric transition (Curie temperature, $T_c$), whereby the samples exhibit piezoelectric properties below the $T_c$ and lose polarization above the $T_c$.[33] The primary peak at 142 °C corresponds to the melting of the polymer ($T_m$), with the enthalpy ($\Delta H_m$) correlating to the crystallinity ($\chi_c$) following Equation S7, where the enthalpy of melting for completely crystalline PVDF-TrFE ($\Delta H_0$) is given as 45 J g$^{-1}$, shown in Fig. 4c of the main text.[34,36]

$$\chi_c = \frac{\Delta H_m}{\Delta H_0} \qquad (S7)$$

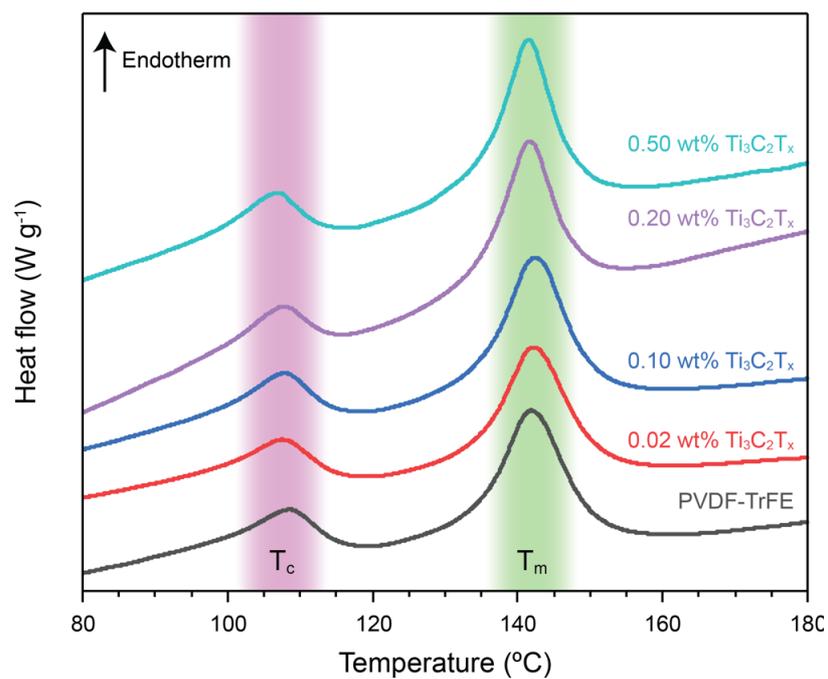

**Fig. S19:** DSC thermograms of the SEA extrusion printed $Ti_3C_2T_x$/PVDF-TrFE films, offset for clarity.

# Piezoresponse force microscopy (PFM) of Ti$_3$C$_2$T$_x$/PVDF-TrFE films

PFM was carried out on the Ti$_3$C$_2$T$_x$ nanosheets, SEA extrusion printed Ti$_3$C$_2$T$_x$/PVDF-TrFE (0.00 wt%, 0.02 wt%, 0.10 wt%, 0.20 wt% and 0.50 wt%) films and solvent cast Ti$_3$C$_2$T$_x$/PVDF-TrFE (0.00 wt% and 0.50 wt%) films in order to probe the trends in the out-of-plane polarization at the nanoscale. An atomic force microscope (Cypher ES, Oxford Instruments) with a high voltage accessory, equipped with a solid platinum cantilever (12Pt400A, Rocky Mountain Nanotechnology) was used for this testing.

The PFM of the Ti$_3$C$_2$T$_x$ nanosheets was carried out in dual AC resonance tracking (DART) mode at a bias of 1 V to confirm the absence of out-of-plane piezoelectricity. The scans were taken with 512 pixels per line and frequency at 0.5 Hz. To prepare the samples, the Ti$_3$C$_2$T$_x$ nanosheets were cast from solvent (DMF) onto gold (Au) coated silicon (Si) wafers and dried under vacuum. The DART-PFM scans were taken at three scales, 5 μm (Fig. S20a, d, g), 2 μm (Fig. S20b, e, h), and 1 μm (Fig. S20c, f, i and Fig. 5a). No discernible difference between the Ti$_3$C$_2$T$_x$ nanosheets (visible in the topography trace, Fig. S20a-c) and the Si substrate was found in the phase (Fig. S20d-f) or amplitude (Fig. S20g-i) traces. Therefore, the Ti$_3$C$_2$T$_x$ nanosheets were not observed to exhibit out-of-plane polarization and assumed as a non-piezoelectric additive in these experiments.

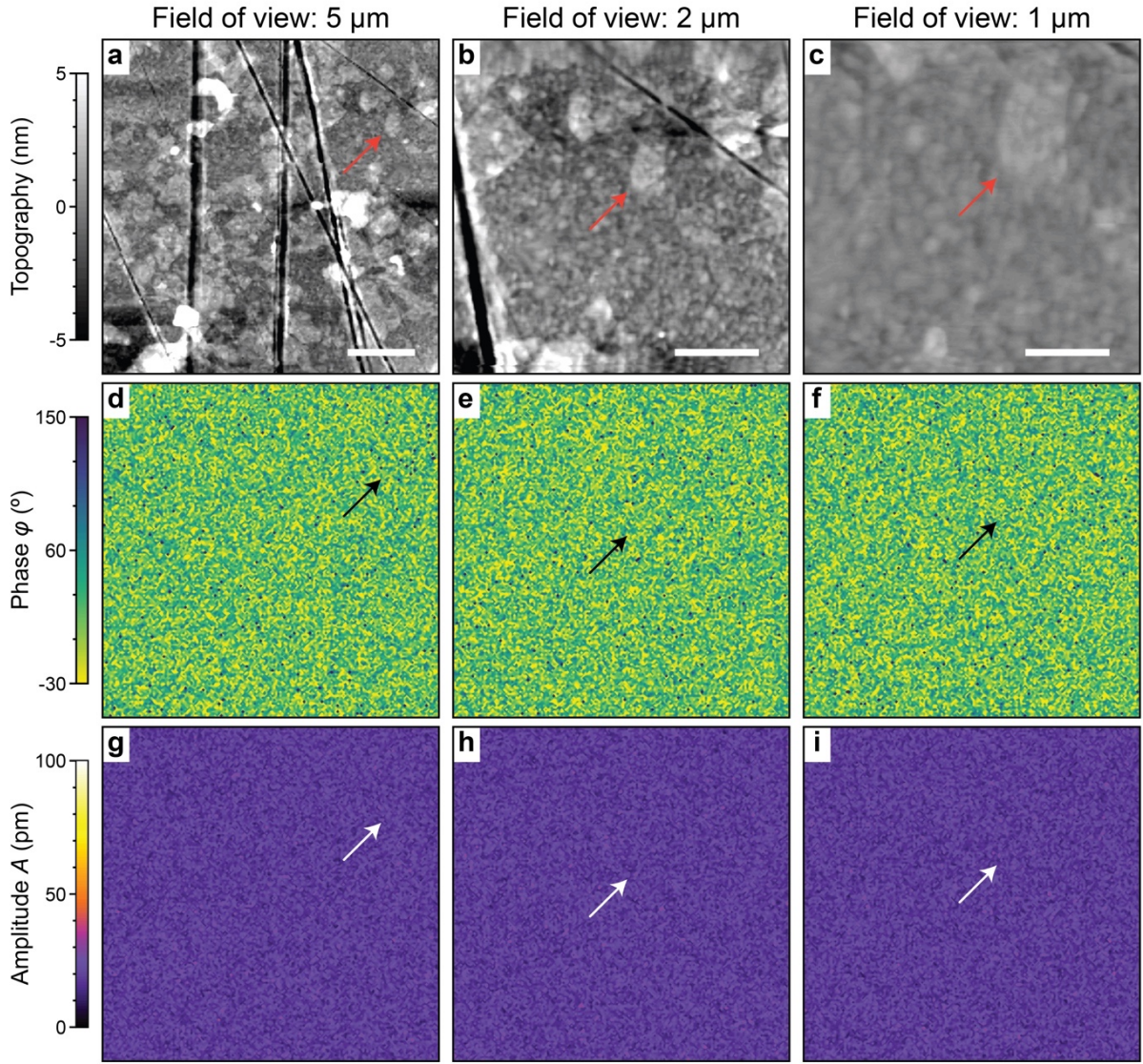

**Fig. S20:** DART-PFM maps of $Ti_3C_2T_x$ nanosheets adsorbed on a gold (Au)-coated silicon (Si) wafer. **a-c** The topography traces. **d-f** The piezoelectric phase traces. **g-i** The piezoelectric amplitude traces. **a, d, g** Field of view corresponding to 5 μm. Scale bar represents 1 μm. **b, e, h** Field of view corresponding to 2 μm. Scale bar represents 500 nm. **c, f, i** Field of view corresponding to 1 μm. Scale bar represents 250 nm. The arrow in each panel points to the location of the same $Ti_3C_2T_x$ nanosheet.

The PFM of the $Ti_3C_2T_x$/PVDF-TrFE films was carried out in lithography mode,[10] whereby a bias was applied to individual regions, monitoring the piezoelectric response through the converse piezoelectric effect ($\gamma_3 = d_{33}E_3$, whereby $\gamma_3$ is the out-of-plane strain, $d_{33}$ is the piezoelectric coefficient and $E_3$ is the out of plane electric field).[16,36,37] The lithography mode was chosen to obtain data below the poling field, where typical ferroelectric hysteresis loops cannot be formed, such that the poling state of the material would be minimally altered.[10] The applied voltage was between -20 V and +20 V, in increments of 2 V (Fig. S21), imaged over an area with lateral dimensions at 5 μm and 256 lines per scan, corresponding to a resolution of approximately 19 nm per pixel, with a scan rate of 0.2 Hz.

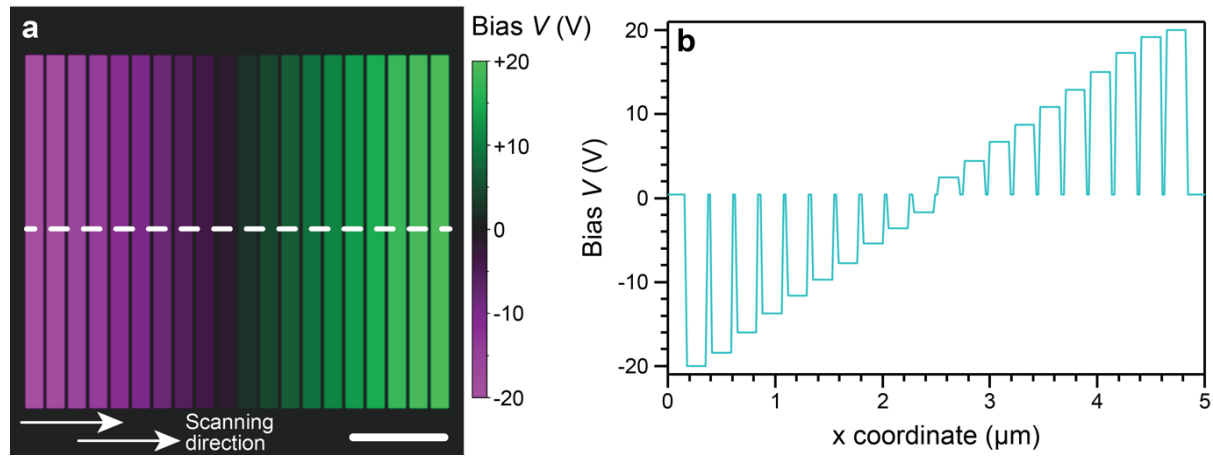

**Fig. S21:** Applied bias characteristics during the PFM measurements on the $Ti_3C_2T_x$/PVDF-TrFE films. **a** Map of applied bias. Scale bar represents 1 μm. **b** Bias as a function of the x coordinate for a single line in the scan direction, which is signified by the dashed line in **a**.

Each voltage was applied to the sample in a rectangular pattern, with the length at 200 pixels and width at 10 pixels, such that each scanned line contains the PFM data for all applied voltages (Fig. S21b). The measured data contained the topography, piezoelectric amplitude ($A$) and piezoelectric phase ($\varphi$). The data was obtained both for SEA extrusion printed (Fig. S22) and solvent-cast (Fig. S23) $Ti_3C_2T_x$/PVDF-TrFE films.

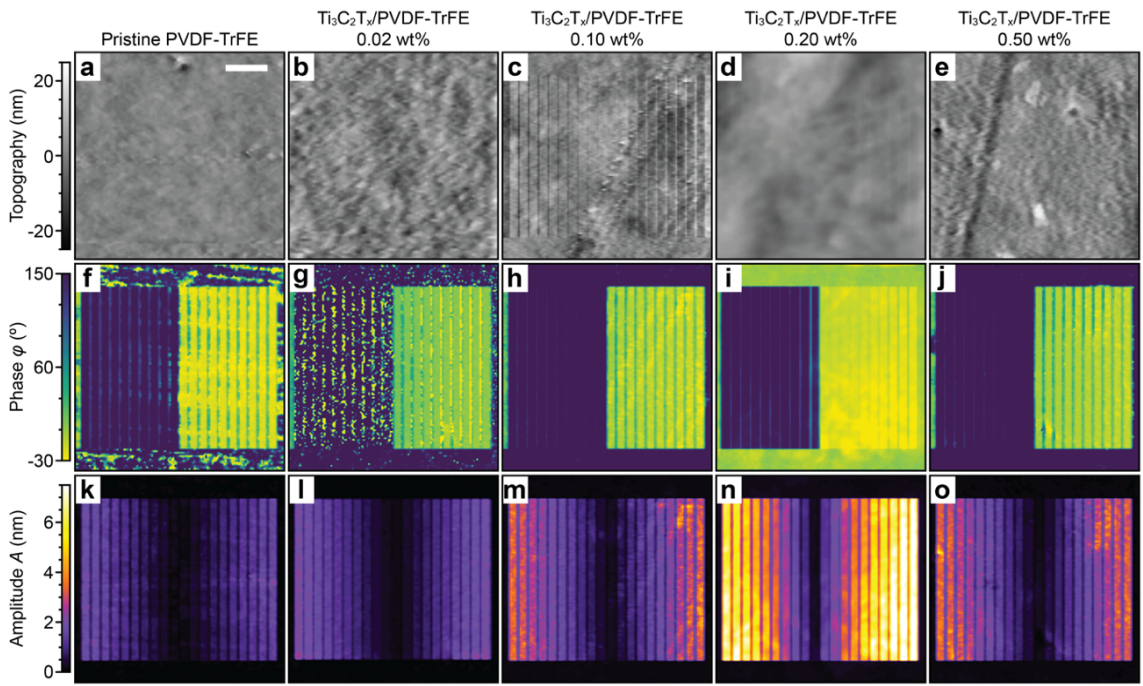

**Fig. S22:** Raw PFM data obtained for the SEA extrusion printed Ti$_3$C$_2$T$_x$/PVDF-TrFE films. **a-e** The topography. **f-j** The piezoelectric phase ($\varphi$). **k-o** The piezoelectric amplitude ($A$). The data was obtained for **a, f, k** pristine PVDF-TrFE (0.00 wt% Ti$_3$C$_2$T$_x$), **b, g, l** Ti$_3$C$_2$T$_x$/PVDF-TrFE (0.02 wt%), **c, h, m** Ti$_3$C$_2$T$_x$/PVDF-TrFE (0.10 wt%), **d, i, n** Ti$_3$C$_2$T$_x$/PVDF-TrFE (0.20 wt%), **e, j, o** Ti$_3$C$_2$T$_x$/PVDF-TrFE (0.50 wt%). The scale bar represents 1 μm.

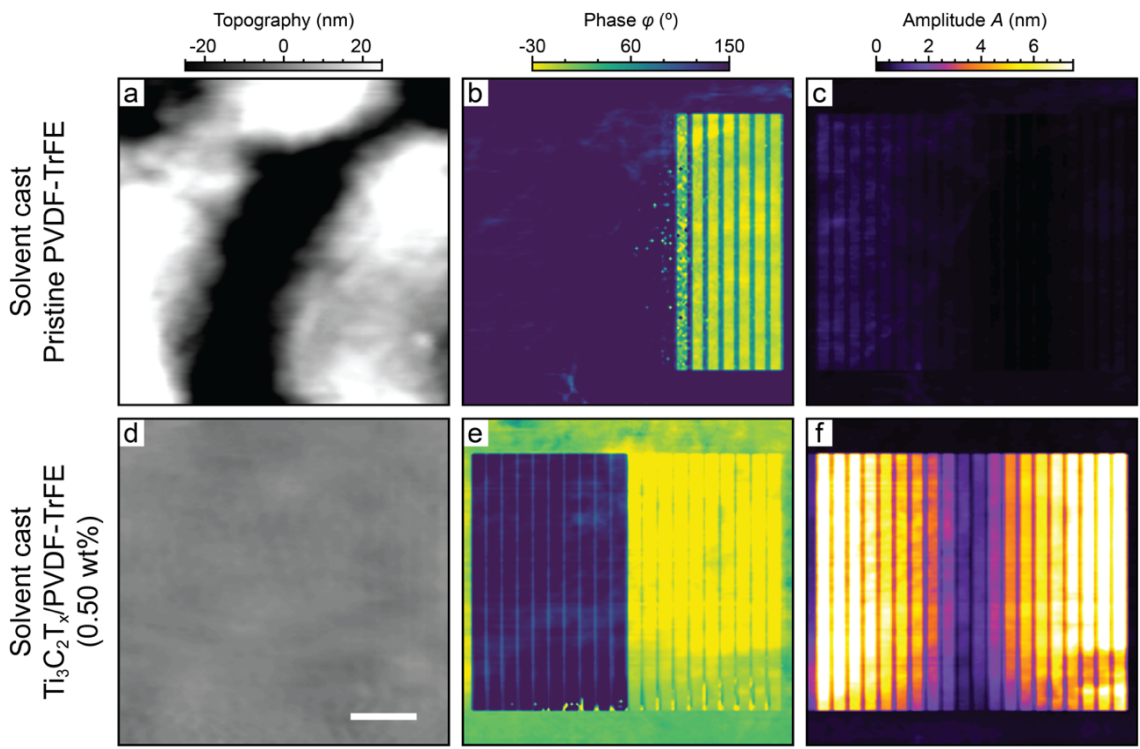

**Fig. S23:** Raw PFM data obtained for the solvent cast **a-c** pristine PVDF-TrFE and **d-f** Ti$_3$C$_2$T$_x$/PVDF-TrFE (0.50 wt%) films. **a, d** The topography. **b, e** The piezoelectric phase ($\varphi$). **c, f** The piezoelectric amplitude ($A$). The scale bar represents 1 μm.

The $A$ and $\varphi$ data (Fig. S22f-o, Fig. S23b, c, e, f) was processed using purpose-built Matlab code, which undertook pixel-by-pixel operations (Fig. S24) to form the data obtained in Fig. 5b by multiplying the $A$ signal (Fig. S24b) by the cosine of the $\varphi$ signal (Fig. S24a,c) and dividing by the Q factor of the cantilever ($Q_f$) for each applied bias (Fig. S24d). The $Q_f$ was measured in the tuning stage directly prior to the measurement (Table S2). To calculate the effective $d_{33}$, the $Acos(\varphi)/Q_f$ data was separated by the applied bias and averaged, obtaining a plot for $Acos(\varphi)/Q_f$ as a function of the applied bias for the samples (Fig. S24d).

**Table S2:** Experimentally determined Q factor ($Q_f$) for each PFM scan.

|  | Q factor ($Q_f$, a.u.) | |
| --- | --- | --- |
|  | SEA extrusion printed | Solvent-cast |
| Pristine PVDF-TrFE (0.00 wt% $Ti_3C_2T_x$) | 39.389 | 41.244 |
| $Ti_3C_2T_x$/PVDF-TrFE (0.02 wt%) | 36.083 | - |
| $Ti_3C_2T_x$/PVDF-TrFE (0.10 wt%) | 37.185 | - |
| $Ti_3C_2T_x$/PVDF-TrFE (0.20 wt%) | 73.754 | - |
| $Ti_3C_2T_x$/PVDF-TrFE (0.50 wt%) | 31.188 | 75.479 |

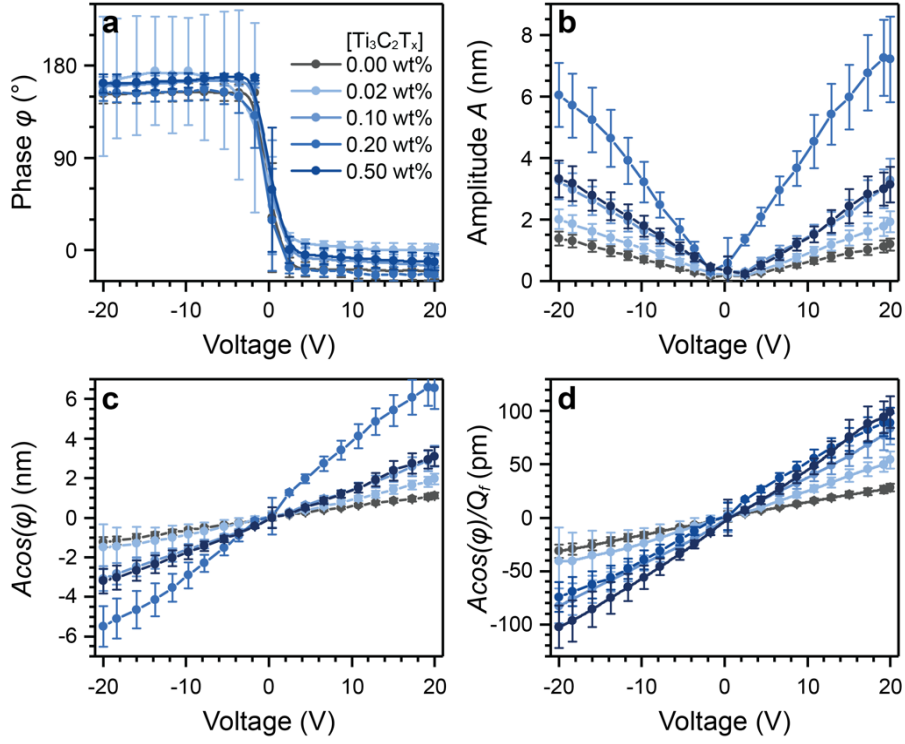

**Fig. S24:** Processed PFM results as a function of the input bias (*V*) for the SEA extrusion printed $Ti_3C_2T_x$/PVDF-TrFE films. **a** The phase ($\varphi$). **b** The amplitude (*A*). **c** The $Acos(\varphi)$. **d** The $Acos(\varphi)/Q_f$.

In the converse piezoelectric effect, the $d_{33}$ is given as shown in Equation S8a:[37]

$$d_{33} = \left(\frac{\partial \gamma_3}{\partial E_3}\right)^{\sigma} \tag{S8a}$$

Here, the superscript $\sigma$ denotes constant stress. In order to minimize the stress on the sample, the cantilever is required to possess a sufficiently low spring constant, in this case approximately 0.3 N m$^{-1}$. It should be noted, however, that the cantilever will nonetheless apply stress to the sample, therefore restricting the expansion in the measured material and providing an underestimate to the calculated $d_{33}$. The out-of-plane strain is given as $\gamma_3 = L/L_0$, whereby $L$ is the magnitude of the expansion or contraction, and $L_0$ is the material thickness. In PFM, $L$ corresponds to the normalized amplitude, shown in Equation S8b:

$$L = \frac{Acos(\varphi)}{Q_f} \tag{S8b}$$

Therefore, the out-of-plane strain then takes on the form shown in Equation S8c:

$$\gamma_3 = \frac{A\cos(\varphi)}{Q_f L_0} \tag{S8c}$$

Moreover, the $E_3$ is given as the $V$ applied per unit distance. In the case of PFM, as the material expands upon applied $V$, the expansion should include the distance of expansion, shown in Equation S8d:

$$E_3 = \frac{V}{L + L_0} \tag{S8d}$$

Hence, substituting Equation 8b for $L$, the expression becomes as shown in Equation S8e:

$$E_3 = \frac{V}{(A\cos(\varphi)/Q_f) + L_0} \tag{S8e}$$

Finally, substituting Equation S8c and Equation S8e into Equation S8a, it takes on the form as shown in Equation S8f:

$$d_{33} = \frac{A\cos(\varphi)}{Q_f V} + \frac{(A\cos(\varphi))^2}{Q_f^2 L_0 V} \tag{S8f}$$

Notably, this expression still does not account for the stress applied to the sample by the cantilever, however it becomes a more accurate equation to obtain the $d_{33}$. Nonetheless, when a 10 nm $A$ (1.00 x 10$^{-8}$ m) is observed at 180° $\varphi$ (corresponding to $\cos(\varphi) = 1$) in a sample with 40 μm (4.00 x 10$^{-5}$ m) $L_0$ under 20 V applied bias and $Q_f$ at 30, the additional term corresponds to an increase in the effective $d_{33}$ of 1.39 x 10$^{-4}$ pm V$^{-1}$, a minute increase relative to an effective $d_{33}$ at 16.67 pm V$^{-1}$. Therefore, for these experiments, the effective $d_{33}$ was taken as $A\cos(\varphi)/Q_f V$.

As the applied $V$ was below the poling voltage (maximum $V$ at 20 V, minimum thickness at 37 μm, corresponding to a maximum $E_3 = 0.54$ MV m$^{-1}$, minimum poling $E_3$ at 50 MV m$^{-1}$),[18] the slope of the plot for $A\cos(\varphi)/Q_f$ as a function of $V$ was expected to be linear for all samples.

Indeed, the data shown in Fig. S24d was found to be linear over the measured range, taking into account the deviation over the measured scan area. The data was fit with a linear trendline for each sample, whereby the slope of the trendline, accounting for the error in each sample, was the effective $d_{33}$ value.

The accuracy of the nanoscale polarization measurements via PFM has been widely debated in recent literature, demonstrating the values can underrepresent or overrepresent the macroscale $d_{33}$ both due to an empirical calculation methodology and the localized measurement approach.[36,38] Nonetheless, these PFM experiments are able to show trends of the $d_{33}$ in samples with similar composition, as has been demonstrated here. Notably, the most accurate methodology is to utilize a single cantilever, as has been undertaken in these experiments. The utilization of multiple cantilevers has the potential to vary in the spring constant and therefore dampen the amplitude signal, subsequently changing the observed trends.

## Macroscale energy harvesting characteristics

### Piezoelectric generator (PEG) fabrication

PEGs were produced by sputter coating a metal electrode layer on both sides of the SEA extrusion printed $Ti_3C_2T_x$/PVDF-TrFE (0.00 wt% and 0.50 wt%) films, attaching wires onto each electrode, and subsequently encapsulating the entire device in insulating adhesive. The sputter coated (Nanochrome, Intlvac) electrodes consisted of a seeding chromium (Cr) layer and a gold (Au) layer, with a total thickness of 60 nm, deposited onto both sides through a shadow mask. The two electrode layers were deposited sequentially, without outgassing the chamber. The chamber pressure was maintained at 2 mTorr, with an argon (Ar) gas flow of 20 $cm^3$ $min^{-1}$. The Cr layer deposition utilized an AC dual magnetron source at 500 W power, with a 10 s deposition time, corresponding to a Cr layer with thickness between 2 nm and 3 nm. The Au layer deposition utilized a DC magnetron source at 100 W power, with a 210 s deposition time, corresponding to an Au layer with thickness at 70 nm. The thickness of both layers was measured by AFM at 60 nm. The shadow mask was laser-cut from a 1 mm thick polymer sheet, pre-set for an opening with 1.5 cm length and 1.6 cm width, corresponding to an active area of 2.4 cm$^2$. Additional tabs were laser-cut adjacent to the width axis, with 0.5 cm length and 0.8 cm width, which were used for the attachment of wires and did not overlap between the two electrodes, therefore did not provide additional active area. The shadow mask design is shown in Fig. S25a and the sputter coated SEA extrusion printed Ti$_3$C$_2$T$_x$/PVDF-TrFE (0.50 wt%) films are shown in Fig. S25b.

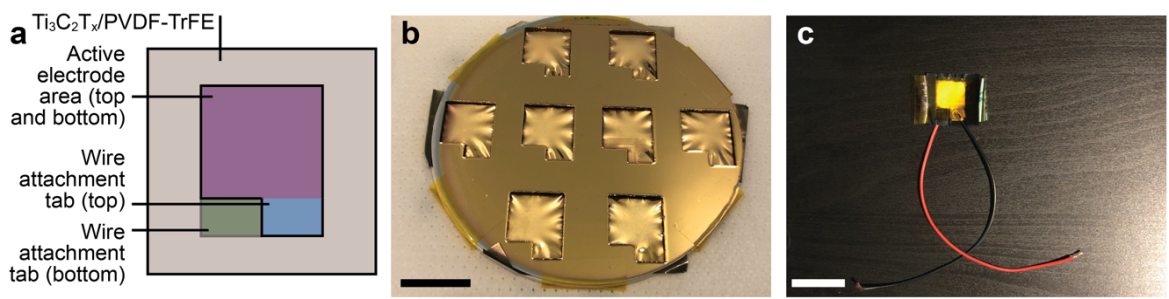

**Fig. S25:** Piezoelectric generator layout. **a** Schematic showing the layout of the electrodes and the wire attachment tabs on both surfaces of the Ti$_3$C$_2$T$_x$/PVDF-TrFE films. **b** Photograph of the layout of the laser-cut shadow mask, with sputter-coated SEA extrusion printed Ti$_3$C$_2$T$_x$/PVDF-TrFE (0.50 wt%) films as the active layer of the PEGs. Scale bar represents 2 cm. **c** Photograph showing the final SEA extrusion printed Ti$_3$C$_2$T$_x$/PVDF-TrFE (0.50 wt%) PEG. Scale bar represents 3 cm.

The wires (FLEXI-E 0.15, Stäubli Electrical Connectors AG) were cut to a length of 15 cm, with 0.6 cm stripped and exposed at each end. One end of each wire was soldered to the non-adhesive side of copper (Cu) foil adhesive (1181, 3M), after cleaning the surface with propan-2-ol. Subsequently, the adhesive side of the Cu foil adhesive was attached to the designated tabs (Fig. S25a) on each side of the sputter coated SEA extrusion printed Ti$_3$C$_2$T$_x$/PVDF-TrFE (0.00 wt% and 0.50 wt%) films. The final step in the PEG fabrication was the encapsulation in insulating polyimide (Kapton) adhesive, which was adhered to both sides of the film to ensure no external electrical influences affect the macroscale energy harvesting experiments. A photograph of the completed PEG is shown in Fig. S25c.

**Macroscale displacement field measurement under compressive stress**

The macroscale energy harvesting experiments were undertaken by the application of cyclic compression force and monitoring of the generated surface charge, configured to replicate the quasi-static Berlincourt method.[39] The cyclic compression was applied by a dynamic mechanical tester (ElectroForce 5500, TA Instruments) following a sinusoidal force pattern (Fig. S29a, b). The surface charge was measured and converted to a voltage signal by a charge amplifier (Nexus 2692, Brüel & Kjær). The resultant voltage signal was logged directly to file by a data acquisition instrument (9223, National Instruments). The PVDF-TrFE PEGs possess a large source impedance,[10] between 1 MΩ and 10 MΩ, therefore the majority of voltage measurement techniques will introduce error arising from the capacitance of the connection cables and the input impedance of the measurement device. For these experiments, the charge amplifier was chosen as it eliminates these errors.[40,41]

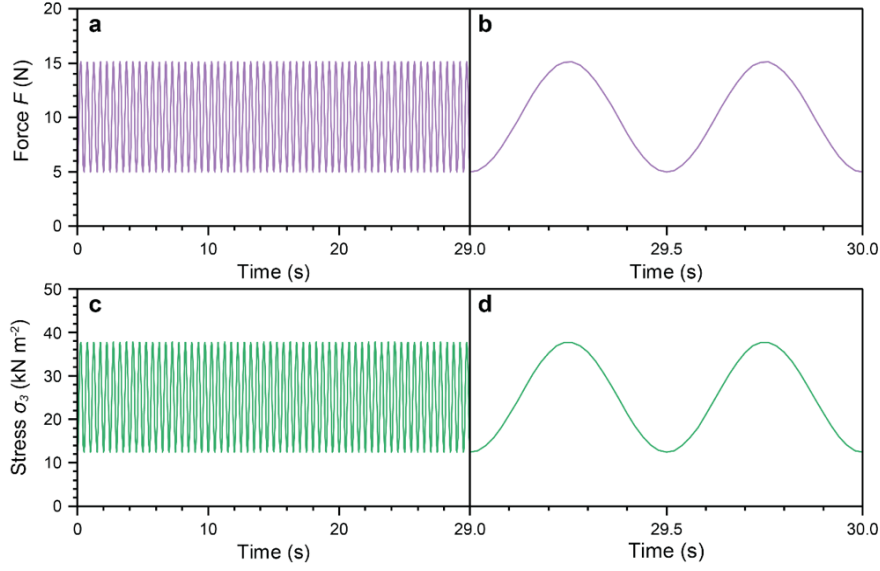

**Fig. S26:** Representative mechanical excitation input. **a** The input force ($F$) over 30 s with frequency at 2 Hz and **b** the enlarged view of the final 1 s. **c** The input stress ($\sigma_3$), calculated from **a**, with **d** the enlarged view of the final 1 s.

The generated surface charge as a result of input force follows the constitutive equation for the direct piezoelectric effect,[16,37] given in Equation S9:

$$D_3 = d_{33}\sigma_3 + \varepsilon_{33}^\sigma E_3 \tag{S9}$$

Here, $D_3$ is the electric displacement field, $d_{33}$ is the piezoelectric charge coefficient, $\sigma_3$ is the applied stress, $\varepsilon^\sigma_{33}$ is the dielectric permittivity at constant stress, $E_3$ is the electric field and the subscripts correspond to the directionality, in this instance all parallel to the thickness axis. Notably, in short circuit conditions where the input impedance of the load (in this instance the charge amplifier) is significantly lower than the output impedance of the PEG, the charge is transferred with no resistance.[41] In this instance, minimal voltage is generated and therefore $E_3 \approx 0$ V m$^{-1}$. Hence, the $\varepsilon^\sigma_{33}$ can be ignored and the expression takes on the form shown in Equation S10:

$$D_3 = d_{33}\sigma_3 \tag{S10}$$

The $d_{33}$ can then be directly calculated from the input stress and the resultant electric displacement field, as shown in Equation S11:

$$d_{33} = \left(\frac{\partial D_3}{\partial \sigma_3}\right)^E \tag{S11}$$

Here, the superscript $E$ denotes a constant electric field ($\partial E_3 = 0$). The stress is calculated from Equation S2 and is shown as a function of time in Fig. S26c, d. The load cell of the mechanical tester, used to apply the stress, was cylindrical with a radius ($r$) of 12.5 mm. The compressive area ($A_\sigma$) (Fig. S27, blue line, Fig. S28, blue dashed line) was calculated individually for the samples, as the placement of the sputter coated electrodes was variable relative to the dimensions of the SEA extrusion printed films. The height ($h$) from the top of the Cu adhesive foil to the top of the sample (Fig. S27) was measured (Table S3) and the $A_\sigma$ was calculated following Equation S12 for a circular segment:[42]

$$A_\sigma = \pi r^2 - \left( r \cos^{-1}\left(\frac{r-h}{r}\right) - (r-h)\sqrt{2rh - h^2} \right) \quad (S12)$$

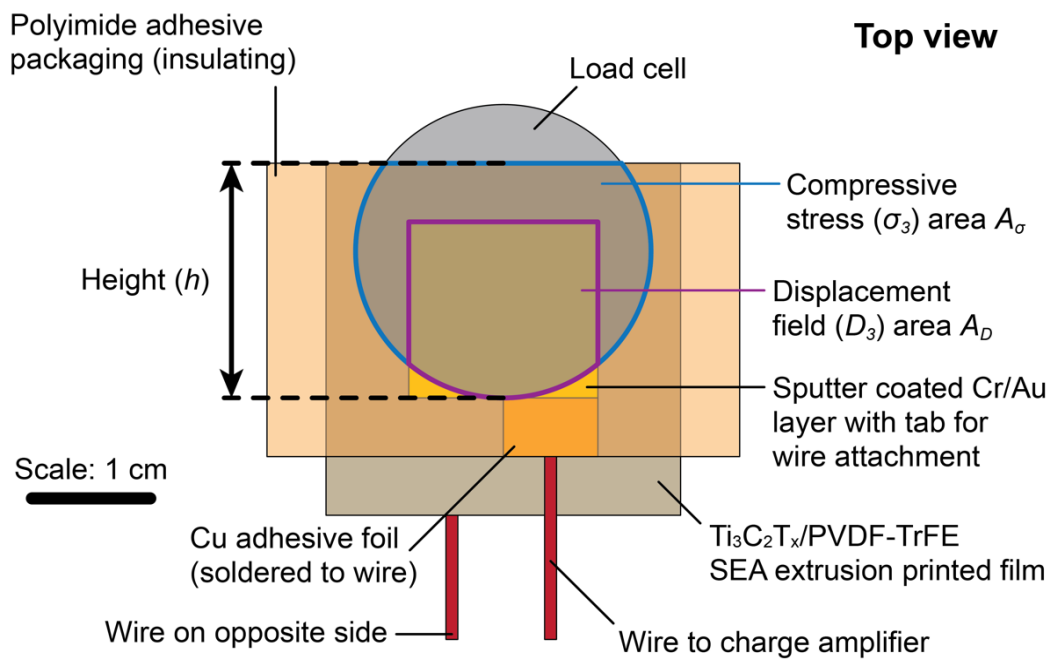

**Fig. S27:** Schematic showing the layout of the Ti$_3$C$_2$T$_x$/PVDF-TrFE PEG for energy harvesting tests.

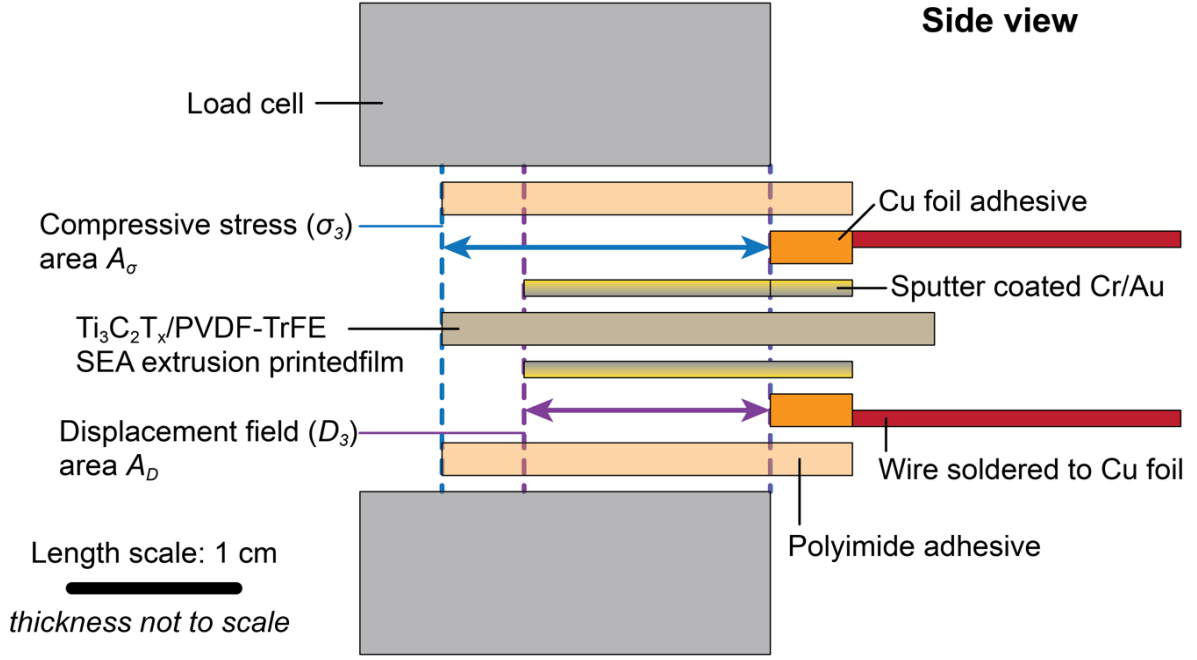

**Fig. S28:** Schematic showing the side view of the components of the Ti$_3$C$_2$T$_x$/PVDF-TrFE PEG during energy harvesting characterization.

**Table S3:** Values for the height ($h$) and the area under stress ($A_\sigma$) for energy harvesting experiments.

| Ti$_3$C$_2$T$_x$ nanosheet concentration in PVDF-TrFE | $h$ (mm) | $A_\sigma$ ($10^{-4}$ m$^2$) |
|---|---|---|
| 0.00 wt% | 6.0 | 4.00 |
| 0.50 wt% | 5.3 | 4.15 |

Similarly, the electric displacement field is a value normalized to the active area ($A_D$), shown in Equation S12:

$$D_3 = \frac{Q}{A_D} \tag{S12}$$

In this instance, the $A_D$ corresponded to the area with sputter coated electrodes on both sides, which is under impact (Fig. S27, purple line, Fig. S28, dashed purple line). The load cell was placed on the PEG such that the load cell did not make contact with the Cu foil adhesive (Fig. S27, Fig. S28). The $A_D$ was measured as 2.25 cm$^2$ (2.25 x $10^{-4}$ m$^2$), based on the active electrode dimensions at 15 mm length and 16 mm width, whereby the major part of the electrode was compressed (Fig. S27).

In order to ensure the generated charge arose only from the piezoelectric effect, the dependence of $D_3$ on $\sigma_3$ was investigated. In piezoelectric materials, as demonstrated in Equation S11, the slope must be linear, corresponding to the $d_{33}$. In the instances where the slope is not linear, either the constant $E$ requirement is not satisfied, or contributions from contact electrification and/or flexoelectricity are present.[43–46] A representative SEA extrusion printed $Ti_3C_2T_x$/PVDF-TrFE (0.50 wt%) PEG was tested for this dependence. The minimum force was set at 5 N, to minimize the effects from contact electrification. The resultant data is shown in Fig. S29. The slope was found to be linear, confirming the sole contribution of the direct piezoelectric effect to the measured surface charge.

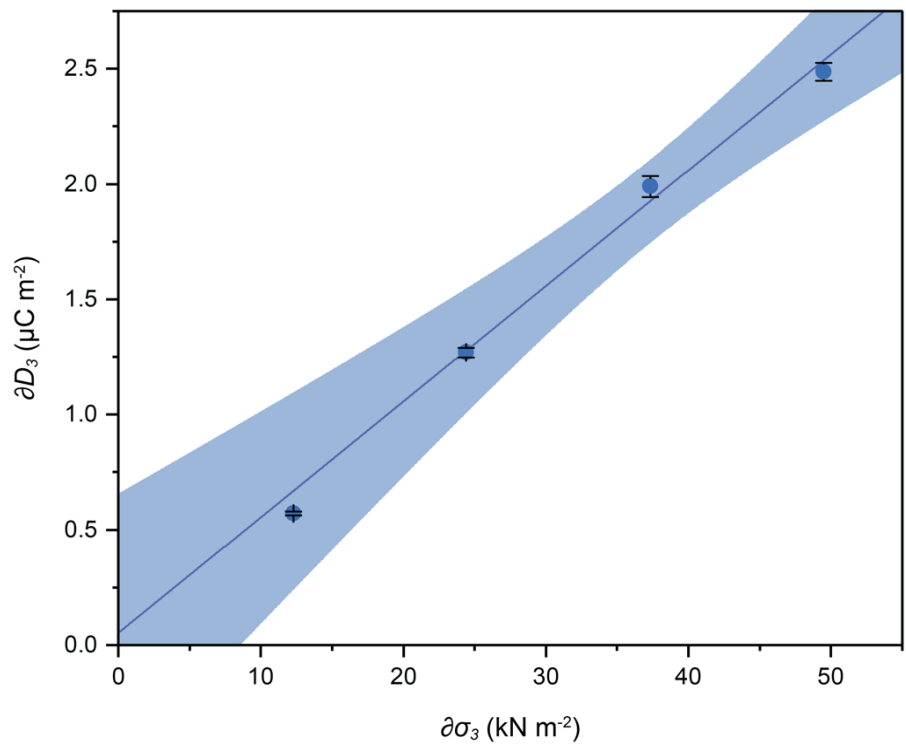

**Fig. S29:** The generated electric displacement field ($\partial D_3$) as a function of the input stress ($\partial \sigma_3$) for the SEA extrusion printed $Ti_3C_2T_x$/PVDF-TrFE (0.50 wt%) PEG. The solid line represents a linear fit to the data and the band represents a 95% confidence interval.

Upon the analysis of the generated surface charge ($Q$) in the SEA extrusion printed pristine PVDF-TrFE co-polymer PEG and the SEA extrusion printed $Ti_3C_2T_x$/PVDF-TrFE (0.50 wt%) PEG (Fig. 5d), it was evident that a significant enhancement in the energy harvesting was observed upon the incorporation of the $Ti_3C_2T_x$ nanosheets into the PVDF-TrFE co-polymer. In this instance, the $D_3$ is related to the polarization ($P_3$) following equation S13:[37]

$$D_3 = \varepsilon_{33}^{\sigma} E_3 + P_3 \qquad (S13)$$

Here, we were able to neglect the previously reported enhancements in the dielectric permittivity ($\varepsilon$) in $Ti_3C_2T_x$/fluoropolymer composites[44] due to the absence of external electric fields ($E \approx 0$ V m$^{-1}$), therefore the enhancements in $Q$ (and subsequently $D_3$) were attributed directly to enhancements in the $P_3$. The polarization is given as the sum of the individual dipole moment vectors ($\mu_3$) within a given volume ($V$), as shown in Equation S14, supporting the enhanced dipole moment alignment within the materials, as the dipole moment magnitude and the volume were constant.

$$P_3 = \frac{\sum \mu_3}{V} \qquad (S14)$$

The $d_{33}$ of the $Ti_3C_2T_x$/PVDF-TrFE (0.50 wt%) PEG (at -52.0 pC N$^{-1}$) was found to be higher than that of completely poled PVDF-TrFE in literature (at approximately -38 pC N$^{-1}$),[16,48] suggesting that the electrical poling technique commonly utilized in literature does not completely polarize the pristine PVDF-TrFE co-polymer.[18,49] The presence of dielectric breakdown at a high poling electric field strength is hypothesized as the limiting factor in achieving completely polarized PVDF-TrFE for their utilization as PEGs.[50] Overcoming the limitation posed by the dielectric breakdown has profound opportunities in a multitude of fields where piezoelectric materials are used. The dipole locking mechanism from a nanomaterial template, as described in this study, has tremendous potential to unlock new applications for flexible piezoelectric materials, where the cost and energy input during manufacture is currently limiting commercial adoption.

**Measurement of dielectric properties**

The investigation of the dielectric properties of the SEA extrusion printed $Ti_3C_2T_x$/PVDF-TrFE was undertaken on the fabricated PEGs. An LCR meter (4284A, Hewlett Packard) was swept between 20 Hz and 1 MHz frequency at 0.5 V, with the probes connected directly to the wires of the PEG. Three individual PEG samples were measured at each $Ti_3C_2T_x$ nanosheet concentration. The measured capacitance ($C$) was normalized *via* Equation S15 to the thickness ($t$) and the overlapping electrode area ($A_D$) of the $Ti_3C_2T_x$/PVDF-TrFE film and the permittivity of free space ($\varepsilon_0 \approx 8.854 \times 10^{-12}$ F m$^{-1}$) in order to obtain the dielectric constant ($\varepsilon_r$):

$$\varepsilon_r = \frac{tC}{A\varepsilon_0} \tag{S15}$$

The resultant dependence of the $\varepsilon_r$ on the frequency is shown in Fig. S30a for the pristine PVDF-TrFE and the 0.50 wt% $Ti_3C_2T_x$/PVDF-TrFE films.

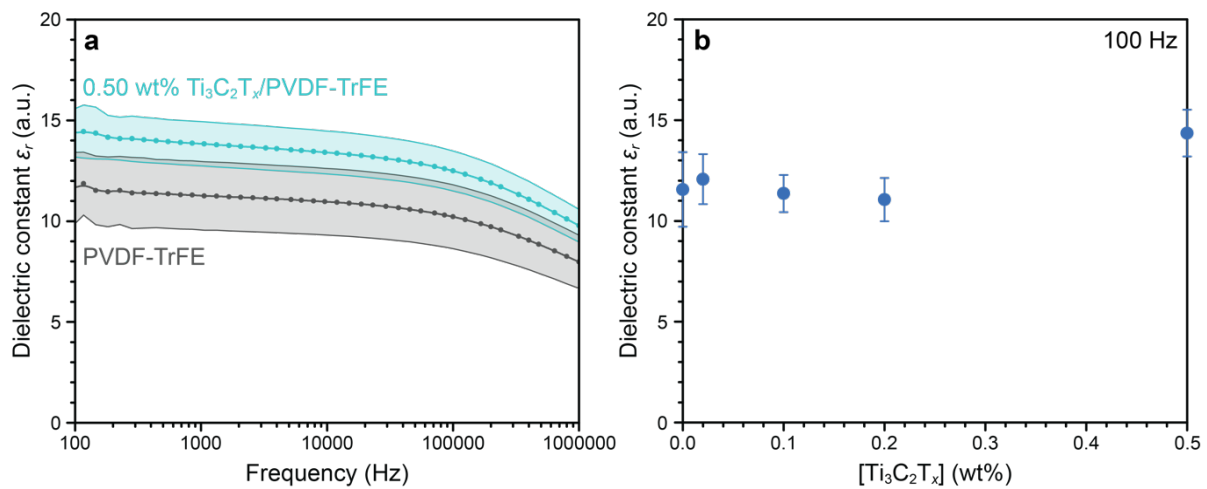

**Fig. S30:** The dielectric constant for the SEA extrusion printed $Ti_3C_2T_x$/PVDF-TrFE composites, shown as a function of **a** frequency and **b** the $Ti_3C_2T_x$ concentration (100 Hz frequency). The error bars were obtained by testing three separate films at each $Ti_3C_2T_x$ concentration.

The $\varepsilon_r$ was found to increase slightly between the pristine PVDF-TrFE film and the Ti$_3$C$_2$T$_x$/PVDF-TrFE film, exhibiting a similar response to an increasing frequency. Notably, the error between the samples was observed to overlap at all frequencies, therefore it was concluded that the increase was not significant. This is in accordance with the data recently presented by Tu et al.[47], demonstrating a 76% increase in the $\varepsilon_r$ at 3.5 wt% of Ti$_3$C$_2$T$_x$ nanosheets (nanosheet size between 1 μm and 2 μm) in poly(vinylidene fluoride-trifluoroethylene-chlorofluoroethylene) (PVDF-TrFE-CFE), a ter-polymer of the PVDF-TrFE co-polymer. Notably, the nanosheet dimensions in this study are significantly smaller in size (approximately 300 nm) and the maximum Ti$_3$C$_2$T$_x$ nanosheet concentration is significantly lower (0.5 wt%), thus the lower increase in the $\varepsilon_r$ of the composites presented in this manuscript is expected. Furthermore, the trend of the $\varepsilon_r$ with increasing Ti$_3$C$_2$T$_x$ nanosheet concentration (Fig. S30b) exhibits little correlation between the two parameters, and the $\varepsilon_r$ of the 0.20 wt% Ti$_3$C$_2$T$_x$/PVDF-TrFE film is equal to that of the pristine PVDF-TrFE film. This data confirms that the increase in the $d_{33}$ of the Ti$_3$C$_2$T$_x$/PVDF-TrFE composites does not arise from an increased $\varepsilon_r$, as discussion in the previous section.

**Piezoelectric voltage coefficient and piezoelectric figure of merit**

The measurement of the $d_{33}$ by the macroscale (Berlincourt) method and the determination of the $\varepsilon_r$ enables the subsequent calculation of the piezoelectric voltage coefficient ($g_{33}$) and consequently the piezoelectric figure of merit (FOM).[16,39] The $g_{33}$ is calculated following Equation S16, corresponding to the generated $Q$ data presented in Fig. 5d.

$$g_{33} = \frac{d_{33}}{\varepsilon^\sigma_{33}} = \frac{d_{33}}{\varepsilon_r \varepsilon_0} \tag{S16}$$

The average $g_{33}$ of the SEA extrusion printed pristine PVDF-TrFE PEG, measured from 60 individual compression cycles, was found to be 341 mV m N$^{-1}$. As expected, the partial polarization from the shear stress at the nozzle wall during the SEA extrusion printing process results in a $g_{33}$ value lower than that of literature reports for electrically poled PVDF-TrFE (approximately 380 mV m N$^{-1}$).[16] More importantly, the SEA extrusion printed 0.50 wt% Ti$_3$C$_2$T$_x$/PVDF-TrFE PEG exhibited a $g_{33}$ value of 402 mV m N$^{-1}$, larger in comparison to both the pristine PVDF-TrFE film prepared in this manuscript and the electrically poled PVDF-TrFE found in literature.

Moreover, the FOM for the SEA extrusion printed Ti$_3$C$_2$T$_x$/PVDF-TrFE was subsequently calculated using Equation S17:

$$\text{FOM} = d_{33} g_{33} = \frac{d_{33}^2}{\varepsilon_{33}^\sigma} \tag{S17}$$

The FOM was calculated for each of the 60 mechanical compression cycles corresponding to the $Q$ measurements in Fig. 5d, shown in Fig. S31 for the SEA extrusion printed PEGs containing 0.00 wt% and 0.50 wt% Ti$_3$C$_2$T$_x$ nanosheets, as well as the literature value for poled PVDF-TrFE films.

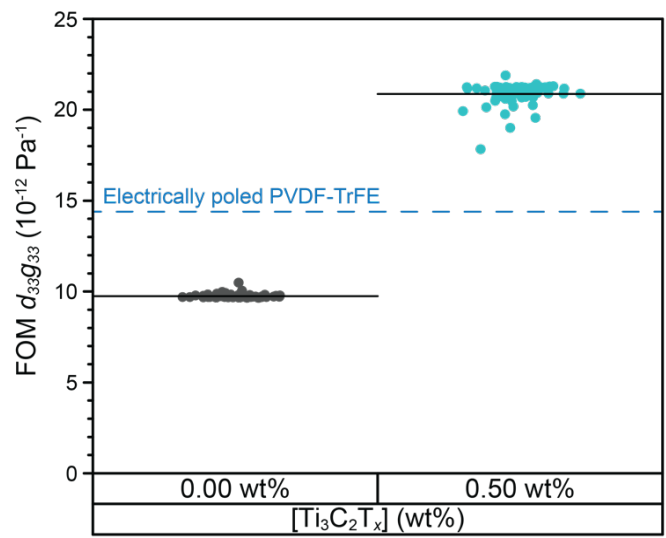

**Fig. S31:** The piezoelectric figure of merit (FOM) of the SEA extrusion printed Ti$_3$C$_2$T$_x$/PVDF-TrFE PEGs, demonstrated for the pristine PVDF-TrFE (0.00 wt% Ti$_3$C$_2$T$_x$, gray circles) and the 0.50 wt% Ti$_3$C$_2$T$_x$ (teal circles) PEGs. Each data point corresponds to one mechanical compression cycle for the data presented in Fig. 5d. The horizontal bars represent the average

for 60 compression cycles. The dashed blue line corresponds to the literature value for the FOM of electrically poled PVDF-TrFE.

The average FOM for the SEA extrusion printed pristine PVDF-TrFE PEG was calculated as 9.7 x $10^{-12}$ $Pa^{-1}$. As expected, the lower $d_{33}$ and $g_{33}$ of the pristine PVDF-TrFE PEG relative to literature values for electrically poled PVDF-TrFE (14.4 x $10^{-12}$ $Pa^{-1}$) results in a 33% lower FOM. Conversely, the FOM of the 0.50 wt% $Ti_3C_2T_x$/PVDF-TrFE PEG at 20.9 x $10^{-12}$ $Pa^{-1}$ is significantly higher (45%) than electrically poled PVDF-TrFE, which is largely attributed to the strong electrostatic interactions between the PVDF-TrFE co-polymer and the $Ti_3C_2T_x$ nanosheets. Further, the FOM of the 0.50 wt% $Ti_3C_2T_x$/PVDF-TrFE PEG is 115% higher than the SEA extrusion printed pristine PVDF-TrFE PEG. Unlike the pristine PVDF-TrFE PEG, which is restricted to the SEA extrusion printing process in order to exhibit partial polarization from induced shear stresses during deposition, the polarization-locked $Ti_3C_2T_x$/PVDF-TrFE solution possesses the flexibility for processing *via* conventional polymer film deposition techniques (e.g., solvent casting) while retaining the high piezoelectric properties (as demonstrated by PFM in Fig. 5c). Importantly, the FOM of the $Ti_3C_2T_x$/PVDF-TrFE PEG is 45% higher than electrically poled PVDF-TrFE, demonstrating a viable low energy deposition technique to produce effective flexible piezoelectric energy harvesting devices on a mass-produced scale.

The comparison of the polarization-locked $Ti_3C_2T_x$/PVDF-TrFE PEGs presented in this work to piezoelectric materials reported in literature (Table S4) demonstrates that the polarization-locked PEG possesses the highest $g_{33}$ and FOM values reported to date, taking into account the analysis of Deutz *et al.*[39] and van den Ende *et al.*[51], which investigate perovskite structure materials and polymer-perovskite composites, respectively.

Table S4: Comparison of the piezoelectric charge coefficient ($d_{33}$), piezoelectric voltage coefficient ($g_{33}$) and the piezoelectric figure of merit (FOM) between the SEA extrusion printed films presented in the manuscript and fluoropolymer-based materials found in literature for which the $d_{33}$ and $g_{33}$ was presented.

| Material | Processing | Poling | $|d_{33}|$ (pC N$^{-1}$) | $g_{33}$[†] (mV m N$^{-1}$) | FOM[‡] (x 10$^{-12}$ Pa$^{-1}$) | Ref. |
|---|---|---|---|---|---|---|
| **PVDF-TrFE** | **Extrusion printing** | None | 29 | 341 | 9.7 | This work |
| **0.50 wt% Ti$_3$C$_2$T$_x$/PVDF-TrFE** | **Extrusion printing** | None | 52 | 402 | 20.9 | This work |
| ZnO/PVDF* | Drop casting | None | 50 | 219 | 11.0 | 52 |
| MnO$_2$/PVDF | Electrospinning, hot pressing and rolling | 80 MV m$^{-1}$, 80 °C, 2 h | 38 | 318 | 12.1 | 53 |
| BCZT50/PVDF* | Melt mixing | 20 MV m$^{-1}$, 70 °C, 0.5 h | 27 | 114 | 3.1 | 54 |
| PVDF-TrFE | Spin coating | 200 MV m$^{-1}$ | 22 | 187 | 4.0 | 55 |
| PVDF | Solvent casting + double rolling | 170 MV m$^{-1}$, 70 °C, 1.2 h | 29 | 234 | 6.8 | 56 |
| MWCNT/PVDF | Solvent casting + double rolling | 100 MV m$^{-1}$, 70 °C, 1.5 h | 33 | 169 | 5.6 | 57 |
| PVDF-HFP | Blade coating | 0.5 MV m$^{-1}$, 100 °C, 0.5 h | 11 | 113 | 1.2 | 58 |
| PVDF-TrFE | Solvent casting | 105 MV m$^{-1}$, 100 °C | 23 | 236 | 5.4 | 59 |
| BT/PVDF-TrFE* | Solvent casting | 10 MV m$^{-1}$, 110 °C, 0.5 h | 34 | 37 | 1.3 | 60 |

[†] Calculated *via* Equation S16;
[‡] Calculated *via* Equation S17;
* Additive has been reported to exhibit piezoelectric properties;

# Supplementary videos

**Video S1:** Temporal evolution of the PVDF-TrFE co-polymer polarization vector direction adsorbed on a $Ti_3C_2T_x$ nanosheet substrate (left) and graphene substrate (right) over a 1.6 ns timespan. The PVDF-TrFE film consists of 70 co-polymer chains.

**Video S2:** Temporal evolution of the individual-chain polarization vector direction in the PVDF-TrFE co-polymer (5 chains shown from a total of 70 chains), adsorbed on a $Ti_3C_2T_x$ nanosheet substrate (left) and graphene substrate (right) over a 1.6 ns timespan.